%% file: main.tex
\documentclass[sigconf]{acmart}
\acmConference[ICSE '24]{International Conference on Software Engineering}{April 12--21, 2024}{Lisbon, Portugal}

\input{packages}
\input{macros}

\begin{document}
\title{Optimistic Prediction of Synchronization-Reversal Data Races}

\author{Zheng Shi}
\orcid{0000-0001-5021-7134}
\affiliation{%
  \institution{National University of Singapore}
  \city{Singapore}
  \country{Singapore}
}
\email{shizheng@u.nus.edu}

\author{Umang Mathur}
\orcid{0000-0002-7610-0660} 
\affiliation{%
  \institution{National University of Singapore}
  \city{Singapore}
  \country{Singapore}
}
\email{umathur@comp.nus.edu.sg}

\author{Andreas Pavlogiannis}
\orcid{0000-0002-8943-0722}
\affiliation{%
  \institution{Aarhus University}
  \city{Aarhus}
  \country{Denmark}}
\email{pavlogiannis@cs.au.dk}


\input{abstract}

\setlength{\belowcaptionskip}{-10pt}
\setlength{\textfloatsep}{15pt}

\maketitle

\input{intro}
\input{prelims}
\input{osr}
\input{osr-algo}

\input{experiments}

\input{related}
\input{conclusions}

\begin{acks}
This work is partially supported by the National Research Foundation, Singapore, and Cyber Security Agency of Singapore under its National Cybersecurity R\&D Programme (Fuzz Testing <NRF-NCR25-Fuzz-0001>) and by a research grant (VIL42117) from VILLUM FONDEN. Any opinions, findings and conclusions, or recommendations expressed in this material are those of the author(s) and do not reflect the views of National Research Foundation, Singapore, and Cyber Security Agency of Singapore.
\end{acks}

\clearpage
\newpage

\bibliographystyle{ACM-Reference-Format}
\bibliography{references}

\clearpage
\newpage

\appendix
\input{appendix}

\end{document}

%% file: packages.tex
\usepackage{amsmath}
\usepackage{amsfonts}
\usepackage{xcolor}
\usepackage{geometry} 
\usepackage[ruled,commentsnumbered,linesnumbered,noend]{algorithm2e}
\usepackage{amsthm}
\usepackage[inline]{enumitem}
\usepackage{longtable}
\usepackage{multirow}
\usepackage{graphicx}
\usepackage{float}
\usepackage{array} 
\usepackage{tikz}
\usepackage{subcaption}
\usepackage{xparse} 
\usepackage{booktabs}
\usepackage{balance}
\usepackage{url}

%% file: macros.tex
\theoremstyle{definition}
\newtheorem{theorem}{Theorem}[section]

\newtheorem{lemma}{Lemma}[section]

\newtheorem{definition}{Definition}

\newtheorem{example}{Example}

\newtheorem*{rep@theorem}{\rep@title}
\newcommand{\newreptheorem}[2]{%
\newenvironment{rep#1}[1]{%
 \def\rep@title{#2 \ref{##1}}%
 \begin{rep@theorem}}%
 {\end{rep@theorem}}}

\newreptheorem{theorem}{Theorem}
\newreptheorem{lemma}{Lemma}

\newcommand{\figlabel}[1]{\label{fig:#1}}
\newcommand{\figref}[1]{Figure~\ref{fig:#1}}

\newcommand{\seclabel}[1]{\label{sec:#1}}
\newcommand{\secref}[1]{Section~\ref{sec:#1}}
\newcommand{\exlabel}[1]{\label{ex:#1}}
\newcommand{\exref}[1]{Example~\ref{ex:#1}}
\newcommand{\deflabel}[1]{\label{def:#1}}
\newcommand{\defref}[1]{Definition~\ref{def:#1}}

\newcommand{\thmlabel}[1]{\label{thm:#1}}
\newcommand{\thmref}[1]{Theorem~\ref{thm:#1}}

\newcommand{\lemlabel}[1]{\label{lem:#1}}
\newcommand{\lemref}[1]{Lemma~\ref{lem:#1}}
\newcommand{\applabel}[1]{\label{app:#1}}
\newcommand{\appref}[1]{Appendix~\ref{app:#1}}
\newcommand{\algolabel}[1]{\label{algo:#1}}
\newcommand{\algoref}[1]{Algorithm~\ref{algo:#1}}
\newcommand{\tablabel}[1]{\label{tab:#1}}
\newcommand{\tabref}[1]{Table~\ref{tab:#1}}

\newenvironment{myparagraph}[1]{\vspace{0.05in}\noindent{\bf #1.}}{}

\newcommand{\set}[1]{\{#1\}}
\newcommand{\setpred}[2]{\{#1 \,|\, #2\}}
\newcommand{\tuple}[1]{\langle #1 \rangle}

\renewcommand{\emptyset}{\varnothing}
\newcommand{\s}[1]{\mathsf{#1}}
\newcommand{\proj}[2]{{#1}|_{#2}}

\newcommand{\lk}{\ell}
\newcommand{\rd}{\s{r}}
\newcommand{\wt}{\s{w}}
\newcommand{\acq}{\s{acq}}
\newcommand{\rel}{\s{rel}}

\newcommand{\trace}{\sigma}
\newcommand{\ev}[1]{\tuple{#1}}
\newcommand{\reordering}{\rho}
\newcommand{\traceOrder}{\s{tr}}
\newcommand{\thOrder}{\s{TO}}
\newcommand{\trAfter}{\leq_{\traceOrder}^{\trace}}
\newcommand{\trAfterReorder}{\leq_{\traceOrder}^{\reordering}}
\newcommand{\thAfter}{\leq_{\thOrder}^{\trace}}
\newcommand{\thAfterReorder}{\leq_{\thOrder}^{\reordering}}
\newcommand{\cfAfter}{\leq_{\s{CF}}^{\trace}}
\newcommand{\ThreadOf}[1]{\s{th}(#1)}
\newcommand{\cfwith}[2]{#1\bowtie#2}
\newcommand{\OpOf}[1]{\s{op}(#1)}
\newcommand{\prev}[2]{\s{prev}_{#1}(#2)}
\newcommand{\match}[2]{\s{match}_{#1}(#2)}
\newcommand{\op}{op}

\newcommand{\acqO}[1]{\s{acqO}(#1)}

\newcommand{\writeOp}[1]{\s{w}(#1)}

\newcommand{\readOp}[1]{\s{r}(#1)}

\newcommand{\lwsingle}[1]{\s{rf}_{#1}}
\newcommand{\lw}[2]{\s{rf}_{#1}(#2)}
\newcommand{\acqSubs}[2]{\s{acq}_#1(#2)}
\newcommand{\relSubs}[2]{\s{rel}_#1(#2)}
\newcommand{\acqUps}[2]{\s{acq}^#1(#2)}
\newcommand{\relUps}[2]{\s{rel}^#1(#2)}
\newcommand{\readSubs}[2]{\s{r}_#1(#2)}
\newcommand{\writeSubs}[2]{\s{w}_#1(#2)}

\newcommand{\NumEvents}{\mathcal{N}}
\newcommand{\NumThreads}{\mathcal{T}}
\newcommand{\NumLocks}{\mathcal{L}}
\newcommand{\NumVars}{\mathcal{V}}
\newcommand{\NumReads}{\mathsf{Reads}}
\newcommand{\NumWrites}{\mathsf{Writes}}
\newcommand{\NumAcqs}{\mathsf{Acq}}
\newcommand{\NumRels}{\mathsf{Rel}}

\newcommand{\events}[1]{\s{Events(#1)}}
\newcommand{\threads}[1]{\s{Threads(#1)}}
\newcommand{\vars}[1]{\s{Vars(#1)}}
\newcommand{\locks}[1]{\s{Locks(#1)}}

\newcommand{\acqs}[1]{\s{Acqs(#1)}}

\newcommand{\last}[1]{\s{last(#1)}}
\newcommand{\lastrel}[1]{\s{lastRel(#1)}}
\newcommand{\oacqs}[1]{\s{OAcqs(#1)}}

\newcommand{\TLC}[1]{\s{TRClosure(#1)}}

\newcommand{\SSP}[1]{\s{OLClosure(#1)}}

\newcommand{\lkfeas}{\mathsf{lockFeasible}}
\newcommand{\optgraph}{G^{\mathsf{Opt}}}
\newcommand{\optvertices}{V^{\mathsf{Opt}}}
\newcommand{\optedges}{E^{\mathsf{Opt}}}
\newcommand{\redgraph}{G^\mathsf{Abs}}
\newcommand{\redvertices}{V^\mathsf{Abs}}
\newcommand{\rededges}{E^\mathsf{Abs}}
\newcommand{\EIS}{\mathsf{EIS}}
\newcommand{\succOpt}[2]{{\mathsf{succ}}^{#1}_{#2}}
\newcommand{\rmq}[1]{\mathsf{rangeMin}(#1)}

\newcommand{\POrder}{<_P}

\newcommand{\rfposet}{\textsf{RF-poset}}
\newcommand{\IND}{\textsf{INDEPENDENT-SET(c)}}

\newcommand{\syncp}{\textsf{SyncP}\xspace}
\newcommand{\ssp}{\textsf{OSR}\xspace} 

\newcommand{\NP}{\textsf{NP}}

\newcommand{\tsan}{\textsc{ThreadSanitizer}\xspace}
\newcommand{\rapid}{\textsc{Rapid}\xspace}
\newcommand{\rvpredict}{\textsc{RVPredict}\xspace}
\newcommand{\raceinjector}{\textsc{RaceInjector}\xspace}
\newcommand{\SSPAlgo}{\textsf{OSR}\xspace}
\newcommand{\grpJava}{\texttt{Category-1}\xspace}
\newcommand{\grpC}{\texttt{Category-2}\xspace}

\input{exec-macros}

\input{pseudocode-macros}

%% file: exec-macros.tex
\newcommand{\execution}[2]{
\scalebox{0.85}{
  \begin{tikzpicture}%
    \foreach \x in {1,...,#1}
    \node[right] at (1.01*\x+0.3,0.25) {$t_{\x}$};
    \draw (1.01,0) -- (#1*1.01+1.1,0);%
    \pgfmathsetmacro{\y}{1};%
    #2%
    \draw (1.01,0) -- (1.01,-0.4*\y);%
    \draw (#1*1.01+1.1,0) -- (#1*1.01+1.1,-0.4*\y);%
    \foreach \x in {2,...,#1}
    \draw (1.01,-0.4*\y) -- (#1*1.01+1.1,-0.4*\y);%
  \end{tikzpicture}
}
}

\newcommand{\executionThin}[2]{
\scalebox{0.7}{
  \begin{tikzpicture}%
    \foreach \x in {1,...,#1}
    \node[right] at (1.01*\x+0.3,0.25) {$t_{\x}$};
    \draw (1.01,0) -- (#1*1.01+1.1,0);%
    \pgfmathsetmacro{\y}{1};%
    #2%
    \draw (1.01,0) -- (1.01,-0.4*\y);%
    \draw (#1*1.01+1.1,0) -- (#1*1.01+1.1,-0.4*\y);%
    \foreach \x in {2,...,#1}
    \draw (1.01,-0.4*\y) -- (#1*1.01+1.1,-0.4*\y);%
  \end{tikzpicture}
}
}

\newcommand{\figev}[2]{
\pgfmathsetmacro{\y}{\y+1};
\pgfmathsetmacro{\y}{\y-1};
\node [left] at (1.01,-0.4*\y)  {\pgfmathprintnumber{\y}};%
\node at (#1.01*1 + 0.6,-0.4*\y) {$ #2 $};%
\pgfmathsetmacro{\y}{\y+1};
}

%% file: pseudocode-macros.tex
\SetKwProg{myfun}{function}{}{}
\SetKwProg{myproc}{procedure}{}{}
\SetKwProg{myhandler}{handler}{}{}
\SetKwFunction{sspclosure}{ComputeOLClosure}
\SetKwFunction{absCycle}{cyclicAbstractGraph}
\SetKwFunction{iteratedrmq}{getSuccessors}
\SetKwFunction{detectinc}{incrementalRaceDetection}
\SetKwFunction{osr}{OSR}
\SetKwFunction{maxLB}{maxLowerBound}
\SetKwFunction{checkAbsDeadP}{CheckAbsDdlck}
\SetKwFunction{checkDeadlock}{checkDeadlock}
\SetKwFunction{rdhandler}{read}
\SetKwFunction{wthandler}{write}
\SetKwFunction{acqhandler}{acquire}
\SetKwFunction{relhandler}{release}
\SetKwInput{Input}{Input}
\SetKwInOut{Output}{Output}
\SetKw{Let}{let}
\SetKw{Break}{break}
\SetKw{NOT}{not}
\SetKw{declare}{declare}
\SetKw{report}{report}
\SetKw{exit}{exit}
\SetKwFor{Foreach}{for each}{}{}%
\SetKwFor{RepTimes}{repeat}{times}{end}
\DontPrintSemicolon



%% file: abstract.tex

\begin{abstract}
Dynamic data race detection has emerged as a key technique 
for ensuring reliability of concurrent software in practice.
However, dynamic approaches can often miss data races 
owing to non-determinism in the thread scheduler.
Predictive race detection techniques cater to this shortcoming
by inferring alternate executions that may expose data races
without re-executing the underlying program.
More formally, the dynamic data race prediction 
problem asks, given a trace $\trace$ of an execution of a concurrent
program, can $\trace$ be correctly reordered to expose a data race?
Existing state-of-the art techniques for data race prediction either do not scale
to executions arising from real world concurrent software,
or only expose a limited class of data races, such as those that can be exposed
without reversing the order of synchronization operations.

In general, exposing data races by reasoning about synchronization reversals
is an intractable problem.
In this work, we identify a class of data races, called 
Optimistic Sync(hronization)-Reversal races that can be detected
in a tractable manner and often include non-trivial data races that cannot
be exposed by prior tractable techniques.
We also propose a sound algorithm \ssp for detecting all optimistic sync-reversal
data races in overall quadratic time, and show that the algorithm
is optimal by establishing a matching lower bound.
Our experiments demonstrate the effectiveness of \ssp ---
on our extensive suite of benchmarks, \ssp reports the largest number of data races,
and scales well to large execution traces.
\end{abstract}

%% file: intro.tex

\section{Introduction}
\seclabel{intro}

Concurrency bugs such as data races and deadlocks often escape in-house testing
and manifest only in production~\cite{Chabbi2022,Sadowski2014},
making the development of  reliable concurrent software a challenging task.
Automated data race detection has emerged as a first line of
defense against undesired behaviors caused by data races,
has been actively studied over multiple decades,
and is also the subject of this paper.
In particular, our focus is on dynamic analyses,
which, unlike static techniques, are the preferred 
class of techniques for detecting data races for
industrial scale software applications~\cite{Sadowski2014}.

\input{figures/fig_ssp_race}

A dynamic data race detector observes an 
execution of a concurrent program $P$ and infers 
the presence of a data race by analysing the trace of the observed execution.
A key challenge in the design of such a technique is
sensitivity to non-deterministic thread schedules --- 
even for a fixed program input,
a data race may be observed under a very specific thread schedule, but not under
other thread schedules. 
This means that a simplistic race detector that, say, only checks
for two conflicting events appearing simultaneously in the execution trace,
is likely going to miss many bugs. 
This is where \emph{predictive analysis} techniques shine --- instead
of looking for bugs only in the execution that was observed,
they additionally also detect bugs in executions that,
while not explicitly observed during testing, can nevertheless be inferred
from the observed execution, without rerunning the underlying program $P$~\cite{said2011generating,huang2014maximal,smaragdakis2012sound,wcp,syncp,Roemer2018}.
Predictive techniques identify the space of
executions or \emph{reorderings} that can
provably be inferred from a given observed execution $\trace$, 
and then look for a reordering $\reordering$ in this space,
that can serve as a witness to a bug such as a data race.
Consider the execution $\trace_1$ in \figref{ssp-race-trace} consisting of events
$e_1, e_2, \ldots, e_{12}$ where $e_i$ denotes the $i^{\text{th}}$
event from the top.
The two write events on variable $x$, $e_1$ and $e_{12}$,
are far apart and not witnessed as a data race
in $\trace_1$.
However, the correct reordering $\reordering_1$ of $\trace_1$, in which
the two write events appear consecutively,
shows that it is nevertheless, a predictable data race of $\trace_1$.
Indeed any program $P$ that generates $\trace_1$ will also generate $\reordering_1$
albeit with a different thread interleaving.

In general, sound (no false positives) and 
complete (no false negatives) data race prediction is 
known to be an intractable problem~\cite{mathur2020complexity}.
Soundness is a key desired property,
since false positives need to be otherwise vetted manually, a task which is particularly 
challenging in the case of concurrent programs.
Consequently, many recent works counter the intractability
by proposing incomplete (but nevertheless sound) 
predictive race detection algorithms that work in polynomial time
and have high precision in practice.
The main contribution of this paper is a 
new race prediction algorithm \ssp that is sound,
has higher prediction power than prior algorithms and
achieves high scalability in practice.


The design of our algorithm \ssp stems from the observation that 
often, data races can be exposed only by inverting the relative order of 
(some pairs of) critical sections, or \emph{synchronizations}.
The data race $(e_1, e_{12})$ in \figref{ssp-race-trace}, for instance,
can in fact only be observed in correct reorderings that invert the order
of the two critical sections on lock $\lk$.
However, reversing synchronization (lock/unlock) operations in the reordering
can further force a reversal in the order in which memory access events must appear in the reordering,
and can be intractable to reason about~\cite{mathur2020complexity,syncp}.
This strong tradeoff between precision 
(obtained by virtue of reversing the order of many synchronization operations)
and performance
has materialized on both the extremes.
Algorithms such as those based on the happens-before 
partial order~\cite{pozniansky2003efficient,shb}
or the recently proposed \syncp~\cite{syncp} 
run in linear time
but fail to expose races that mandate
reasoning about synchronization reversals.
On the other extreme, methods that exhaustively search for reversals, 
either resort to expensive constraint solving~\cite{said2011generating,huang2014maximal}
or saturation style reasoning~\cite{m2,seqc},
and do not scale to long execution traces
observed in real world concurrent applications.
Our proposed algorithm \ssp aims to strike a balance ---
it is designed to \emph{\underline{o}ptimistically} reason about 
\emph{\underline{s}ynchronization} \emph{\underline{r}eversals},
and identifies those reversals that do not lead to the reversal of memory operations.
The pair $(e_1, e_{12})$ in \figref{ssp-race} is an example of a race
that \ssp reports.

\ssp reports all \emph{optimistic synchronization-reversal} races
in overall time $\widetilde{O}(\NumEvents^2)$,
spending $\widetilde{O}(\NumEvents)$ time for processing each event 
in the given execution trace $\trace$.
Here, $\NumEvents$ is the number of events in $\trace$ and $\widetilde{O}$
hides polynomial multiplicative factors due to number of locks
and threads which are typically considered constants.
In order to check for the absence of memory reversals,
\ssp constructs a graph (\emph{optimistic reordering graph}) of events and checks if it is acyclic.
Naively, such an acyclicity check would take
$\widetilde{O}(\NumEvents)$ time for every pair of conflicting events,
resulting in a total cubic running time. 
A key technical contribution of our work is to perform this check in
amortized constant time
by constructing a succinct representation of this graph,
called \emph{abstract} optimistic reordering graph, 
of constant size.
We show that this abstract graph preserves acyclicity, 
and can be constructed in an incremental manner in amortized constant time, 
allowing us to perform race prediction for the entire
input execution in overall quadratic (instead of cubic) time.
Finally, we show that the problem of checking the existence of an
optimistic sync-reversal race also admits a matching
quadratic time lower bound, thereby implying that our algorithm is optimal.

We implemented \ssp and evaluate its performance thoroughly.
Our evaluation demonstrates the 
effectiveness of our algorithm on a comprehensive suite of 153 Java and C/C++ benchmarks derived from real-world programs. 
Our results show \ssp has comparable scalability as linear time algorithms \syncp and WCP, 
while it reports significantly more races than the second most predictive one on many benchmarks,
confirming our hypothesis that going beyond the principle of synchronisation preservation allows us to discover significantly more races and with better performance.
\ssp, thus, advances the state-of-the-art in sound predictive race detection.

The rest of the paper is organized as follows.
In~\secref{prelims}, we discuss relevant background.
In~\secref{ssp}, we formally define the notion of optimistic sync-reversal races, 
and present our algorithm 
OSR for 
detecting all optimistic sync-reversal races
in~\secref{osr-algo}.
Our evaluation of \ssp and its comparison with other race prediction algorithms
is presented in~\secref{evaluation}. 
In~\secref{related} we discuss related work and conclude in~\secref{conclusions}.

%% file: figures/fig_ssp_race.tex

\begin{figure}[t]
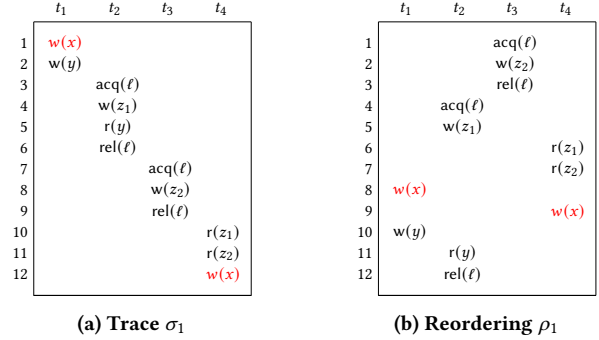

\centering
\begin{subfigure}[b]{0.45\columnwidth}
	\centering
  \executionThin{4}{
    \figev{1}{\textcolor{red}{w(x)}}
    \figev{1}{\wt(y)}
    \figev{2}{\acq(\lk)}
    \figev{2}{\wt(z_1)}
    \figev{2}{\rd(y)}
    \figev{2}{\rel(\lk)}
    \figev{3}{\acq(\lk)}
    \figev{3}{\wt(z_2)}
    \figev{3}{\rel(\lk)}
    \figev{4}{\rd(z_1)}
    \figev{4}{\rd(z_2)}
    \figev{4}{\textcolor{red}{w(x)}}
  }
	\caption{Trace $\trace_1$}
  \figlabel{ssp-race-trace}
\end{subfigure}  
\hfill
\begin{subfigure}[b]{0.45\columnwidth}
  \centering
  \executionThin{4}{
    \figev{3}{\acq(\lk)}
    \figev{3}{\wt(z_2)}
    \figev{3}{\rel(\lk)}
    \figev{2}{\acq(\lk)}
    \figev{2}{\wt(z_1)}
    \figev{4}{\rd(z_1)}
    \figev{4}{\rd(z_2)}
    \figev{1}{\textcolor{red}{w(x)}}
    \figev{4}{\textcolor{red}{w(x)}}
    \figev{1}{\wt(y)}
    \figev{2}{\rd(y)}
    \figev{2}{\rel(\lk)}     
  }
	\caption{Reordering $\reordering_1$}
    \figlabel{ssp-race-witness}
\end{subfigure}
\hfill
\caption{The two conflicting events $e_1 = \ev{t_1, \wt(x)}$ and $e_{12} = \ev{t_4, \wt(x)}$
is a \emph{predictable} data race of $\trace_1$ which is also an optimistic sync-reversal race, witnessed
by the \emph{correct reordering} $\reordering_1$
that reverses critical sections.
}
\figlabel{ssp-race}
\end{figure}

%% file: prelims.tex

\section{Preliminaries}
\seclabel{prelims}

In this section, we discuss preliminary notation
and the formal definition of the problem of dynamic data 
race prediction. 
Next, we 
briefly recall the notion of
\emph{sync-preserving} data races~\cite{syncp}
and discuss some of the limitations of this notion, 
paving the way to our algorithm \ssp.

\input{notations}

\input{syncp}


%% file: notations.tex


\myparagraph{Trace and events}
An execution trace (or simply trace) $\trace$ of a concurrent program is a sequence of events $\trace = e_1 e_2 \ldots e_\NumEvents$.
An event is a tuple $e = \ev{i, t, \op}$, 
where $i$ is a unique identifier for $e$, 
$t$ is the thread that performs $e$ and $\op$ is the operation corresponding to $e$;
often the identifier $i$ will be clear from context and we will drop it.
We use $\ThreadOf{e}$ and $\OpOf{e}$ to denote the thread and operation of $e$.
Operations are
$\rd(x)$, $\wt(x)$ (read or write access of memory location or variable $x$)
or
$\acq(\lk)$, $\rel(\lk)$ (acquire or release of lock $\lk$);
fork and join operations are omitted from presentation but not from our implementation.
For a trace $\trace$, we will use
$\events{\trace}$, $\threads{\trace}$, $\vars{\trace}$, $\locks{\trace}$
to denote respectively the set of all events, 
threads, variables and locks appearing in $\trace$.

\myparagraph{Well-formedness} 
We assume that traces are well-formed, in that they do not violate
\emph{lock semantics}.
In particular, for a well formed trace $\trace$, we require that
for each lock $\lk \in \locks{\trace}$, 
the sequence of operations on $\lk$ alternate between acquires and releases,
where each release event is preceded by a matching
acquire event of the same thread.
For an acquire (resp. release) event $e$,
we use the notation $\match{\trace}{e}$ to denote the matching 
release (resp. acquire) event of $e$ in $\trace$ if one exists;
otherwise we say $\match{\trace}{e} = \bot$.


\myparagraph{Trace order, thread order and reads-from} 
The trace order $\trAfter$ of a trace $\trace$ 
is the total order induced by the sequence of events in $\trace$,
i.e., $e_1 \trAfter e_2$ iff either $e_1 = e_2$ or $e_1$ appears earlier than $e_2$ in $\trace$.
The thread order $\thAfter$ is a partial order on $\events{\trace}$
such that for any two events $e_1, e_2$,
we have $e_1 \thAfter e_2 $ iff $\ThreadOf{e_1}$ = $\ThreadOf{e_2}$ and 
$e_1 \trAfter e_2$. 
%
When looking for predictable data races, 
we often look for reorderings of a given trace that preserve its control flow,
and determine this using the \emph{reads-from} function.
For a read event $r \in \events{\trace}$
with $\OpOf{r} = \rd(x)$ for some variable $x$, 
the \emph{writer} of $r$, denoted  
$w = \lw{\trace}{r}$ is the last write event on $x$
before $r$, i.e.,
$\OpOf{w} = \wt(x)$,
$w \trAfter r$ and 
$\neg (\exists w' \neq w, \OpOf{w'} = \wt(x) \land w \trAfter w' \trAfter r)$.
Without loss of generality, 
we will assume that $\lw{\trace}{e}$ is always defined for each read event $e$.
%
%
%
Given a set $S \subseteq \events{\trace}$, we say that $S$
is $(\thAfter, \lwsingle{\trace})$-closed if
\begin{enumerate*}[label=(\alph*)]
    \item for all events $e_1, e_2 \in \events{\trace}$
    if $(e_1 \thAfter e_2 \land e_2 \in S)$, then $e_1 \in S$, and
    \item for all events $\forall e_1, e_2 \in \events{\trace}$,
    if $(e_1 = \lw{\trace}{e_2} \land e_2 \in S)$, then $e_1 \in S$.
\end{enumerate*}
We use $\TLC{S}$ to denote the smallest set $S'$ such that $S \subseteq S'$ and $S'$ is $(\thAfter, \lwsingle{\trace})$-closed.

\myparagraph{Correct reordering}
Predictive race detection, given a trace $\trace$, asks if an 
alternate execution trace $\reordering$
witnesses a data race, and more importantly,
$\reordering$ can be \emph{inferred} from $\trace$.
The notion of correct reorderings precisely formalizes this.
Given well-formed traces $\trace$ and $\reordering$,
with $\events{\reordering} \subseteq \events{\trace}$,
we say that $\reordering$ is a \emph{correct reordering}
of $\trace$ if $\reordering$ respects the
thread order and reads-from relations of $\trace$.
This means that
\begin{enumerate*}
    \item $\events{\reordering}$ is $(\thAfter, \lwsingle{\trace})$-closed,
    \item for any two events $e_1, e_2 \in \events{\reordering}$,
    if $e_1 \thAfter e_2$, then $e_1 \thAfterReorder e_2$, and
    \item for any two events $e_1, e_2 \in \events{\reordering}$,
    if $e_1 =\lw{\trace}{e_2}$, then $e_1 =\lw{\reordering}{e_2}$.
\end{enumerate*}
 

\myparagraph{Data races and predictable data races} 
A pair of events $(e, e')$ in $\trace$ is said to be a conflicting pair,
denoted $\cfwith{e}{e'}$, if
both are access events to the same variable, and at least one of them is a write event,
i.e., $(\OpOf{e}, \OpOf{e'}) \in \set{(\wt(x), \wt(x)), (\wt(x), \rd(x)), (\rd(x), \wt(x))}$
for some $x \in \vars{\trace}$.
For a trace $\pi$ with $\events{\pi} \subseteq \events{\trace}$,
we say that event $e$ is $\trace$-enabled in $\pi$ 
if $e \not\in \events{\pi}$
but all thread-predecessors of $e$ are in $\pi$, i.e.,
$\setpred{e' \in \events{\trace}}{e' \neq e, e' \thAfter e} \subseteq \events{\pi}$.
A conflicting pair $(e, e')$ is said to be a \emph{data race}
of $\trace$ if there is a prefix $\pi$ of $\trace$ such that
both $e$ and $e'$ are $\trace$-enabled in $\pi$.
Finally, a conflicting pair $(e, e')$ is a 
\emph{predictable data race} of $\trace$ if there is a correct reordering
$\reordering$ of $\trace$ such that both $e$ and $e'$ are $\trace$-enabled
in some prefix of $\reordering$.
In this case, we say that $\reordering$ witnesses the data race $(e, e')$.

\input{figures/fig_demo_data_race}

\begin{example} 
\exlabel{predictable-race}
Consider trace $\trace_2$ in \figref{trace-no-race} 
containing $6$ events 
performed by two threads $t_1$ and $t_2$.
As before, we use $e_i$ to denote the $i^\text{th}$ event of $\trace_2$.
The two events $e_1 = \ev{t_1, \wt(x)}$
and $e_5 = \ev{t_2, \wt(x)}$ are conflicting (i.e., $\cfwith{e_1}{e_5}$).
The pair $(e_1, e_5)$ is not a data race in $\trace_2$ as no prefix of $\trace_2$
has both these events simultaneously enabled.
Consider the trace $\reordering_2$ in~\figref{reordering-race};
it is a correct reordering of $\trace_2$ because
it preserves both the thread order and reads-from relation of $\trace_2$.
For the same reason, $\reordering'_2$ is a correct reordering of
$\trace_2$ (and also of $\reordering_2$).
Now, observe that $(e_1, e_5)$ is a data race in $\reordering_2$
(and also in $\reordering'_2$) because in the prefix $\pi = \ev{t_2, \acq(\lk)}$,
both $e_1$ and $e_5$ are $\reordering_2$-enabled
(resp. $\reordering'_2$-enabled) and thus $\trace_2$-enabled.
Thus, while $(e_1, e_5)$ is not a data race in $\trace_2$,
it is a predictable data race of $\trace_2$.
\end{example}




The problem of predicting data races --- given an execution trace $\trace$,
determine if there is a predictable data race of $\trace$ --- has been studied
before~\cite{huang2014maximal,said2011generating,wcp,smaragdakis2012sound,Roemer2018,m2}
and is known to be an intractable problem~\cite{mathur2020complexity}.
This means that any sound and complete 
algorithm for predicting data races is unlikely to scale
to real world software applications whose execution traces can have billions of events.
To cater to this, practical data race predictors resort to
incomplete but sound algorithms that run in polynomial time.
In the next section, we discuss the recently proposed \syncp algorithm
that employs the principle of \emph{synchronization preservation}
for predicting data races whose theoretical complexity is linear.

%% file: figures/fig_demo_data_race.tex
\begin{figure}[t]
\centering
\begin{subfigure}[b]{0.15\textwidth}
\centering
\execution{2}{
  \figev{1}{\textcolor{red}{\wt(x)}}
	\figev{1}{\acq(\lk)}
	\figev{1}{\rel(\lk)}
	\figev{2}{\acq(\lk)}
	\figev{2}{\textcolor{red}{\rd(x)}}
	\figev{2}{\rel(\lk)}
}
\caption{Trace $\trace_2$}
\figlabel{trace-no-race}
\end{subfigure}
\hfill
\begin{subfigure}[b]{0.15\textwidth}
\centering
\execution{2}{
	\figev{2}{\acq(\lk)}
  \figev{1}{\textcolor{red}{\wt(x)}}
	\figev{2}{\textcolor{red}{\rd(x)}}
  \figev{2}{\rel(\lk)}
	\figev{1}{\acq(\lk)}
	\figev{1}{\rel(\lk)}
}
\caption{Reordering $\reordering_2$}
\figlabel{reordering-race}
\end{subfigure}
\hfill
\begin{subfigure}[b]{0.15\textwidth}
\centering
\execution{2}{
	\figev{2}{\acq(\lk)}
  \figev{1}{\textcolor{red}{\wt(x)}}
	\figev{2}{\textcolor{red}{\rd(x)}}
  \figev{2}{\rel(\lk)}
}
\caption{Reordering $\reordering'_2$}
\figlabel{syncp-reordering-race}
\end{subfigure}
\caption{The two write events $e_1 = \ev{t_1, \wt(x)}$ and 
$e_5 = \ev{t_2, \wt(x)}$ in $\trace_2$
are conflicting. $(e_1, e_5)$ is not a data race but a predictable data race of $\trace_2$,
witnessed by correct reorderings $\reordering_2$ and $\reordering'_2$.}
\figlabel{demo_data_race}
\end{figure}

%% file: syncp.tex

\subsection{Sync-Preserving Data Races}



Our work is closer in spirit to the work of~\cite{syncp} which
presents the \syncp algorithm that works in linear time and is the current
state-of-the-art race prediction algorithm.
The principle employed by \syncp is to focus on a special class
of reorderings and the data races witnessed by such reorderings;
we discuss these next.

\myparagraph{Sync-preserving reorderings and data races}{
A correct reordering $\reordering$ of a trace
$\trace$ is said to be \emph{sync(hronization)}-preserving
if for any two critical sections of $\trace$ (on the same lock)
 that are both present in $\reordering$, their relative order is the same,
 That is, for every lock $\lk \in \locks{\trace}$ and for any two 
 acquire events $a_1, a_2 \in \events{\trace}$ such that
 $\OpOf{a_1} = \OpOf{a_2} = \acq(\lk)$,
 if $a_1, a_2 \in \events{\reordering}$, then we have:
 $a_1 \trAfterReorder a_2$ iff $a_1 \trAfter a_2$.
 A pair of conflicting events $(e, e')$ in 
 $\events{\trace}$ is said to be a sync-preserving data race
 of $\trace$ if there is a sync-preserving correct reordering
 $\reordering$ of $\trace$ that witnesses this race.
 }

\begin{example}
Consider again, the trace $\trace_2$ and
recall from \exref{predictable-race} that
the pair $(e_1, e_5)$ is not a data race of $\trace_2$ but a predictable race
witnessed by the correct reordering $\reordering_2$.
Observe however that $\reordering_2$ is not a sync-preserving reordering
of $\trace_2$ because it flips the order of the two critical sections on lock $\lk$.
Nevertheless, $(e_1, e_5)$ is a sync-preserving race of $\trace_2$.
This is because the reordering $\reordering'_2$ is, in fact,
a sync-preserving reordering of $\trace_2$
(even though it is a prefix of the non-sync-preserving reordering $\reordering_2$);
there is only one critical section in $\reordering'_2$ and
thus vacuously, the relative order on critical sections is the same as in $\trace_2$.
\end{example}

\myparagraph{Limited predictive power of \syncp}{
While the \syncp algorithm runs in overall linear time,
it can miss data races which are not synchronization-preserving.
These are precisely those conflicting pairs $(e, e')$
such that any correct reordering that witnesses a race on $e$ and $e'$
necessarily reverses the relative order of two 
critical sections on a common lock.
We illustrate this next, and remark that, in general, reasoning about
even a single reversal is intractable~\cite{syncp}.

\begin{example}
\exlabel{not-syncp-race}
Let us again consider the trace $\trace$ in \figref{ssp-race-trace} (\secref{intro}).
The two conflicting events $e_1 = \ev{t_1, \wt(x)}$ and $e_{12} = \ev{t_4, \wt(x)}$,
are a predictable data race of $\trace_1$ as witnessed
by the correct reordering $\reordering_1$ in \figref{ssp-race-witness},
which is not a sync-preserving correct reordering of $\reordering_1$.
In fact, consider any correct reordering $\alpha$ of $\trace_1$ that witnesses 
the race $(e_1, e_{12})$.
Then $\alpha$ must include the events $e_{10}$
and $e_{11}$, and thus the corresponding write events $e_4$ and $e_8$,
together with the thread predecessors
$e_3 = \ev{t_2, \acq(\lk)}$ and $e_7 = \ev{t_3, \acq(\lk)}$.
Next, for well-formedness, at least one of the matching releases 
$e_6 = \ev{t_2, \rel(\lk)}$ as well as $e_9 = \ev{t_3, \rel(\lk)}$
must also be present in $\alpha$.
However, including $e_6$ in $\alpha$ would enforce that $e_5 =\ev{t_2, \rd(z)}$,
and its write event $e_2 = \ev{t_1, \wt(y)}$ are present in $\alpha$,
and then, the event $e_1$ must also be present in the reordering making it no longer enabled in $\alpha$.
This, therefore, means that $e_6 \not\in \events{\alpha}$,
and thus, the only other available release event
$e_9$ must be present in $\alpha$ (for well-formedness).
Further, to ensure well-formedness, $e_3$ must appear after $e_9$ in $\alpha$.
Thus, any reordering $\alpha$ witnessing the race between $e_1$and $e_{12}$ 
must reverse the order of the critical sections.
\end{example}
}

%% file: osr.tex

\section{Optimistic Reasoning for Reversals}
\seclabel{ssp}

Given that reasoning about synchronization reversals is computationally hard,
how do we identify such races efficiently?
At a high level, the intractability in data race prediction
arises because a search for a correct reordering
entails (1) a search for an appropriate set of events (amongst exponentially many sets)
and further, 
(2) given an appropriate set of events, 
a search for a linear order (amongst exponentially many linear orders)
on this set which is well-formed, is a correct reordering and witnesses the race.
We propose (1) a new notion of data races called
\emph{optimistic sync(hronization) reversal} races which can be
predicted by opting for an \emph{optimistic} approach to resolve
both these steps, and (2) an algorithm \ssp to detect all such data races in $\widetilde{O}(\NumEvents^2)$ time.
In this section, we discuss this notion of data races 
 and discuss our algorithm in \secref{osr-algo}.

\input{osr-race}

\input{cmp-osr-m2-syncp}

%% file: osr-race.tex
\subsection{Optimistic Sync-Reversal Races}
\seclabel{osr-race-def}



A crucial aspect of choosing the correct set of
events is to ensure that multiple acquire events
on the same lock do not stay unmatched;
otherwise, the set cannot be linearized to a well-formed trace.
In general, adding a matching release event
may lead to recursive addition of further events.
Some choices may (recursively) at times lead to the 
addition of one of the two focal events $e, e'$
(candidate data race), leading to them being no longer enabled.
We define a simple and tractable notion of \emph{optimistic lock-closure},
which, instead of considering all choices,
simply includes all matching release events
as long as the two focal events are not included.
In the following, we fix a trace $\trace$.

\myparagraph{Optimistic lock-closure}
Let $e_1, e_2 \in \events{\trace}$.
We say that a set $S \subseteq \events{\trace}$ is \emph{optimistically lock-closed}
with respect to $(e_1, e_2)$
if 
\begin{enumerate*}[label=(\alph*)]
	\item $e_1, e_2 \not\in S$ and $\prev{\trace}{e_1}, \prev{\trace}{e_2} \in S$, 
	\item $S$ is $(\thAfter, \lwsingle{\trace})$-closed, and
	\item for every acquire event $a \in S$, if
	$e_1, e_2 \not\in$ $\TLC{\match{\trace}{a}}$,
	then $\match{\trace}{a} \in S$.
\end{enumerate*}
We denote the smallest set that contains $S$ 
and is optimistically lock-closed set, as $\SSP{S, e_1, e_2}$

\begin{example}
\exlabel{olcosed-set}
Let us recall trace $\trace_1$ from \figref{ssp-race}
and consider the set $S_1 = \set{e_3, e_4, e_7, e_8, e_9, e_{10}, e_{11}}$.
Observe that $S_1$ is optimistically lock-closed with respect to $(e_1, e_{12})$, because 
\begin{enumerate*}
\item $S_1$ doesn't include either of $e_1, e_{12}$, 
\item $S_1$ is $(\thAfter, \lwsingle{\trace})$-closed, and finally,
\item $e_1, e_{12}$ $\notin \TLC{e_9}$.
\end{enumerate*}
Note that $e_1 \in \TLC{e_6}$ but $e_6 \not\in S_1$. 
\end{example}

Even though the notion of optimistically lock-closed set
is simple, in general, checking if such a set can be
linearized into a correct reordering
that witnesses a data race, is an intractable problem, as we show next (\thmref{np-hardness-lock-closed}).

\begin{theorem}
	\thmlabel{np-hardness-lock-closed}
    Let $\trace$ be a trace, let $e_1, e_2$ be conflicting events 
    and let $S \subseteq \events{\trace}$ be an optimistically lock-closed 
    set with respect to $(e_1, e_2)$.
    The problem of determining whether there is a correct reordering
    $\reordering$ such that $\events{\reordering} = S$ is \NP-hard.
\end{theorem}

The proof of \thmref{np-hardness-lock-closed} is 
presented in appendix \ref{prf:np-hardness}. 
Given the above result, we also define 
the following more tractable notion of \emph{optimistic reordering}
that ensures that there are no memory reversals, and moreover,
critical sections are reversed only when absolutely 
required, i.e., that  unmatched critical sections appear
later than matched ones.

\myparagraph{Optimistic correct reordering}
A trace $\reordering$ is said to be
an optimistic correct reordering of $\trace$ if
\begin{enumerate*}[label=(\alph*)]
	\item $\reordering$ is a correct reordering of $\trace$, 
	\item for all pairs of conflicting memory access 
	events $\cfwith{e_1}{e_2}$ in $\events{\reordering}$,
	$e_1 \trAfterReorder e_2$ iff 
	$e_1 \trAfter e_2$, and
	\item for any lock $\lk$ and for any two acquire events
	$a_1 \neq a_2$ (with $\OpOf{a_1} = \OpOf{a_2} = \acq(\lk)$), 
	if $a_1$ and $a_2$ are both matched in $\reordering$
	(i.e., $\match{\trace}{a_i} \in \events{\reordering}$ for both $i \in \set{1, 2}$),
	then we must have $a_1 \trAfterReorder a_2$ iff 
	$a_1 \trAfter a_2$.
\end{enumerate*}

We now formalize 
\emph{optimistic sync-reversal} data races.

\begin{definition}[Optimistic Sync-Reversal Race]
\deflabel{osr-race}
Let $\trace$ be a trace and let $(e_1, e_2)$ be a pair of conflicting events in $\trace$.
We say that $(e_1, e_2)$ is an optimistic sync-reversal data race
if there is an optimistic correct reordering $\reordering$ of $\trace$
such that $\events{\reordering}$ is optimistically lock-closed
with respect to $(e_1, e_2)$
and both $e_1$ and $e_2$ are $\trace$-enabled in $\reordering$.
\end{definition}

\begin{example}
\exlabel{osr-race}
In \figref{ssp-race}, the pair $(e_1, e_{12})$ 
is an optimistic sync-reversal race,
because the prefix
$\reordering'_1$ with first $7$ events of $\reordering_1$
is an optimistic reordering of the 
optimistically lock closed set $S_1$,
outlined in \exref{olcosed-set},
(in which $e_1$ and $e_{12}$ are $\trace_1$-enabled).
This is because, all conflicting accesses of $\reordering'_1$ 
have the same relative order as in $\trace_1$, and further,
the unmatched acquire event is positioned after all closed critical sections. 
Similarly, for the trace $\trace_2$ of \figref{demo_data_race}, 
the linearization $\reordering'_2 = \ev{t_2, \acq(\lk)}$
of the set $S_2$ (outlined in \exref{olcosed-set}) is trivially an
optimistic correct reordering.
\end{example}

%% file: cmp-osr-m2-syncp.tex

\subsection{Comparison with other techniques}
\seclabel{qualitative-comparison}

Here, we qualitatively compare our proposed class of races 
with those reported by other sound predictive race detection
techniques proposed in the literature, namely
\syncp~\cite{syncp} and M2~\cite{m2} and illustrate how the set of
races reported by \ssp
is neither a strict subset, nor a strict super set of those detected by 
each.


\begin{example}
    Recall again the execution trace $\trace_1$ in \figref{ssp-race}.
    In \exref{osr-race} we established that the pair $(e_1, e_{12})$ 
    is an optimistic sync-reversal race,
    while in \exref{not-syncp-race}, we showed that it is not a sync-preserving data race.
    When determining if $(e_1, e_{12})$ can be declared a predictive data race,
    the M2 algorithm computes the set $S = \set{e_1, e_2, e_3, e_4, e_5, e_6, e_7, e_8, e_{10}, e_{11}}$
    to be the candidate set that witnesses the race.
    Observe however, this set contains the event $e_1$ and thus cannot witness the race $(e_1, e_{12})$
    since one of these events is not enabled in $S$.
    Thus, some optimistic sync-reversal races are neither sync-preserving races,
    nor can be detected by M2.
\end{example}

\input{figures/non-osr-race-supp}

\begin{example}
    Consider the trace in \figref{syncp-only-race-supp}.
    The pair $(e_1, e_{21})$ is a sync-preserving data race
    as witnessed by the correct reordering shown in~\figref{syncp-only-race-supp-witness}.
    This pair, however is not an optimistic sync-reversal data race
    since the smallest optimistically lock-closed set capable of witnessing the race
    is the set $S_{\ssp} = \set{e_{[3,9]}, e_{[12,14]}, e_{17,20}}$,
    where $e_{i,j}$ is shorthand for $e_i, e_{i+1}, \ldots, e_{j-1}, e_j$.
    Observe that $S_{\ssp} $contains two unmatched acquire events of lock $\lk_2$,
    and adding either matching release will bring $e_1$ in the set.
    Likewise, M2 computes the set containing all events but $e_{21}$, and thus contains $e_1$.
    Thus, there are sync-preserving races which are neither optimistic sync-reversal
    races, nor can be detected by M2.
\end{example}

\begin{example}
    Finally, consider the trace in~\figref{m2-only-race-supp}, derived
    from~\cite{m2}. Here, the pair $(e_{10}, e_{19})$ 
    is a data race that M2 can predict (also
    see \figref{m2-only-race-supp-witness} for the witnessing execution). 
    We remark that any correct reordering witnessing this race 
    must reverse the order of
    the two acquire events $e_8$ and $e_{13}$, as well as
    the order of conflicting memory access events $e_9$ and $e_{14}$.  
    Consequently, this is an example
    of a race reported by M2 that is neither a sync-preserving race, nor
    an optimistic sync-reversal race.
\end{example}

%% file: figures/non-osr-race-supp.tex

\begin{figure*}[t]
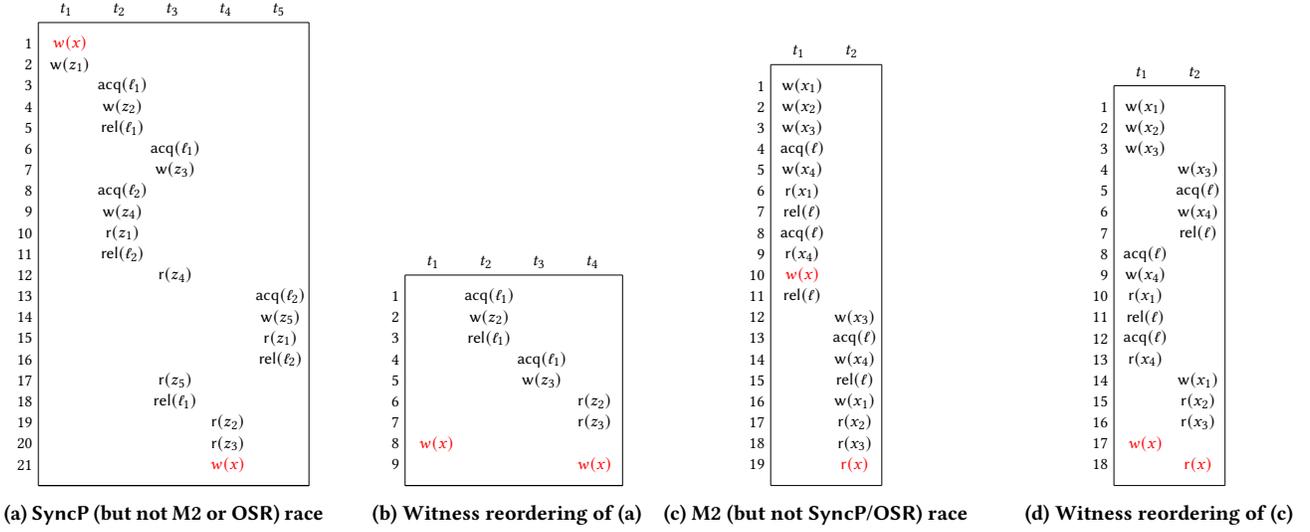

\centering
\begin{subfigure}[b]{0.5\columnwidth}
	\centering
  \executionThin{5}{
    \figev{1}{\textcolor{red}{w(x)}}
    \figev{1}{\wt(z_1)}
    \figev{2}{\acq(\lk_1)}
    \figev{2}{\wt(z_2)}
    \figev{2}{\rel(\lk_1)}
    \figev{3}{\acq(\lk_1)}
    \figev{3}{\wt(z_3)}
    \figev{2}{\acq(\lk_2)}
    \figev{2}{\wt(z_4)}
    \figev{2}{\rd(z_1)}
    \figev{2}{\rel(\lk_2)}
    \figev{3}{\rd(z_4)}
    \figev{5}{\acq(\lk_2)}
    \figev{5}{\wt(z_5)}
    \figev{5}{\rd(z_1)}
    \figev{5}{\rel(\lk_2)}
    \figev{3}{\rd(z_5)}
    \figev{3}{\rel(\lk_1)}
    \figev{4}{\rd(z_2)}
    \figev{4}{\rd(z_3)}
    \figev{4}{\textcolor{red}{w(x)}}
  }
	\caption{\syncp (but not M2 or \ssp) race}
  \figlabel{syncp-only-race-supp}
\end{subfigure}  
\hfill
\begin{subfigure}[b]{0.45\columnwidth}
  \centering
  \executionThin{4}{
    \figev{2}{\acq(\lk_1)}
    \figev{2}{\wt(z_2)}
    \figev{2}{\rel(\lk_1)}
    \figev{3}{\acq(\lk_1)}
    \figev{3}{\wt(z_3)}
    \figev{4}{\rd(z_2)}
    \figev{4}{\rd(z_3)}
    \figev{1}{\textcolor{red}{w(x)}}
    \figev{4}{\textcolor{red}{w(x)}}
  }
    \caption{Witness reordering of (a)}
    \figlabel{syncp-only-race-supp-witness}
\end{subfigure}
\vspace{0.5cm}
\begin{subfigure}[b]{0.5\columnwidth}
	\centering
  \executionThin{2}{
    \figev{1}{\wt(x_1)}
    \figev{1}{\wt(x_2)}
    \figev{1}{\wt(x_3)}
    \figev{1}{\acq(\lk)}
    \figev{1}{\wt(x_4)}
    \figev{1}{\rd(x_1)}
    \figev{1}{\rel(\lk)}
    \figev{1}{\acq(\lk)}
    \figev{1}{\rd(x_4)}
    \figev{1}{\textcolor{red}{w(x)}}
    \figev{1}{\rel(\lk)}
    \figev{2}{\wt(x_3)}
    \figev{2}{\acq(\lk)}
    \figev{2}{\wt(x_4)}
    \figev{2}{\rel(\lk)}
    \figev{2}{\wt(x_1)}
    \figev{2}{\rd(x_2)}
    \figev{2}{\rd(x_3)}
    \figev{2}{\textcolor{red}{\rd(x)}}
  }
	\caption{M2 (but not \syncp/\ssp) race}
  \figlabel{m2-only-race-supp}
\end{subfigure}  
\hfill
\begin{subfigure}[b]{0.45\columnwidth}
  \centering
  \executionThin{2}{
    \figev{1}{\wt(x_1)}
    \figev{1}{\wt(x_2)}
    \figev{1}{\wt(x_3)}
    \figev{2}{\wt(x_3)}
    \figev{2}{\acq(\lk)}
    \figev{2}{\wt(x_4)}
    \figev{2}{\rel(\lk)}
    \figev{1}{\acq(\lk)}
    \figev{1}{\wt(x_4)}
    \figev{1}{\rd(x_1)}
    \figev{1}{\rel(\lk)}
    \figev{1}{\acq(\lk)}
    \figev{1}{\rd(x_4)}
    \figev{2}{\wt(x_1)}
    \figev{2}{\rd(x_2)}
    \figev{2}{\rd(x_3)}
    \figev{1}{\textcolor{red}{w(x)}}
    \figev{2}{\textcolor{red}{\rd(x)}}
  }
    \caption{Witness reordering of (c)}
    \figlabel{m2-only-race-supp-witness}
\end{subfigure}
\hfill
\hfill
\caption{Two traces containing two predictable races. One of them (a) can be detected by \syncp, but not M2 nor OSR. 
(b) is the witness of race in (a). 
The other one (c) can be detected by M2, but not \syncp nor OSR. 
Trace in (c) is directly cited from M2 paper \cite{m2} without modification. 
(d) is the witness of race in (c).
}
\figlabel{non-osr-races-supp}
\end{figure*}

%% file: osr-algo.tex

\section{The \ssp Algorithm}
\seclabel{osr-algo}

We now describe our algorithm \ssp that detects
optimistic sync-reversal data races.
For ease of presentation, we will first discuss how
to check if a given pair $(e_1, e_2)$ of conflicting events
is an optimistic sync-reversal data race (\secref{check-single-race}),
in $\widetilde{O}(\NumEvents)$ time, where $\NumEvents$ is the number of events in the given trace.
Naively, it can be used to report all 
optimistic sync-reversal data races in
$\widetilde{O}(\NumEvents^3)$ time,
by enumerating all $O(\NumEvents^2)$ pairs of conflicting events
and checking each of them in 
$\widetilde{O}(\NumEvents)$ time.
Instead, \ssp runs in overall $\widetilde{O}(\NumEvents^2)$ time
and is based on interesting insights that enable it to perform incremental computation
over the entire trace (\secref{incremental-race}).
We present our overall algorithm and its optimality in \secref{overall-algo}.

\input{check-single-race}
\input{incremental-race}

\input{overall-algo}

%% file: check-single-race.tex

\subsection{Checking Race On A Given Pair Of Events}
\seclabel{check-single-race}

Based on \defref{osr-race}, 
the task of checking 
if a given pair $(e_1, e_2)$
of conflicting events is an optimistic sync-reversal data race
entails examining all optimistic lock-closed
sets and checking if any of these can be linearized.

\myparagraph{Constructing optimistically lock-closed set}{
Our algorithm, however, exploits the following
observation (\lemref{osr-race-closure-set}), 
and focuses on only a single set, namely the smallest such set.
In the following, we will abuse the notation and use
$\SSP{e_1, e_2}$ to denote the set $\SSP{S_{e_1, e_2}, e_1, e_2}$,
where $S_{e_1, e_2} = \set{\prev{\trace}{e_1}} \cup \set{\prev{\trace}{e_2}}$.
Here, 
$\prev{\trace}{e}$ is the last event $f$ such that $f \thAfter e$;
if no such event exists, we say $\prev{\trace}{e} = \bot$, in which case
$\set{\prev{\trace}{e}} = \emptyset$.

\begin{lemma}
\lemlabel{osr-race-closure-set}
\label{lemma:osr-race-closure-set}
Let $e_1, e_2$ be conflicting
events in trace $\trace$. 
If $(e_1, e_2)$ is an optimistic sync-reversal race, then
it can be witnessed in an optimistic correct reordering $\reordering$
such that $\events{\reordering} = \SSP{e_1, e_2}$.
\end{lemma}

\input{algos/algo-closure}
In \algoref{computessp}, we outline our algorithm to compute
the smallest set that we identified in \lemref{osr-race-closure-set}.
It takes $3$ arguments --- the two events $e_1, e_2$
and a set $S_0$;
for computing $\SSP{e_1, e_2}$, we must set $S_0 = \emptyset$;
later in \secref{incremental-race} this set will be used to enable incremental computation.
This algorithm performs a fixpoint computation starting
from the set $S_0 \cup \TLC{\prev{\trace}{e_1}} \cup \TLC{\prev{\trace}{e_2}}$,
and identifies an unmatched acquire event $a$ and checks if its 
matching release $r$ can be added without
adding $e_1$ or $e_2$; if so, $r$ is added;
$\acqs{S}$ denotes the set of acquire events in the set $S$.
The algorithm ensures that the set is $(\thAfter, \lwsingle{\trace})$-closed
at each step, and runs in $O(\NumThreads^2  \NumEvents) = \widetilde{O}(\NumEvents)$ time.
}

\myparagraph{Checking optimistic reordering}{
First, we check if the set $S$ constructed by \algoref{computessp} is
\emph{lock-feasible}, i.e., the set of unmatched acquires 
$\oacqs{S, \lk} = \setpred{a \in \acqs{S}}{\match{\trace}{a} \not\in S}$
for each lock $\lk$ is either singleton or empty:
\begin{align*}
\lkfeas(S) \equiv \forall \lk \in \locks{\trace}, |\oacqs{S, \lk}| \leq 1
\end{align*}
Observe that if $\lkfeas(S)$ does not hold, then every linearization of 
$S$ will have more than one critical sections (on some lock) that overlap,
making it a non-well-formed trace.
Next, inspired from the notion of optimistic reordering,
we construct the \emph{optimistic-reordering-graph} 
$\optgraph_S = (\optvertices_S, \optedges_S)$, where
$\optvertices_S = S$, and
$\optedges_S = \optedges_{S, \thAfter} \cup \optedges_{S, \cfwith{}{}} \cup \optedges_{S, \s{match}} \cup \optedges_{S, \s{unmatch}}$.
Here, $\optedges_{S, \thAfter}$ is the set of edges $(e, e')$,
where $e = \prev{\trace}{e'}$.
The set $\optedges_{S, \cfwith{}{}}$ 
consists of all  immediate conflict edges,
i.e., all pairs $(e, e')$ 
in $S$ such that
$\cfwith{e}{e'}$, $e \trAfter e'$ and there is no intermediate event in $\trace$ that conflicts with both.
The set 
$\optedges_{S, \s{match}}$ consists of all pairs
$(r, a')$ such that $r \trAfter a'$ and there is a common lock $\lk$
for which $\OpOf{r} = \rel(\lk), \OpOf{a'} = \acq(\lk)$,
both $r$ and $a'$ are matched in $S$, and there is no intermediate
critical section on $\lk$.
Finally, the remaining set of edges order matched critical sections before
unmatched ones, i.e.,
$\optedges_{S, \s{unmatch}} = \setpred{(r, a')}{\exists \lk, \OpOf{r} = \rel(\lk), \OpOf{a'} = \acq(\lk),
\match{\trace}{a'} \not\in S}$.
Since optimistic reorderings forbid reversal in the order
of conflicting memory accesses, as well as in the order of
same-lock critical sections that are completely matched, 
it suffices to check the acycliclity of $\optgraph$, 
so that the existence of witness is guaranteed.
}

\begin{lemma}
\lemlabel{opt-reordering-graph}
\label{lemma:opt-reordering-graph}
Let $\trace$ be a trace and let $S \subseteq \events{\trace}$ such
that $S$ is $(\thAfter, \lwsingle{\trace})$-closed and also lock-feasible.
Then, there is an optimistic reordering $\reordering$ of $\trace$ on the set $S$
iff the graph $\optgraph_S$ is acyclic.
\end{lemma}


\input{figures/fig_sync_reversal_graph}

\begin{example}
For trace $\trace_1$ in \figref{ssp-race-trace}, 
we have $\SSP{e_1, e_{12}} =$ $\set{e_3, e_4, e_7, e_8, e_9, e_{10}, e_{11}}$. 
The optimistic-reordering-graph over 
$S_1 = \SSP{e_1, e_{12}}$ is shown in \figref{ssp_race_full_graph};
Observe that there is no cycle. 
Indeed, as guaranteed by \lemref{opt-reordering-graph}, there is an 
optimistic reordering, namely the $7$ length prefix of $\reordering_1$
from \figref{ssp-race-witness} that witnesses the race $(e_1, e_{12})$.
Let us now consider $\trace_3$, \figref{no_race_small_trace}.
The optimistic lock-closure with respect to $(e_4, e_9)$
is $S_3 = \SSP{e_4, e_9} = \set{e_1, e_2, e_3, e_6, e_7, e_8}$.
The optimistic reordering graph over $S_3$,
shown in \figref{no_race_opt_graph},
contains a cycle.
Indeed, $(e_4, e_9)$ is not a predictable race.
\end{example}

We remark that $\optgraph$ can be constructed
and checked for cycles in time $O(\NumThreads\NumEvents) = \widetilde{O}(\NumEvents)$.
Thus the overall algorithm for checking if given $(e_1, e_2)$
is an optimistic sync-reversal race is --- first compute $\SSP{e_1, e_2}$
in $\widetilde{O}(\NumEvents)$ time, check lock-feasibility in $O(\NumLocks\NumThreads) = \widetilde{O}(1)$
time and perform graph construction and cycle detection in $\widetilde{O}(\NumEvents)$ time.
We thus have the following theorem.

\begin{theorem}
Let $\trace$ be a trace and let $e_1, e_2$ be conflicting events in $\trace$.
The problem of determining if $(e_1, e_2)$ is an optimistic sync-reversal race
can be solved in time $O\big(\NumThreads(\NumThreads\NumEvents + \NumLocks)\big) = \widetilde{O}(\NumEvents)$ time.
\thmlabel{osr-race-given-two-events}
\label{theorem:osr-race-given-two-events}
\end{theorem}

%% file: algos/algo-closure.tex

\begin{algorithm}
    \caption{Computing optimistic lock closure}
    \algolabel{computessp}
    \myproc{\sspclosure{$S_0$, $e_1$,$e_2$}}{
        $S \gets S_0 \cup \TLC{\prev{\trace}{e_1}} \cup \TLC{\prev{\trace}{e_2}}$ \\
        \While{$S$ changes}{
	        \If{\big($\exists a \in \acqs{S}, \match{\trace}{a} \notin S$ $\land$ $e_1, e_2 \notin \TLC{\match{\trace}{a}}$ \big)
            }{
                $S \gets S \cup \TLC{\match{\trace}{a}}$
	        }
        }
        \Return $S$
    }
\end{algorithm}

%% file: figures/fig_sync_reversal_graph.tex

\begin{figure}[t]
\centering
\begin{subfigure}[b]{0.2\textwidth}
	\centering
  \includegraphics[scale=0.06]{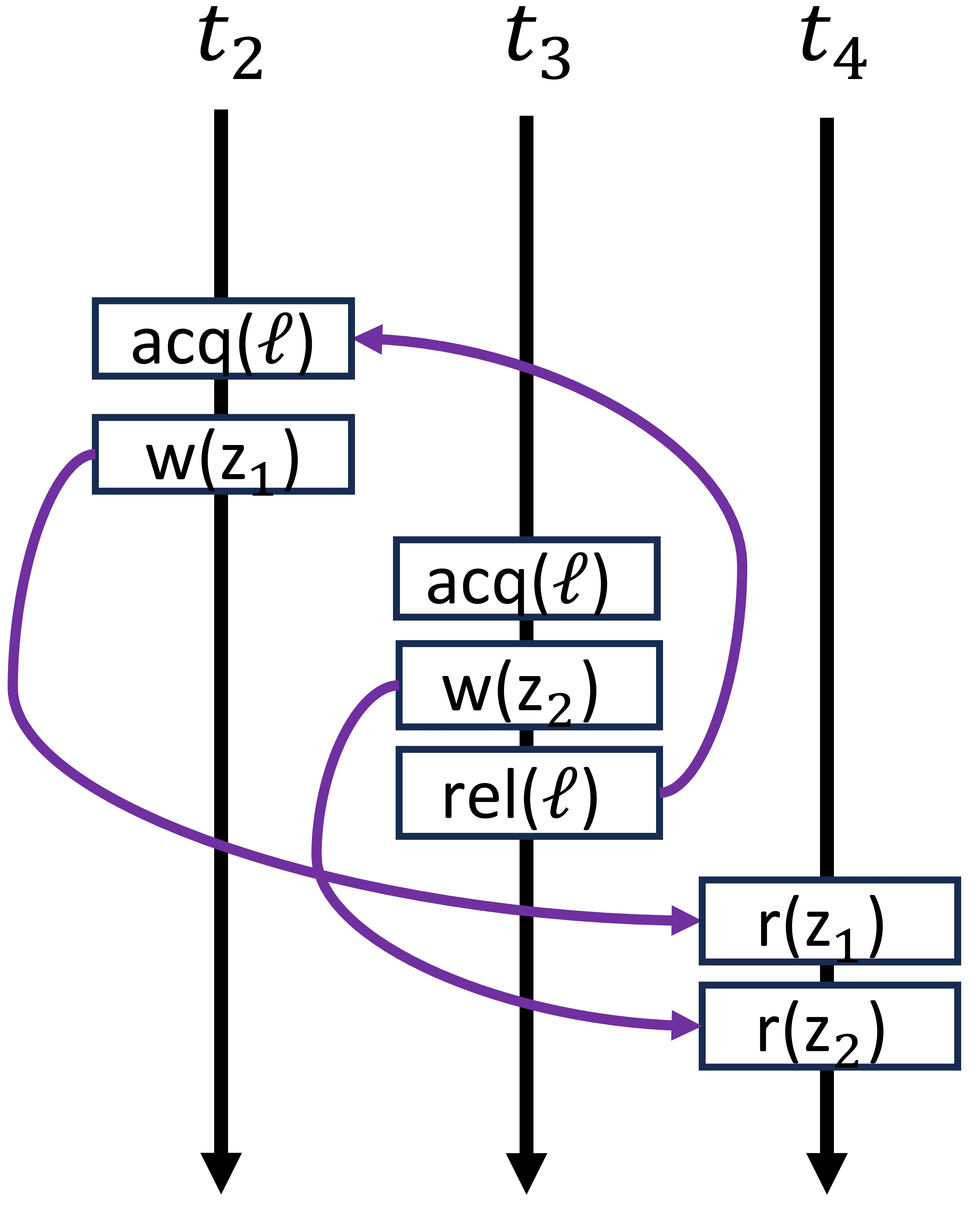}
	\caption{Opt. reord. graph of $\trace_1$}
	\figlabel{ssp_race_full_graph}
\end{subfigure}
\hfill
\begin{subfigure}[b]{0.23\textwidth}
  \centering
  \includegraphics[scale=0.06]{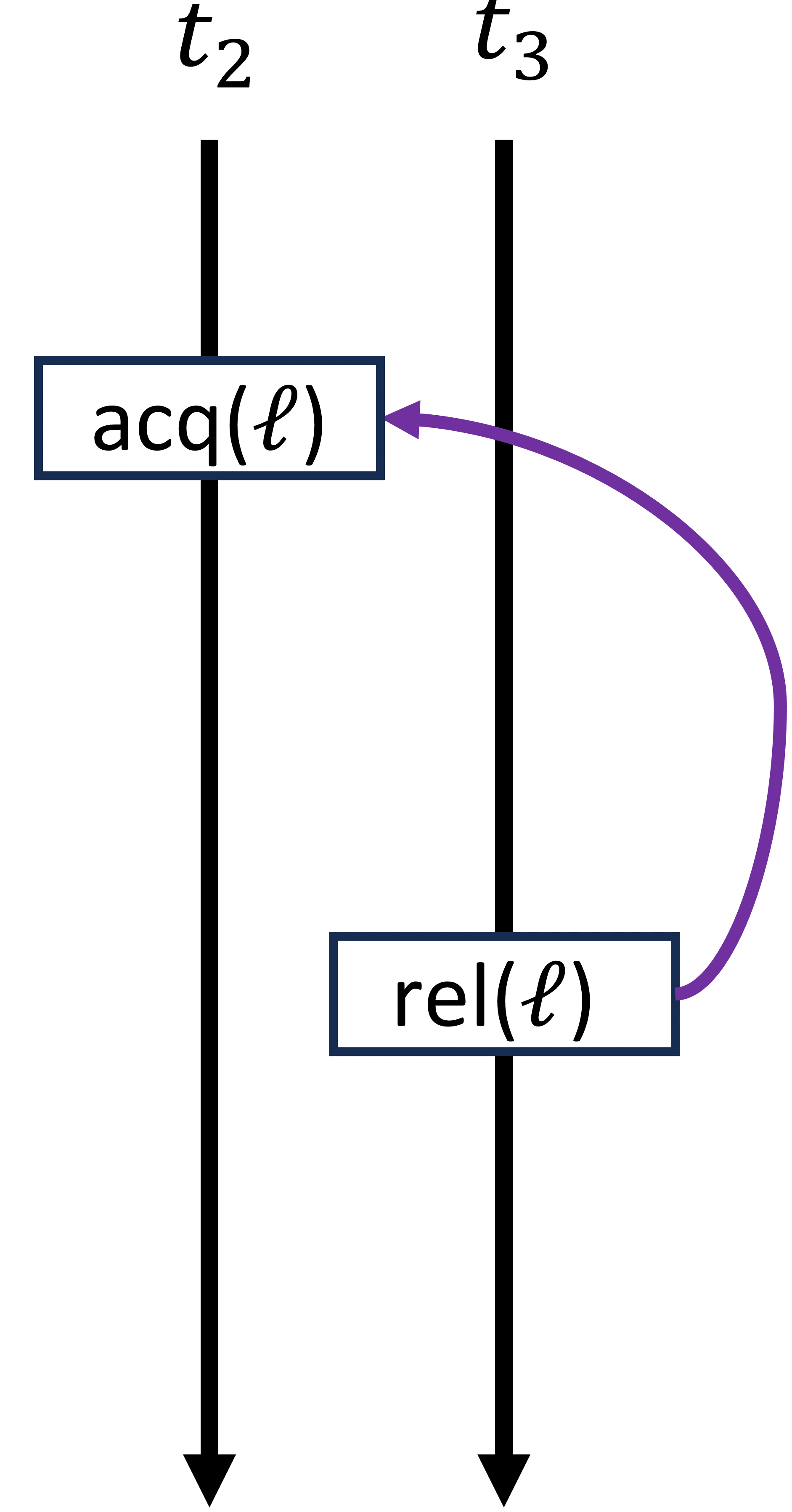}
	\caption{Abst. opt. reord. graph of $\trace_1$}
        \figlabel{ssp_race_small_graph}
\end{subfigure}
\caption{Optimistic and abstract optimistic reordering graphs of $\trace_1$ from \figref{ssp-race-trace}
are acyclic}
\figlabel{ssp_race_graph}
\end{figure}

%% file: incremental-race.tex

\subsection{Incremental Race Detection}
\seclabel{incremental-race}

\myparagraph{Overview}
Recall that there are $O(\NumEvents^2)$ pairs of conflicting events,
and instead of naively examining each of them, we develop
an incremental algorithm that determines the existence of an optimistic sync-reversal
race in total $\widetilde{O}(\NumEvents^2)$ time.
We achieve this by spending  $\widetilde{O}(\NumEvents)$ time
per (read/write) event $e \in \events{\trace}$, and determine
in overall $\widetilde{O}(\NumEvents)$ time if there is
some event $e'$ such that $(e', e)$ is a race, 
by scanning the trace from earliest to latest events.
To do so, our algorithm exploits several novel insights.
Let us fix one of the events $e$.
First, we show that the optimistic lock closure can be computed incrementally
from previously computed sets, instead of computing it from scratch
for each $e'$.
Even though the closure sets can be computed incrementally, the
optimistic-reordering-graph $\optgraph$ (\secref{check-single-race})
cannot be computed in an incremental fashion, because
the edges in this graph depend upon precisely which events are present in the set.
In particular, a previously unmatched acquire event may 
become matched in a larger set, and thus, we may have fewer edges in the larger graph.
Our second insight caters to this --- we represent the graph
succinctly as an \emph{abstract optimistic-reordering-graph} which has 
$\widetilde{O}(1)$ (instead of $\widetilde{O}(\NumEvents)$) nodes,
and moreover, can be computed by pre-populating an appropriate data structure
and performing \emph{range minima queries} over it,
to determine reachability information in the abstract graph
in $\widetilde{O}(1)$ time.

\input{incremental-closure}
\input{incremental-reordering}

%% file: incremental-closure.tex


\myparagraph{Incrementally constructing optimistic lock closure}{
The incremental closure computation relies on the observation that
the closure is monotonic with respect to thread-order (\lemref{incremental-closure-comp}).
Thus, if we fix a thread $t$, and scan the events of $t$ 
from earliest to latest events,
then we can reuse prior computations. 
In fact, \algoref{computessp} already works in this fashion --- it builds
on top of the given input set $S$.
\lemref{incremental-closure-comp} establishes the correctness and time complexity
of closure computation.
\begin{lemma}
\lemlabel{incremental-closure-comp}
\label{lemma:incremental-closure-comp}
Let $e_1, e_2, e'_2 \in \events{\trace}$ be events in trace $\trace$
with $e_2 \thAfter e'_2$. 
Let  $S = \SSP{e_1, e_2}$ and let $S' = \SSP{e_1, e'_2}$.
We have the following:
\begin{enumerate*}
	\item $S \subseteq S'$.
	\item $S =$ \sspclosure{$e_1, e_2, \emptyset$}, and further this call (in \algoref{computessp}) takes $\widetilde{O}(|S|)$ time.
	\item $S' =$ \sspclosure{$e_1, e'_2, S$}, and further this call (in \algoref{computessp}) takes $\widetilde{O}(|S'| - |S|)$ time.
\end{enumerate*}
\end{lemma}
}

%% file: incremental-reordering.tex



\myparagraph{Abstract optimistic-reordering-graph}{
For a set $S \subseteq \events{\trace}$, the abstract optimistic-reordering-graph
is a tuple $\redgraph_S = (\redvertices_S, \rededges_S)$, where the vertices
and edges are defined as follows.
\begin{enumerate*}
    \item $\redvertices_S =$ $\bigcup_{\lk \in \locks{\trace}} \set{\lastrel{S, \lk}} \cup \oacqs{S, \lk}$,
    where $\lastrel{S, \lk}$ is the last release event on lock $\lk$
    (according to $\trAfter$) which is present in $S$.
    \item $(e, e') \in \rededges$ if there is a path from $e$ to $e'$ in the graph $\optgraph_S$.
\end{enumerate*}
In other words, $\redgraph_S$ only contains
$O(\NumLocks)$ vertices, corresponding to the last release events,
and acquire events that are unmatched in $S$,
and preserves the reachability information between these events.
\lemref{abstract-graph-preserves-cycle}
 formalizes the intuition behind this graph --- it preserves
the cyclicity information of the larger graph $\optgraph_S$,
because any cycle in $\optgraph_S$ must involve a `backward' edge from
a matched release and an unmatched acquire event.
$\redgraph_S$ can thus be used to check for the
existence of an optimistic reordering
using an $\widetilde{O}(1)$ check instead of an $\widetilde{O}(\NumEvents)$
check based on \lemref{opt-reordering-graph}.
\begin{lemma}
\lemlabel{abstract-graph-preserves-cycle}
\label{lemma:abstract-graph-preserves-cycle}
Let $\trace$ be a trace and let $S \subseteq \events{\trace}$
be a $(\thAfter, \lwsingle{\trace})$-closed set.
$\optgraph_S$ has a cycle iff $\redgraph_S$ has a cycle.
\end{lemma}
\input{figures/fig_smallGraphDemo}
} 
 
\begin{example} 
\figref{ssp_race_small_graph} shows the abstract optimistic reordering graph
for trace $\trace_1$ in \figref{ssp-race-trace}, corresponding
to the set $S_1 = \SSP{e_1, e_{12}}$, and contains
the last release of lock $\lk$ in $S_1$ as well as the only open acquire in $S_1$.
This graph, like the graph in \figref{ssp_race_full_graph}
is acyclic.
In \figref{no_race}, the abstract graph (\figref{no_race_abs_graph})
captures the path
$e_2 \rightarrow e_3 \rightarrow e_7 \rightarrow e_8$ of \figref{no_race_opt_graph}
with a direct edge $e_2 \rightarrow e_8$, 
thereby preserving the cycle.
\end{example}

\myparagraph{Constructing vertices and backward edges of $\redgraph_S$}{
Recall that $S$ is a $(\thAfter, \lwsingle{\trace})$-closed subset of $\events{\trace}$.
The set of vertices of this graph can be determined
in $O(\NumLocks)$ time by maintaining the last event of every thread present in $S$.
This information can be inductively maintained
as $S$ is being computed incrementally. 
The `backward' edges --- namely those pairs
$(r, a)$ where $a \in S$ is an unmatched acquire on some lock $\lk$,
and $r = \lastrel{S, \lk}$ but $a \trAfter r$ --- can be computed in $O(\NumLocks)$ time.
}

\myparagraph{Pre-computing earliest immediate successor}{
For constructing forward edges,
we first pre-compute a map (for each pair of threads $t_1, t_2$),
$\EIS_{t_1, t_2}$ such that, 
for every $e_1 \in \proj{\events{\trace}}{t_1} = \setpred{e \in \events{\trace}}{\ThreadOf{e} = t}$, 
the event $\EIS_{t_1, t_2}(e_1)$ is the \underline{e}arliest
\underline{i}mmediate \underline{s}uccessor of $e_1$ in thread $t_2$,
in the full graph $\optgraph_{\events{\trace}}$;
observe the subscript $\events{\trace}$ instead of an arbitrary set $S$.
$\EIS_{t_1, t_2}$ can be computed as a pre-processing step
in $O(\NumThreads\NumEvents) = \widetilde{O}(\NumEvents)$ time and stored as an array,
indexed by the events of thread $t_1$.
}

\input{algos/algo-rmq-successor}

\myparagraph{Determining forward edges of $\redgraph_S$}{
The forward edges of $\redgraph_S$ summarize paths in $\optgraph_S$ and are computed as follows.
Recall that we are given a $(\thAfter, \lwsingle{\trace})$-closed
subset $S$ of $\events{\trace}$, and the path between two events must only be contained
with the events of $S$, thus the arrays $\set{\EIS_{t_1, t_2}}_{t_1, t_2 \in \threads{\trace}}$
cannot be used as is to efficiently determine paths.
However, a combination of range minima queries~\cite{rmqsolution} and shortest path
computation can nevertheless still be used to determine path information efficiently.
Let us use $\succOpt{S}{e, t}$ to denote the earliest event in thread $t$
that has a path from event $e$, using only forward edges of $\optgraph_S$.
The event $\succOpt{S}{e, t}$ can be computed using a 
Bellman-Ford-Moore~\cite{Bellman1958,ford1956,Moore1959} style shortest path computation,
as shown in \algoref{rmq-successors}.
This algorithm performs $\rmq{A}[a, b]$ queries which return
the earliest event (according to $\thAfter$) in the segment of the array
$A$ starting at index $a$ and ending at index $b$.
With  $\widetilde{O}(\NumEvents)$ time and space pre-processing,
each range minimum query takes $O(1)$ time~\cite{rmqsolution, gabow1984scaling}, 
Thus, the task of determining $\set{\succOpt{S}{e, t}}_{t \in \threads{\trace}}$ 
takes $O(\NumThreads^2)$ time.
Now, in the graph $\redgraph_S$, we add an edge from $e$ to $e'$ if $\succOpt{S}{e, \ThreadOf{e'}} \thAfter e'$.
Thus, we add all forward edges of the graph in overall $O(\NumThreads^2\NumLocks)$ time.
}

\input{algos/algo-detect-races-given-one-event}

\myparagraph{Checking if a given event $e$ is in race with some event}{
    We now have all the ingredients to describe our overall incremental
    algorithm to check if event $e$ is in optimistic-sync-reversal
    race with some event of a given thread $t$ (\algoref{detect-races-given-one-event}).
    For this, we first initialize all
    the arrays $\set{\EIS_{t_1, t_2}}_{t_1, t_2 \in \threads{\trace}}$ using a linear
    scan of the trace $\trace$, and also do pre-processing for fast performing range minima queries,
    spending overall time $O(\NumThreads\NumEvents)$.
    Then, we iterate over each event $e'$ of thread $t$ that conflict with $e$, 
    starting from the earliest to the latest.
    For each event, we incrementally update the optimistic lock-closure
    set $S$ and check if it is lock-feasible.
    If so, we construct the abstract optimistic-reordering-graph $\redgraph_S$
    and check if it is acyclic, and report a race if so.
}

\begin{theorem}
\thmlabel{osr-race-given-event}
\label{theorem:osr-race-given-event}
Let $\trace$ be an execution, $e \in \events{\trace}$
be a read or write event and let $t \in \threads{\trace}$.
The problem of checking if there is an event $e'$ 
with $\ThreadOf{e'} = t$ such that $(e, e')$ is an optimistic-sync-reversal race, can be solved
in time $O\big((\NumThreads^2 + \NumLocks) \NumLocks \NumEvents \big)$.
\end{theorem}

%% file: figures/fig_smallGraphDemo.tex

\begin{figure}[t]
\centering
\begin{subfigure}[b]{0.28\columnwidth}
  \centering
  \execution{2}{
    \figev{1}{\wt(z)}
    \figev{2}{\acq(\lk)}
    \figev{2}{\rd(z)}
    \figev{2}{\textcolor{red}{w(x)}}
    \figev{2}{\rel(\lk)}
    \figev{1}{\acq(\lk)}
    \figev{1}{\wt(z)}
    \figev{1}{\rel(\lk)}
    \figev{1}{\textcolor{red}{w(x)}}
  }
  \caption{Trace $\trace_3$}
  \figlabel{no_race_small_trace}
\end{subfigure}
\begin{subfigure}[b]{0.37\columnwidth}
  \centering
  \includegraphics[scale=0.06]{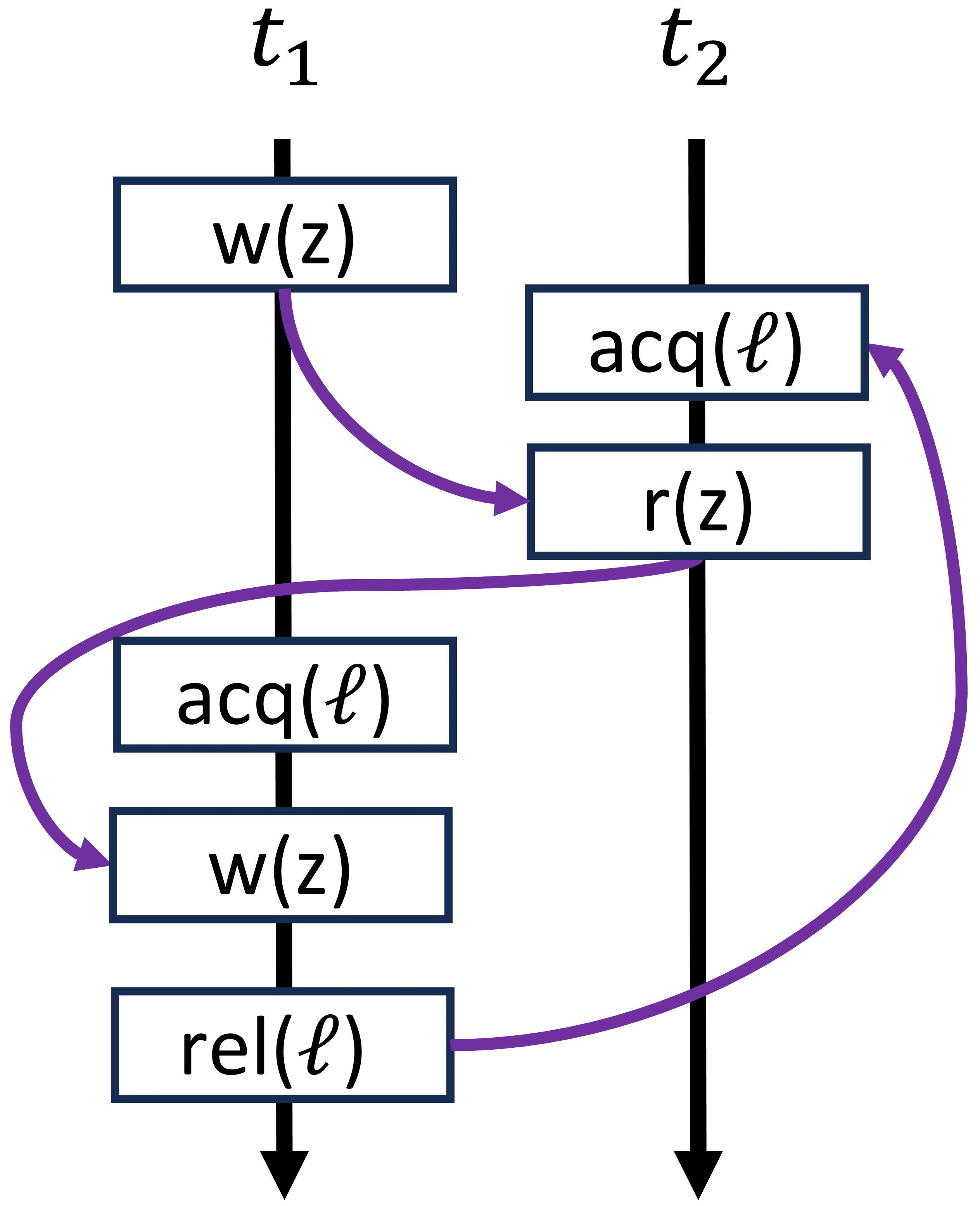}
  \caption{Reordering graph}
  \figlabel{no_race_opt_graph}
\end{subfigure}
\begin{subfigure}[b]{0.33\columnwidth}
  \centering
  \includegraphics[scale=0.06]{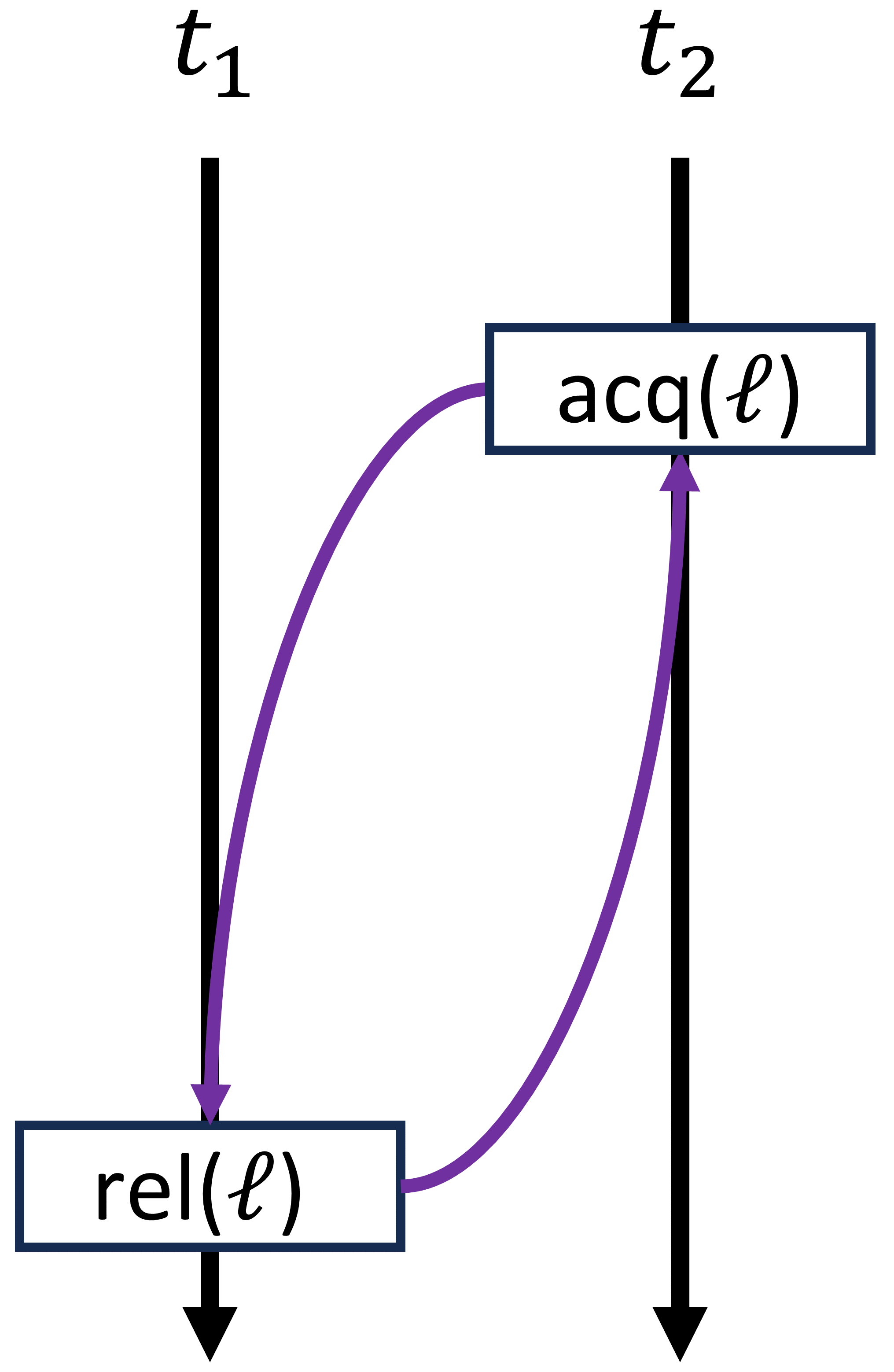}
  \caption{Abstract graph}
  \figlabel{no_race_abs_graph}
\end{subfigure}
\caption{In $\trace_3$, $(e_4, e_9)$ is not a predictable race.
The optimistic reordering graph and the abstract optimistic reordering graph are cyclic.
}
\figlabel{no_race}
\end{figure}

%% file: algos/algo-rmq-successor.tex

\begin{algorithm}[t]
\caption{Earliest successors of event $e$ within set $S$}
\algolabel{rmq-successors}
\myproc{\iteratedrmq{$e$, $S$}}{
    \Let $t_e = \ThreadOf{e}$,  $\mathsf{visitedThr}\gets \emptyset$\\
    \Let $\textsf{last}^{S}_t$ be the last event by $t$ in S, for $t \in \threads{\trace}$ \\
    \For{$t \in \threads{\trace}$}{
        $\succOpt{S}{e, t} \gets \rmq{\EIS_{t_e, t}}[e, \textsf{last}^{S}_{t_e}]$
    }

    \While{$\mathsf{visitedThr} \neq \threads{\trace}$}{
        \Let $t_1$ be s.t. $t_1\not\in \mathsf{visitedThr}$ and $\succOpt{S}{e, t_1}$ is the earliest in $\trAfter$ from $\set{\succOpt{S}{e, t}}_{t \not\in \mathsf{visitedThr}}$
        
        \For{$t_2 \in \threads{\trace}$}{
            $newSucc \gets \rmq{\EIS_{t_1, t_2}}[\succOpt{S}{e, t_1}, \textsf{last}^{S}_{t_1}]$ \\
            \If{$newSucc \thAfter \succOpt{S}{e, t_2}$}{
                $\succOpt{S}{e, t_2} \gets newSucc$ \\
            }
        }

        $\mathsf{visitedThr} \gets \mathsf{visitedThr} \cup \set{t_1}$
    }
    \Return $\set{\succOpt{S}{e, t}}_{t \in \threads{\trace}}$
    }
\end{algorithm}

%% file: algos/algo-detect-races-given-one-event.tex
\begin{algorithm}[t]
    \caption{Detecting races between $e$ and thread $t$}
    \algolabel{detect-races-given-one-event}
    \myproc{\detectinc{$e$, $t$}}{
        $S \gets \emptyset$ \;
        \For{$e' \in \proj{\events{\trace}}{t}$ s.t. $\cfwith{e}{e'}$ and $e' \trAfter e$}{
            $S \gets \sspclosure{e, e', S}$ \\
            \If{$\lkfeas(S)$ and $\redgraph_S$ is acyclic}{
                    \declare ($e'$, $e$) as race.
            }
        }
    }
\end{algorithm}

%% file: overall-algo.tex

\subsection{Detecting All Optimistic Sync-Reversal Races}
\seclabel{overall-algo}

Given a trace $\trace$, 
all the optimistic sync-reversal races in $\trace$
can now be detected by enumerating all events $e$ and threads $t$
and checking if \detectinc{$e$, $t$} reports a race.
Our resulting algorithm \ssp (\algoref{detect-races})
runs in time $O\big(\NumThreads \NumLocks (\NumThreads^2+\NumLocks)\NumEvents^2\big)$.

\input{algos/algo-detect-races}

\begin{theorem}
\thmlabel{osr-race-all}
\label{theorem:osr-race-all}
Given a trace $\trace$, the problem of checking if
$\trace$ has an optimistic sync-reversal data race, can be solved in
time  $O\big(\NumThreads \NumLocks (\NumThreads^2+\NumLocks)\NumEvents^2\big) = \widetilde{O}(\NumEvents^2)$ time.
\end{theorem}

\myparagraph{Hardness of detecting optimistic sync-reversal races}{
	We have, thus far, established that the problem of checking
	the existence of optimistic sync-reversal data races can be solved in quadratic time.
	In the following, we also show a matching quadratic time lower bound,
	thus establishing that our algorithm \ssp is indeed optimal.
	The lower bound is conditioned on the 
	Strong Exponential Time Hypothesis (SETH), which is a widely believed conjecture.
	We use fine-grained reductions to establish a reduction 
	from the \emph{orthogonal vectors problem} which holds true under SETH~\cite{Williams2005}.
	The full proof of the following result is presented in Appendix \ref{appendix:n2-hardness}. 

\begin{theorem}
    \thmlabel{n2-hardness}
    \label{theorem:n2-hardness}
    Assume SETH holds.
    Given an arbitrary trace $\trace$, the problem of determining if $\trace$ has an \ssp race cannot
    be solved in time $O(\NumEvents^{2-\epsilon})$ (where  $\NumEvents = |\events{\trace}|$) for every $\epsilon>0$.
\end{theorem}
}

%% file: algos/algo-detect-races.tex

\begin{algorithm}[t]
\caption{Detecting optimistic sync-reversal races in $\trace$}
\algolabel{detect-races}
\myproc{\osr{$\trace$}}{
    \For{$e \in \events{\trace}$ s.t. $e$ is a memory access event}{
            \For{$t' \in \threads{\trace}$}{$\detectinc{e, t'}$}
    }
} 
\end{algorithm}

%% file: experiments.tex

\section{Evaluation}
\seclabel{evaluation}
We implemented our algorithm \SSPAlgo in Java, 
using the \rapid dynamic analysis framework~\cite{rapidproject}. 
We evaluate the performance and precision of \SSPAlgo, 
on 153
benchmarks and compare it with prior state-of-the-art sound predictive race detection algorithms. 
We discuss our experimental set up in~\secref{setup} and our evaluation results in
\secref{grpJava-eval}, \secref{grpC-eval} and \secref{scalability-eval}.

\input{setup}

\input{evaluation}

%% file: setup.tex

\subsection{Experimental Setup}
\seclabel{setup}


\myparagraph{Benchmarks} 
Our evaluation subjects are both Java (\grpJava) as well as C/C++/OpenMP (\grpC) benchmarks. 
\grpJava, derived from~\cite{syncp}, contains 30 Java programs from 
the IBM Contest benchmark suite~\cite{farchi2003concurrent}, 
the Java Grande forum benchmark suite~\cite{sen2005detecting},
DaCapo~\cite{blackburn2006dacapo},
SIR~\cite{do2005supporting} and other standalone benchmarks.
\grpC contains 123 benchmarks from
OmpSCR~\cite{dorta2005openmp}, 
DataRaceBench~\cite{liao2017dataracebench} 
DataRaceOnAccelerator~\cite{schmitz2019dataraceonaccelerator}, 
NAS parallel benchmarks~\cite{bailey1991parallel}, 
CORAL~\cite{CORALOne,CORALTwo}, 
ECP proxy applications~\cite{ecpapplications} and the Mantevo project~\cite{mantevoproject}. 
For an apples-to-apples comparison, we
evaluate all compared techniques on the same execution trace
to remove bias due to thread-scheduler.
For this, we generate traces out of these programs using 
\tsan~\cite{serebryany2009threadsanitizer} (for \grpC) and 
using \rvpredict~\cite{rvpredict2010meredith} (for \grpJava). 
For Java programs, we generate one trace per program 
and for C/C++ programs, 
we generate multiple traces of the same program with different thread number and input parameters. 
All compared methods then evaluate each generated trace 3 times.
We did not exclude any traces from the benchmarks, 
except one corrupted trace.

As part of our evaluation,
we also explored synthetically created benchmark traces from 
\raceinjector~\cite{wang2023raceinjector, raceInjectorTraces}, that uses 
SMT solving to inject data races into existing traces.
However, the traces in~\cite{raceInjectorTraces} are
short, could not be used to distinguish most compared methods
and were not useful for a conclusive evaluation.
 Our evaluation on these traces is deferred to \appref{sec-5} (\tabref{race-injector}).
As observed in prior works~\cite{flanagan2009fasttrack,syncp,seqc,m2}, 
a large fraction of events in traces are thread-local, and 
do not affect the precision or soundness of race detection algorithms, 
but can significantly slow down race detection. 
Therefore, we filter out these thread-local events, as with prior work~\cite{syncp,m2,wcp}. 

\myparagraph{Compared methods} 
We compare \SSPAlgo with state-of-the-art sound predictive algorithms:
WCP~\cite{wcp}, SHB~\cite{shb}, M2~\cite{m2} and \syncp~\cite{syncp}. 
Amongst these, SHB and WCP are partial order based methods and run in linear time. 
M2 and \syncp are closer in spirit to ours --- they first
identify a set of events and then a linearization of this set that can witness a data race.
\syncp works in linear time while M2 has higher polynomial complexity of 
$\widetilde{O}(\NumEvents^4 \log(\NumEvents))$~\cite{m2}.
For all these algorithms, we use the publicly available source codes~\cite{shb, wcp, syncp, m2}. 
To achieve fair comparison, we modify each of them, so that 
(1) each algorithm reports on the same criteria (events v/s memory locations v/s program locations)
(2) any redundant operations not relevant to the reporting criteria are removed. 
A comparison with recent work SeqC~\cite{seqc} was not possible 
because the implementation of SeqC is neither publicly available nor 
could be obtained even after contacting the authors.
Our evaluation didn't include comparison with solver-aided race predictors, such as \rvpredict \cite{huang2014maximal}.
Based on prior work \cite{wcp}, such  predictors are known to not scale, have unpredictable race reports and typically have lower predictive power than the simplest of race prediction algorithms, thanks to the windowing strategy they implement.

\myparagraph{Machine configuration and evaluation settings} 
The experiments are conducted on a 2.0GHz 64-bit Linux machine. 
For \grpJava (Java) benchmarks, we set the heap size of JVM to be 60GB and timeout to be 2 hours;
this set up is similar to previous works~\cite{syncp, wcp}, except for the 
larger heap space, mandated by the larger memory requirement of M2.
For \grpC (C/C++) benchmarks, we set the heap size to be 400GB and timeout to be 3 hours,
since these are much more challenging --- the
number of events, locks and variables in these are typically $10-100\times$ more 
than traces in \grpJava.
All experiments are repeated $3$ times and the times reported are averaged over these 3 runs.

\myparagraph{Reported metrics}{
Our evaluation aims to understand the 
prediction power (precision)
as well as the scalability of~\ssp and assess how it compares against 
existing state-of-the-art race prediction techniques.
For each execution trace , we report key characteristics
(number of events, threads, locks, read events, write events, 
acquire events and release events) to estimate how challenging each benchmark is.
Next, we measure and report the following :
\begin{description}
    \item[\it Running time.] For each algorithm, we report the average
    running time (over $3$ trials)
    for processing the entire execution. This is aimed to understand if the
    worst case quadratic complexity of \osr affects its performance in practice,
    or it is on par with other linear time methods such as WCP, SHB and \syncp.

    \item[\it Race reports in \grpJava.] 
    For benchmarks in \grpJava, we report the number of racy events reported;
    an event $e_2$ is racy if there is a conflicting event $e_1$ 
    earlier in the trace, such that $(e_1, e_2)$ is a race. 
    We also report the number of distinct source code lines for these racy events. 
    We note here one racy source code line could correspond to many racy events.

    \item[\it Race reports in \grpC.] 
    For benchmarks in \grpC, we report the number of variables (memory locations) that are racy. 
    A variable $x$ is racy if there is a racy event $e$ that accesses $x$. 
    The number of racy events in the C/C++ benchmarks is typically very large, and
    reporting each racy event throttles nearly all algorithms. 
    If a compared method times out, we report the number of racy variables found before timing out. 
    This enables us to better evaluate their ability to find races in a more reasonable setting.
    Besides, most algorithms report many races before they timeout.

    \item[\it Scaling behavior of \osr.]
    \osr runs in worst case quadratic time.
    We empirically evaluate how \osr scales with trace length, for a small set of benchmarks
    to gauge its in-practice behavior.
\end{description}
}

%% file: evaluation.tex

\input{tables/SPNumRaces}
\subsection{Evaluation Results For Java Benchmarks}
\seclabel{grpJava-eval}

\tabref{java-results} summarizes the results for \grpJava.

\myparagraph{Prediction power} 
\ssp reports the largest number of races on each trace;
it reports about 200 more racy events
and $3$ extra racy locations over the second most predictive method (\syncp);
we remark that any extra data race can be an insidious bug~\cite{boehm2012position} and deserves rigorous attention by developers.
Although WCP can detect sync-reversal races in principle, 
and reports much fewer races than \ssp (and also misses races reported by \syncp).
M2 takes much more memory and time than \ssp,
and times out on two benchmarks (linkedlist and lufact),
while runs out of memory on the benchmark tsp. 
On other benchmarks, \ssp demonstrates the same prediction 
power as M2.
Overall M2 detects 29.2k less races. 
In terms of racy source code locations, 
\ssp also reports 24, 47, 3, 13 more than SHB, WCP, \syncp and M2, respectively.
We remark that this class of benchmarks does not bring out the
full potential of \ssp --- even if \ssp reports
the highest number of races individually for each benchmark,
at least one other method also reports this number of races.
\grpC though does better justice to \ssp.

\myparagraph{Running time} 
SHB and WCP are lightweight partial order-based linear time algorithms
and finish fastest.
On the other hand, M2 performs an expensive computation, times out on some large traces
and takes more than 6 hours to finish.
\syncp runs in linear time, but our algorithm
\ssp outperforms it by about $1.5\times$. 
We note that the linkedlist benchmark is especially challenging, 
with large number of variables, as a result of which  
\syncp allocates a large memory to account for its heavy data structure usage. 

Thus, for \grpJava benchmarks, 
\ssp demonstrates highest race coverage, 
and runs faster than the state-of-the-art \syncp. 

\subsection{Evaluation Results For C/C++ Benchmarks}
\seclabel{grpC-eval}
\input{tables/extraNumRaces}
\tabref{c-results} summarizes our evaluation over \grpC (C/C++) benchmarks. 
In \appref{sec-5}, we present detailed statistics of these benchmarks
(see \tabref{cpp-traces-info} and \tabref{extra-results-detail}).

\myparagraph{Prediction power} \ssp displays high race coverage on this set of traces. 
Overall, \ssp reports $2.5\times$ more races than the second most predictive method (SHB).
On all, except 5, of the 118 benchmarks, 
\ssp reports the highest number of racy variables.
Each of the remaining $5$ benchmark traces have a large number of events,
and only the lightweight algorithms (SHB and WCP)  finish within the 3 hour time limit. 
In terms of total races found, 
\ssp reports $2.5\times$ and $2.7\times$ more races 
than SHB ($2^\text{nd}$ highest) and WCP ($3^\text{rd}$ highest). 
\syncp and M2 time out on most benchmarks.
We speculate that this is because
both these methods have high memory requirement
and result in large time spent in garbage collection.
\ssp, therefore, has the highest race coverage even for the C/C++ benchmarks.

We remark that the number of racy variables in this class of benchmarks is very high.
We speculate this is because our instrumentation using \tsan does not 
explicitly tag atomic operations.
Further many benchmarks perform matrix operations, giving rise to
many distinct memory locations.
Nevertheless, we choose to report all races because 
data races can render these programs potentially non-robust,
and under weak memory consistency, data races can lead to undefined semantics.

\myparagraph{Running time} 
Overall, SHB runs the fastest. 
\syncp and M2, on the other hand, frequently time out. 
The difference in the performance between \syncp, M2 and OSR gets exacerbated on the C/C++ benchmarks because these contain much larger execution traces than Java benchmarks. 
The performance of \ssp (total running time of 42 hours) is close to WCP (30 hours).
\ssp, therefore, achieves an optimal balance between predictive power and scalability --- \ssp has the highest predictive power and outperforms SHB, WCP, \syncp, M2, and often runs faster than more exhaustive  techniques. 

\subsection{Scalability}
\seclabel{scalability-eval}
\input{./figures/ScalabilityFigure}

In this section, we take a closer look at the run-time behavior of \ssp to understand its unexpected high scalability on some benchmarks. 
We select the most challenging benchmarks from each of the following groups: HPCBench, CoMD, DataRaceBench, OMPRacer in \grpC. 
For these benchmarks, we measure the time to process every million events and report it in \figref{scalability}. 
We observe that on these four benchmarks, \ssp scales linearly for a large prefix, while gradually slows down on two of them. 
The near-linear behavior of \ssp is likely an artefact of the fact that, many of these benchmarks traces have large number of data races, thus the race check for a single event succeeds quickly instead of the worst case linear time requirement. 
Therefore, instead of spending overall quadratic time, \ssp spends linear time on average.

%% file: tables/SPNumRaces.tex

\setlength{\tabcolsep}{3pt}
\renewcommand{\arraystretch}{1.1}
\begin{table*}
\centering
\caption{Evaluation on \grpJava (Java benchmarks). Columns 1-3 denote the name, 
number of events and number of threads for each benchmark. 
Columns 4-13 are the number of racy events (and racy program locations) reported and average running time of each algorithm.}
\tablabel{java-results}
\vspace{-0.1in}
\begin{tabular}{ lll ll ll ll ll ll}
\toprule

1 & 2 & 3 & 4 & 5 & 6 & 7 & 8 & 9 & 10 & 11 & 12 & 13 \\

\hline

\multirow{2}{*}{\textbf{Benchmarks}} & \multirow{2}{*}{$\NumEvents$} & \multirow{2}{*}{$\NumThreads$} & 
\multicolumn{2}{c}{SHB} &
\multicolumn{2}{c}{WCP} &
\multicolumn{2}{c}{SyncP} &
\multicolumn{2}{c}{M2} &
\multicolumn{2}{c}{\ssp} \\

& & & Races & Time (s) & Races & Time (s) & Races & Time (s) & Races & Time (s) & Races & Time (s) \\

\hline

array & 11 & 3 & 0(0) & 0.05 & 0(0) & 0.08 & 0(0) & 0.06 & 0(0) & 0.03 & 0(0) & 0.09 \\ 

critical & 11 & 4 & 3(3) & 0.04 & 1(1) & 0.05 & 3(3) & 0.07 & 3(3) & 0.02 & 3(3) & 0.07 \\        
          
account & 15 & 4 & 3(1) & 0.04 & 3(1) & 0.06 & 3(1) & 0.06 & 3(1) & 0.02 & 3(1) & 0.08 \\        
                 
airtickets & 18 & 5 & 8(3) & 0.05 & 5(2) & 0.08 & 8(3) & 0.06 & 8(3) & 0.03 & 8(3) & 0.08 \\     
                      
pingpong & 24 & 7 & 8(3) & 0.04 & 8(3) & 0.07 & 8(3) & 0.06 & 8(3) & 0.03 & 8(3) & 0.08 \\       
       
twostage & 83 & 12 & 4(1) & 0.06 & 4(1) & 0.10 & 4(1) & 0.14 & 8(2) & 0.05 & 8(2) & 0.10 \\         
                
wronglock & 122 & 22 & 12(2) & 0.07 & 3(2) & 0.11 & 25(2) & 0.22 & 25(2) & 0.18 & 25(2) & 0.13 \\ 
           
bbuffer & 9 & 3 & 3(1) & 0.05 & 1(1) & 0.06 & 3(1) & 0.05 & 3(1) & 0.02 & 3(1) & 0.10 \\          
            
prodcons & 246 & 8 &1(1) & 0.07 & 1(1) & 0.13 & 1(1) & 0.16 & 1(1) & 0.06 & 1(1) & 0.12 \\      
           
clean & 867 & 8 & 59(4) & 0.11 & 82(4) & 0.23 & 60(4) & 0.26 & 110(4) & 0.65 & 110(4) & 0.20 \\   

mergesort & 167 & 5 & 1(1) & 0.89 & 1(1) & 0.13 & 3(1) & 0.10 & 5(2) & 0.04 & 5(2) & 0.12 \\

bubblesort & 1.7K & 13 & 269(5) & 0.15 & 100(5) & 0.30 & 269(5) & 2.29 & 374(5) & 8.40 & 374(5) & 0.28 \\

lang & 1.8K & 7 & 400(1) & 0.17 & 400(1) & 0.26 & 400(1) & 0.33 & 400(1) & 0.54 & 400(1) & 0.22 \\

readwrite & 9.8K & 5 & 92(4) & 0.27 & 92(4) & 0.63 & 199(4) & 0.81 & 228(4) & 9.00 & 228(4) & 0.69 \\

raytracer & 526 & 3 & 8(4) & 0.10 & 8(4) & 0.17 & 8(4) & 0.15 & 8(4) & 0.09 & 8(4) & 0.15 \\

bufwriter & 10K & 6 & 8(4) & 0.29 & 8(4) & 0.77 & 8(4) & 0.75 & 8(4) & 0.52 & 8(4) & 0.49 \\

ftpserver & 17K & 11 & 69(21) & 1.18 & 70(21) & 0.99 & 85(21) & 6.01 & 85(21) & 2.43 & 85(21) & 0.79 \\

moldyn & 21K & 3 & 103(3) & 1.03 & 103(3) & 0.73 & 103(3) & 0.79 & 103(3) & 31.43 & 103(3) & 0.46 \\

linkedlist & 910K & 12 & 6.0K(4) & 3.77 & 6.0K(3) & 6.80 & 7.1K(4) & 378.25 & 0(0) & 7200 & 7.1K(4) & 6.56 \\

derby & 75K & 4 & 29(10) & 0.94 & 28(10) & 2.30 & 29(10) & 19.08 & 30(11) & 5.66 & 30(11) & 3.67 \\

jigsaw & 3.2K & 8 & 4(4) & 0.17 & 4(4) & 0.39 & 6(6) & 2.90 & 6(6) & 0.23 & 6(6) & 0.35 \\

sunflow & 3.3K & 17 & 84(6) & 0.17 & 69(6) & 0.39 & 119(7) & 2.53 & 130(7) & 1.10 & 130(7) & 0.35 \\

cryptorsa & 1.3M & 7 & 11(5) & 5.95 & 11(5) & 10.87 & 35(7) & 156.19 & 35(7) & 20.39 & 35(7) & 173.74 \\

xalan & 672K & 7 & 31(10) & 3.22 & 21(7) & 12.07 & 37(12) & 160.62 & 37(12) & 6.56 & 37(12) & 230.03 \\

lufact & 892K & 5 & 22.0K(3) & 3.39 & 22.0K(3) & 7.16 & 22.0K(3) & 62.10 & 0(0) & 7200 & 22.0K(3) & 4.15 \\

batik & 131 & 7 & 10(2) & 0.09 & 10(2) & 0.11 & 10(2) & 0.12 & 10(2) & 0.04 & 10(2) & 0.12 \\

lusearch & 751K & 8 & 232(44) & 2.86 & 119(27) & 7.94 & 232(44) & 9.26 & 232(44) & 50.4 & 232(44) & 3.65 \\

tsp & 15M & 10 & 143(6) & 33.63 & 140(6) & 66.07 & 143(6) & 146.24 & 0(0) & 7200 & 143(6) & 160.39 \\

luindex & 16K & 3 & 1(1) & 0.38 & 2(2) & 0.68 & 15(15) & 0.71 & 15(15) & 0.53 & 15(15) & 0.49 \\

sor & 1.9M & 5 & 0(0) & 4.79 & 0(0) & 9.92 & 0(0) & 13.16 & 0(0) & 10.61 & 0(0) & 38.0 \\

\hline

\textbf{Sum} & & & 29.5K(157) & 64.0 & 29.2K(134) & 129.7 & 
30.9K(178) & 961.9 & 1.9K(168) & 6.0h & 31.1K(181) & 625.7 \\ 

\bottomrule

\end{tabular}
\end{table*}

%% file: tables/ExtraNumRaces.tex

\setlength{\tabcolsep}{4pt}
\begin{table*}
\centering
\caption{Evaluation summary on \grpC (C/C++ benchmarks). 
Benchmarks are grouped based on their source, and each row corresponds to one group.
Column 1 denotes the source and size of each group. 
Columns 2 and 3 respectively denote the range and the total number of events in each group. 
Column 4 denotes the range of number of threads in the benchmarks. 
Column 5-14 denote the total number of racy memory locations,
and average running time (in minutes) reported by each algorithm.}
\label{tab:c-results}
\vspace{-0.1in}
\begin{tabular}{llll ll ll ll ll ll}
\toprule

1 & 2 & 3 & 4 & 5 & 6 & 7 & 8 & 9 & 10 & 11 & 12 & 13 & 14 \\

\hline

\multirow{2}{*}{\textbf{Benchmark Group}} & 
\multicolumn{2}{c}{$\NumEvents$} & 
\multirow{2}{*}{$\NumThreads$} &
\multicolumn{2}{c}{SHB} &
\multicolumn{2}{c}{WCP} &
\multicolumn{2}{c}{SyncP} &
\multicolumn{2}{c}{M2} &
\multicolumn{2}{c}{\ssp} \\


& Range & Total & & Races & Time & Races & Time & Races & Time & Races & Time & Races & Time \\

\hline
CoMD (8) & [2.5M, 117M] & 707M & [16, 56] & 41.6k & 29.1 & 247k & 72.3 & 32 & 1440 & 672 & 1440 & 441k & 32.4 \\
SimpleMOC (1) & [19M, 19M] & 19M & [16, 16] & 380 & 0.1 & 388 & 23.8 & 32 & 180 & 32 & 180 & 32 & 180 \\
OMPRacer (15) & [0.7M, 157M] & 625M & [16, 58] & 1.2M & 17.4 & 0.7M & 84.1 & 3.3k & 2.4k & 1.9k & 2.1k & 1.3M& 35.8 \\
DRACC (13) & [0.5k, 104M] & 694M & [16, 16] & 2247 & 8.4 & 2247 & 105.7 & 2442 & 1440.5 & 361 & 990.1 & 2450 & 66.6 \\
DRB (33) & [0.5k, 900M] & 5.7B & [16, 56] & 50.4k & 169.5 & 54.5k & 0.6k & 1.5k & 5.4k & 0.9k & 4.9k & 47.4k & 1.4k \\
HPC (46) & [1k, 335M] & 3.8B & [16, 56] & 6.5M & 174.2 & 6.4M & 775.4 & 102k & 7.7k & 3305 & 6.8k & 18.3M & 574.8 \\
misc (7) & [1k, 29M] & 49M & [4, 219] & 8548 & 0.9 & 8481 & 182.2 & 479 & 900.9 & 895 & 444.6 & 4289 & 183.4 \\
\hline
\textbf{Total} (123) & & 11.6B & & 7.9M & 6.7h & 7.4M & 30.3h & 109k & 324.4h & 8.1k & 280.5h & 20.1M &41.2h \\
\bottomrule
\end{tabular}
\end{table*}

%% file: figures/ScalabilityFigure.tex

\begin{figure*}
\begin{subfigure}{.45\textwidth}
  \centering
  \includegraphics[scale=0.35]{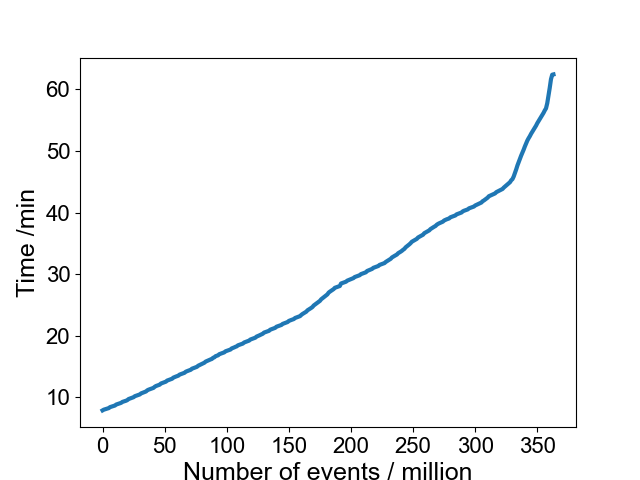}
  \caption{HPCBench / ftt-363M-56th}
  \figlabel{scalability-hpcbench}
\end{subfigure}
\begin{subfigure}{.5\textwidth}
  \centering
  \includegraphics[scale=0.35]{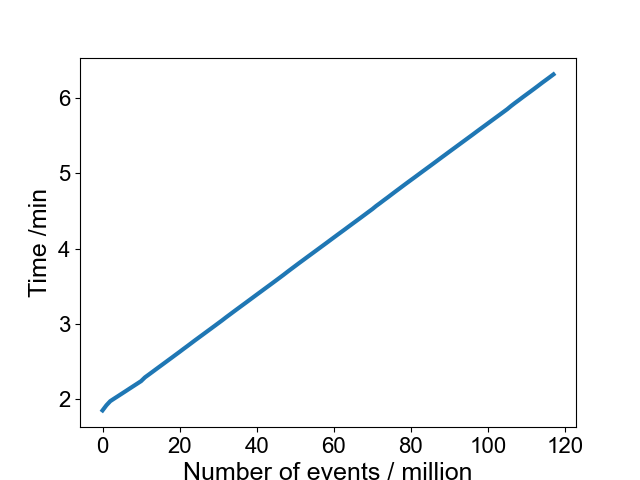}  
  \caption{CoMD / OpenMP-117M-56th}
  \figlabel{scalability-comd}
\end{subfigure}
\newline
\newline
\newline
\begin{subfigure}{.45\textwidth}
  \centering
  \includegraphics[scale=0.35]{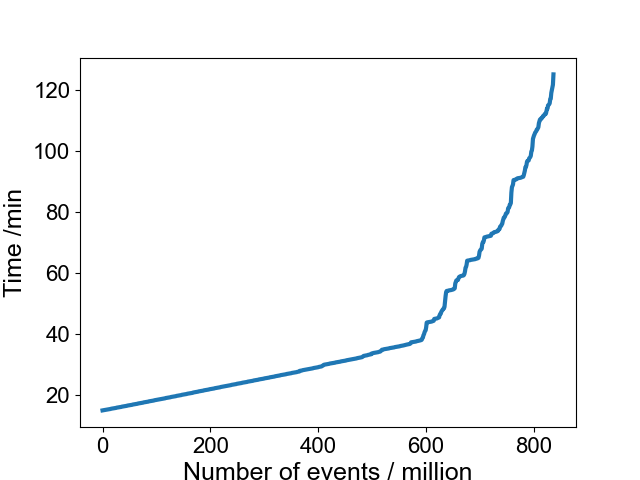}
  \caption{DataRaceBench / DRB177-837M-16th}
  \figlabel{scalability-dataracebench}
\end{subfigure}
\begin{subfigure}{.5\textwidth}
  \centering
  \includegraphics[scale=0.35]{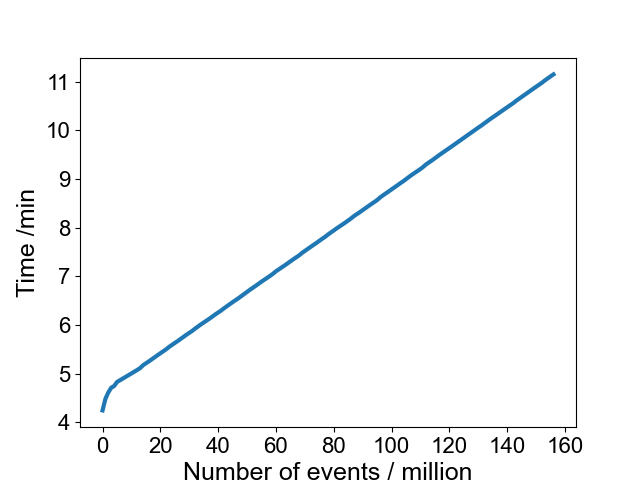}
  \caption{OMPRacer / Lulesh-157M-56th}
  \figlabel{scalability-ompracer}
\end{subfigure}
\caption{Time spent to process every million events for 4 selected traces.}
\figlabel{scalability}
\end{figure*}

%% file: related.tex

\section{Related Work}
\seclabel{related}

\myparagraph{Dynamic predictive analysis}
Happens-before (HB)~\cite{lamport1978hb} based race 
detection~\cite{pozniansky2003efficient,flanagan2009fasttrack} 
has been adopted by mature tools \cite{serebryany2009threadsanitizer, helgrind}, 
and has subsequently been strengthened
to SHB~\cite{shb} so that all races reported are sound.
Causal Precedence (CP) \cite{smaragdakis2012sound} and Weak Causal Precedence (WCP) \cite{wcp} weaken HB in favor of predictive power, and run in polynomial and linear time, respectively. 
Other works such as DC \cite{roemer2018high, roemer2020smarttrack} and SDP \cite{gencc2019dependence} are also partial order based methods that are either sound by design or perform graph-based analysis to regain soundness.
SyncP \cite{syncp}, M2 \cite{m2}, SeqCheck~\cite{seqc} work similar to \ssp, 
by constructing an appropriate set of events and appropriate linearization over this set. 
SMT solver backed approaches~\cite{huang2014maximal,said2011generating}
aim for sound and complete race prediction but do not scale to moderately large execution traces.
The complexity of data race prediction was extensively studied in~\cite{mathur2020complexity}
and was shown to be \NP-hard and also W[1]-hard, implying that an FPT algorithm
(parameterized by the number of threads) for race prediction is unlikely.
The fine-grained complexity of HB and \syncp was studied in~\cite{kulkarni2021dynamic}; in practice,
HB can be sped up using the tree clock data structure~\cite{TreeClock2022}.
Predictive analyses have also been developed for deadlocks~\cite{dirk2018,syncpdeadlock2023}, atomicity violations~\cite{Sorrentino2010,Aerodrome2020}, for more general temporal specifications~\cite{ang2023predictive} and more recently has been investigated from the lens of generalizing trace equivalence ~\cite{grainConcurrency2024}.

\myparagraph{Other concurrency testing approaches}
Static analysis techniques employ forms of lockset style reasoning~\cite{savage1997eraser}
to detect data races~\cite{naik2006effective,blackshear2018racerd,zhan2016echo,li2019sword} to report
data races, but are known to report false positives.
Model checking techniques for concurrent software~\cite{kokologiannakis2021genmc,Norris2013,Abdulla2014}
have been employed to detect concurrency bugs~\cite{godefroid2005software,oberhauser2021vsync}.
Another class of systematic exploration techniques include
controlled concurrency testing~\cite{deligiannis2023industrial,nekara2021}, including those that employ randomization~\cite{burckhardt2010randomized,yuan2018partial,luo2021c11tester}
and state-based learning~\cite{mukherjee2020learning}.
More recently, feedback driven randomized techniques have been employed for
testing concurrent programs~\cite{jeong2019razzer,xu2020krace}
Randomization has also been shown to reduce time overhead of dynamic data race detection~\cite{bond2010pacer,thokair2023dynamic,marino2009literace}.

\vspace{-0.2in}

%% file: conclusions.tex

\section{Conclusions and Future Work}
\seclabel{conclusions}

We propose \ssp, a sound polynomial time
race prediction algorithm that identifies data races
that can be witnessed by \emph{optimistically reversing synchronization operations}.
\ssp significantly advances the state-of-the-art in 
sound dynamic data race prediction.
\ssp-style reasoning can be helpful
for exposing other concurrency bugs such as deadlocks~\cite{syncpdeadlock2023,dirk2018}
and atomicity violations.

%% file: appendix.tex

\input{sub-appendix/appendix-sec3}
\input{sub-appendix/appendix-sec4}

\input{sub-appendix/appendix-sec5}

%% file: sub-appendix/appendix-sec3.tex

\section{Proofs from section 3}


\subsection{Proof of \thmref{np-hardness-lock-closed}}
\label{prf:np-hardness}

\begin{reptheorem}{thm:np-hardness-lock-closed}
    Let $\trace$ be a trace, let $e_1, e_2$ be conflicting events 
    and let $S \subseteq \events{\trace}$ be an optimistically lock-closed set.
    The problem of determining whether there is a correct reordering
    $\reordering$ such that $\events{\reordering} = S$ and 
    both $e_1$ and $e_2$ are $\trace$-enabled in $\reordering$ is \NP-hard.
\end{reptheorem}

\input{NPHard}


%% file: NPHard.tex

We prove this theorem by instead establishing the following stronger \thmref{np-hard-olclosure};
it claims that, the problem of determining if the \emph{smallest} optimistically lock-closed set
can be linearized, is \NP-hard problem.

\begin{theorem}
    \thmlabel{np-hard-olclosure}
    Let $\trace$ be a trace, let $e_1, e_2$ be conflicting events, and let $S = \SSP{e_1, e_2}$. The problem of determining whether there is a correct reordering $\reordering$ s.t. $\events{\reordering} = S$ is \NP-hard.
\end{theorem}

The high level idea behind our proof of \thmlabel{np-hard-olclosure}
is inspired from \cite{mathur2020complexity}, which shows that the problem of
checking if a given pair of conflicting events is a predictable data race, is \NP-hard.
In~\cite{mathur2020complexity}, the proof proceeds by first showing that an intermediate
problem, namely $\rfposet$ realizability, is \NP-hard.
An instance of this problem is a triple $\mathcal{P} = (X, P, RF)$, 
where $X$ is some set of read, write, acquire and release events,
$P \subseteq X \times X$ is a partial order on $X$ and $RF$ 
is a function that maps every read event $r \in X$
to a unique write event $w \in X$ on the same memory location.
$\mathcal{P}$ is a positive intance of $\rfposet$-realixability if 
there is a linearization $\trace$ of $X$ that respects $P \cup \setpred{(w, r)}{w = RF(r)}$ 
and also ensures that between any read (resp. release) event $r$ 
and its corresponding write (resp. matching acquire) event $w = RF(r)$,
there is no other write event of the same memory location (resp. lock) as $r$.
In~\cite{mathur2020complexity}, the \NP-hardness of $\rfposet$ realizability is 
established via a reduction from $\IND$, which is the problem 
of checking if for an input graph $G$, there is an independent set of $G$ of size at least $c$.
Following this, \cite{mathur2020complexity} establishes a reduction from $\rfposet$ realizability to the race prediction problem. 

Our proof is inspired from this, but is a direct reduction
from $\IND$ to our problem ---  given a trace 
$\trace$ and a pair 
$(e_1, e_2)$, determine if there is a correct reordering
containing exactly the events $\SSP{e_1, e_2}$.
Given an input graph $G$ (instance of $\rfposet$ realizability problem),
we construct trace $\trace$ with two events $e_1$ and $e_2$ in $\trace$ as follows.
We first construct an intermediate $\rfposet$ instace $\mathcal{P}$
by slightly modifying the $\rfposet$ instance constructed by \cite{mathur2020complexity},
ensuring that $\mathcal{P}$ is realizable iff the graph $G$ has an independent set of size $\geq c$.
Starting with $\mathcal{P}$, we can then construct a trace $\trace$ with two specific events $e_1, e_2$
such that $\mathcal{P}$ can be realized iff there is a correct reordering
of $\trace$ for which $\SSP{e_1, e_2}$ can be linearized. 

\input{./figures/NPHard-Construction}

\begin{proof}
Given an $\IND$ problem on graph $G$, we encode a $\rfposet$ realizability instance
 $\mathcal{P}$ as following.
 The set of events $X$ belong to $2c+2$ threads $t_1, t_2 \ldots t_{2c+2}$, and we describe the
 total order $\tau_i$ of events in each thread $t_i$ next.

\begin{enumerate}
    \item For $i = 2 \cdot c + 1$, $\tau_i = \writeOp{q}, \writeOp{v_{2c+1}}$ 

    \item For $i = 2 \cdot c + 2$, $\tau_i = \tau_i^1 \circ \tau_i^2$, where 
    \[\tau_i^1 = \readOp{s_1}, \dots, \readOp{s_c}, \acq(\lk_1), \dots, \acq(\lk_c), \readOp{q}, \writeSubs{2}{x}, \rel(\lk_c), \dots, \rel(\lk_1).\] 
    Let $C_i = \readOp{v_{i+c}} \cdot \readOp{v_{i}}$ and $\tau_i^2 = \readOp{v_{2c+1}} \circ C_c \circ \dots \circ C_1$.

    \item For each integer $1 \le i \le c$, $\tau_i = \tau_i^1 \circ \tau_i^2 \circ \dots \tau_i^n \circ \writeOp{v_i}$. 
    For $2 \le j \le n-1$, we have $\tau_i^j = \acqSubs{i}{\ell_{j,l_1}}, \dots, \acqSubs{i}{\ell_{j,l_m}}, \writeOp{y_i^j}, \\ \readOp{z_i^j}, \relSubs{i}{\ell_{j,l_m}}, \dots, \relSubs{i}{\ell_{j,l_1}}$, where $l_1, \dots, l_m$ denotes the neighbors of node $j$ in graph $G$. Let $\tau_i^1 = \acqSubs{i}{\ell_{1,l_1}}, \dots, \\ \acqSubs{i}{\ell_{1,l_m}}, \writeOp{s_i}, \readOp{z_i^1}$ $\relSubs{i}{\ell_{1,l_m}}, \dots, \relSubs{i}{\ell_{1,l_1}}$, and $\tau_i^n = \acqSubs{i}{\ell_{n,l_1}}, \dots, \acqSubs{i}{\ell_{n,l_m}}, \writeOp{y_i^n}, \readSubs{i}{x}, \relSubs{i}{\ell_{n,l_m}},$ $\dots \relSubs{i}{\ell_{n,l_1}}$.

    \item For each integer $c + 1 \le i \le 2c$, $\tau_i = \tau_i^1 \circ \tau_i^2 \circ \dots \tau_i^n \circ \writeOp{v_i}$, where $\tau_{i}^j = \acqUps{j}{\ell_{i-c}}, \writeOp{z_{i-c}^j}, \readOp{y_{i-c}^{j+1}}, \relUps{j}{\ell_{i-c}}$.
\end{enumerate}
See an example of the reduction outlined above in~\figref{NPHard-Construction}.
Each variable in $\mathcal{P}$ is written once, so the read-from $RF$ relation is clear. 
The partial order $P$ is the thread order induced by the threads $t_1, \ldots t_{2c+2}$.

We remark that the poset $\mathcal{P}$ is almost identical to the one
in~\cite{mathur2020complexity} (let's call it $\mathcal{P}'$), except for the
extra read and write events on variables $\set{v_1, \ldots, v_{2c+1}}$ 
at the end of each thread in $\mathcal{P}$.
We will use $X_{\mathcal{P}'}$ to denote the subset of events that belong to $\mathcal{P}'$.

Let us first argue that $\mathcal{P}$ is realizable iff $\mathcal{P}'$
is realizable.
If $\mathcal{P}'$ is realizable, then there is a linearization $\rho'$ of 
$X_{\mathcal{P}'}$ that preserves thread order and the reads-from of events in $X_{\mathcal{P}'}$.
Consider the the trace $\rho = \rho', \wt(v_1) \ldots \wt(v_{2c}), \rd(v_1) \ldots \rd(v_{2c})$,
where we omit the obvious thread identifiers of events on $\set{v_1, \ldots, v_{2c+1}}$.
Clearly, $\rho'$ witnesses the realizability of $\mathcal{P}$.
Now, if $\mathcal{P}'$ is realizable using a linearization $\rho$, it is easy to argue that
the linearization $\rho'$ obtained by removing events of memory locations 
$\set{v_1, \ldots, v_{2c+1}}$ witnesses the realizability of $\mathcal{P}'$.
It thus also follows that $G$ is a positive instance of $\IND$ iff
$\mathcal{P}$ is realizable.

Let us now construct the trace $\trace$ and complete our reduction.
The set of events of trace will be $\events{\trace} = X \uplus \set{\wt_1(a), \wt_2(a)}$,
where $a$ is a fresh memory location and $\wt_1(a)$
writes to $a$ as the new last event of $t_{2c+1}$, while $\wt_2(a)$ writes to $a$
as the new last event of $t_{2c+2}$.
Observe that in $\mathcal{P}$, every write event in thread $\tau_i$ is read by events in $\tau_{c+i}$ for $1 \le i \le c$. 
We let $t_i$ be an arbitrary interleaving of $\tau_i$ and $\tau_{i+c}$ 
that respects the thread order and read-from relation. 

We construct the trace $\trace$ to be the following.
\begin{equation}
    \trace = \tau_{2c+1} \circ \wt_1(a) \circ t_{1} \circ \dots \circ t_{c} \circ \tau_{2c+2}\circ \wt_2(a)  \nonumber
\end{equation}

First, it is clear that $\SSP{w_1(a), w_2(a)} = X$.
Thus, $\SSP{w_1(a), w_2(a)}$ can be linearized iff $\mathcal{P}$ can be realized.
Consequently, the input graph $G$ has an independent set of size $\geq c$ iff $(\wt_1(a), \wt_2(a))$
is witnessed as a race of $\trace$ using $\SSP{w_1(a), w_2(a)}$.
Finally, it is clear that the construction takes polynomial time in the size of the graph $G$.
Thus, it follows that the problem of checking if, for a given trace $\trace$
and a pair of conflicting events $(e_1, e_2)$ in $\trace$,
whether there is a correct reordering $\reordering$ of $\trace$ with $\events{\reordering} = \SSP{e_1, e_2}$, is also $\NP$-hard. 
\end{proof}

%% file: figures/NPHard-Construction.tex
\begin{figure*}
    \centering
    \includegraphics[scale=0.75]{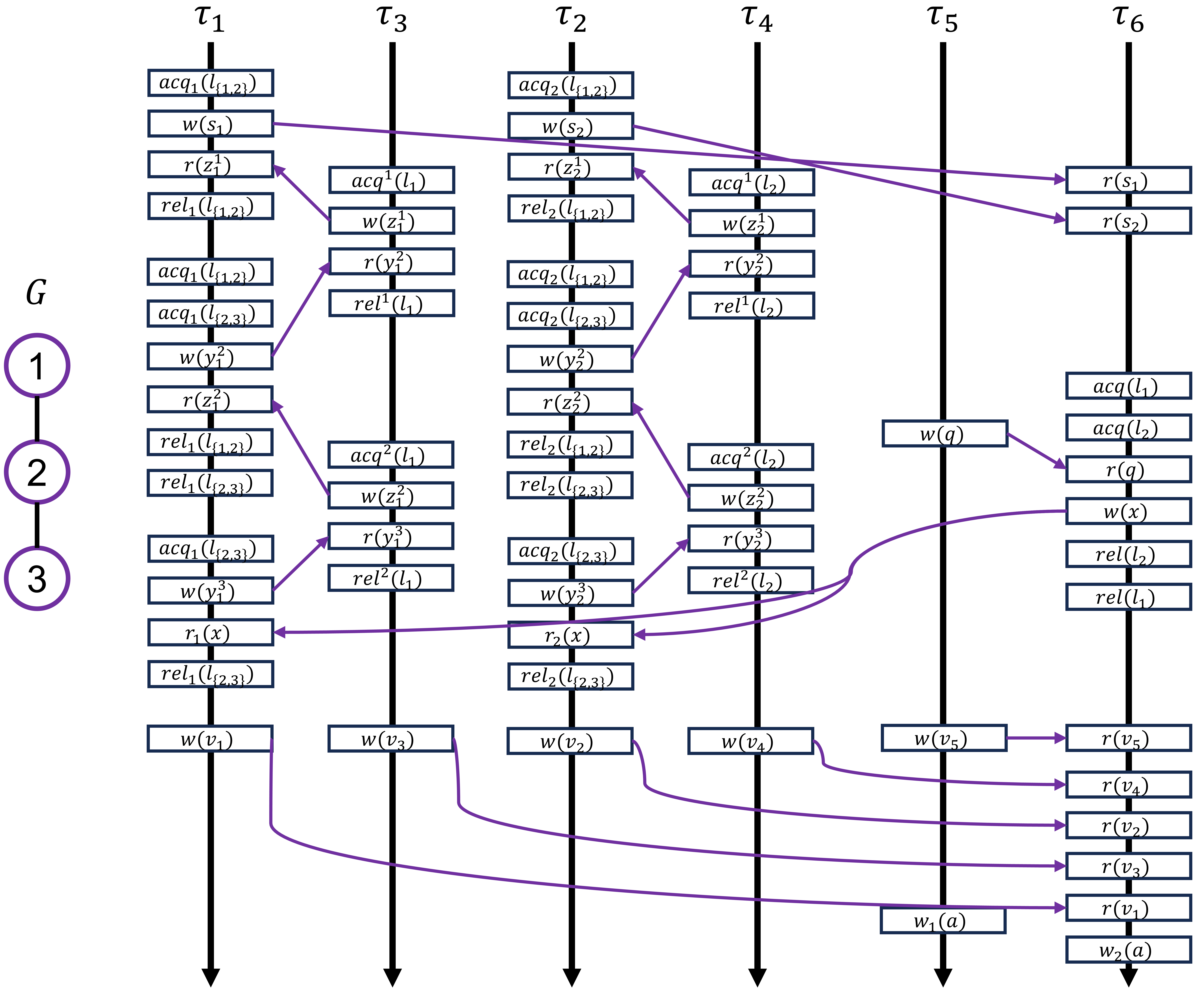}
    \caption{Given a graph $G$ and independent set size of 2, our construction to show NP-Hardness of linearizing $\SSP{e_1, e_2}$}
    \label{fig:NPHard-Construction}
\end{figure*}

%% file: sub-appendix/appendix-sec4.tex

\section{Proofs from section 4}
\subsection{Proof of \lemref{osr-race-closure-set}}

\begin{replemma}{lemma:osr-race-closure-set}
Let $(e_1, e_2)$ be a conflicting pair
of events in trace $\trace$. 
If $(e_1, e_2)$ is an optimistic sync-reversal race, then
it can be witnessed in an optimistic correct reordering $\reordering$
with $\events{\reordering} = \SSP{e_1, e_2}$.
\end{replemma}

\begin{proof}[Proof Sketch]
Given $\trace$ and two conflicting events $\mathsf{e_1, e_2}$, let $S = \SSP{e_1, e_2}$ and $X$ be an arbitrary optimistically lock-closed event set. 
By definition, $S$ is the smallest optimistically lock-closed set and thus $S \subseteq X$. Let $\optgraph_S, \optgraph_X$ be the optimistic-reordering-graph of $S$ and $X$, respectively, 
We now show if there is a cycle in $\optgraph_S$, then $\optgraph_X$ also has a cycle.

First we consider the nodes in $\optgraph_S$ and we have for any node $v$ in $\optgraph_S$, $v$ must be in $\optgraph_X$, because $X$ is a superset over $S$. 
Second we show for all forward edges $u \rightarrow v$ in $\optgraph_S$, this edge also exists in $\optgraph_X$. 
This is because $u, v$ are also nodes in $X$, and the edge $u \rightarrow v$ exists, if both $u, v$ are in $X$. 

Next, we show that for each backward edges $(r, a)$ in $\optgraph_S$, 
there is a path from $r$ to $a$ in $\optgraph_X$;
here, by backward edge we mean that $v \trAfter u$.
Notice that since $(r, a)$ is a backward edge,
it must be that $r = \last{\rel(\lk)}_S$
be the last release of lock $\lk$ and $a = \acqO{\lk}_S$ is an unmatched
acquire in $S$, for some $\lk$. 
We first establish that indeed $a$ must also be unmatched in $X$. 
Since $S$ is optimistically lock closed, it must be that
$\set{e_1, e_2} \cap \TLC{\match{\trace}{a}} \neq \emptyset$. 
If, on the contrary, $\match{\trace}{a} \in X$,
the fact that $X$ is $(\thAfter, \lwsingle{\trace})$-closed,
we must have $\set{e_1, e_2} \cap X \neq \emptyset$. 
Clearly this would contradicts the fact that $X$ is optimistically lock-closed. 
Thus, $a$ is unmatched even in $X$. 
Further, from the definition of $\optgraph_X$, it follows that $(r', a)$ is an edge
of $\optgraph_X$.

Now let $r' = \last{\rel(\lk)}_X$
We have $r \trAfter r'$, because $S \subseteq X$. 
This means that there is a path in $\optgraph_X$
of the form $r \to r' \to a$ in $\optgraph_X$.
In other words, all paths of $\optgraph_S$ are preserved in $\optgraph_X$
and thus, so are the cycles of $\optgraph_S$.

By definition, if $\mathsf{(e_1, e_2)}$ is an optimistic sync-reversal race, 
there is a optimistic correct reordering $\reordering$, 
s.t. the optimistic-reordering-graph of $\events{\reordering}$ has no cycle. 
Then we can conclude $\optgraph_S$ also has no cycle, 
thus the linearization of $\optgraph_S$ can also witness this race (thanks to \lemref{opt-reordering-graph}).
\end{proof}

\subsection{Proof of \lemref{opt-reordering-graph}}
\begin{replemma}{lemma:opt-reordering-graph}
Let $\trace$ be a trace and let $S \subseteq \events{\trace}$ such
that $S$ is $(\thAfter, \lwsingle{\trace})$-closed and also lock-feasible.
Then, there is an optimistic reordering $\reordering$ of $\trace$ on the set $S$
iff the graph $\optgraph$ is acyclic.
\end{replemma}

\begin{proof}[Proof Sketch]
    Let us first assume that $\optgraph_S$ is acyclic.
    Consider a linearization $\reordering$ of $\optgraph_S$.
    We argue that $\reordering$ is an optimistic reordering of $\trace$.
    First, $S$ is $(\thAfter, \lwsingle{\trace})$-closed and $\optgraph_S$
    orders all events of the same thread as in $\trace$.
    Hence $\reordering$ respects $\thAfter$.  
    Second, for every read event $r$, its corresponding writer $\lwsingle{\trace}(r)$ is in $S$,
    and further is ordered before $r$ in the graph $\optgraph_S$, and every other conflicting write $w'$ is either after $r$ or before $w$ in $\trace$ and thus in $\optgraph_S$ and thus in $\reordering$.
    Finally, lock semantics are preserved since $S$ is lock-feasible, the matched critical sections are totally ordered and further the unmatched acquire is ordered after every other release of the same lock.
    Finally, $\reordering$ is an optimistic reordering because the order of matched critical sections
    and the order of conflicting events is preserved because they are explicit edges in $\optgraph_S$.

    Now assume that there is an optimistic reordering $\reordering$ of $\trace$ with $\events{\reordering} = S$. 
    We will argue that for every edge $(e_1, e_2)$ of $\optgraph_S$,
    we have $e_1 \trAfterReorder e_2$.
    This would imply that $\optgraph_S$ is acyclic, since $\reordering$ is acylic.
    First, consider the case when $(e_1, e_2)$ is such that $e_1 \thAfter e_2$.
    Since $\reordering$ is a correct reordering of $\trace$, we must also have $e_1 \thAfterReorder e_2$.
    Second, consider the case when $e_1 = \rel(\lk)$ and $e_2 = \acq(\lk)$.
    If $e_2$ is matched in $S$, then we have $e_1 \trAfterReorder e_2$ since $\reordering$ is an optimistic reordering.
    Otherwise, $e_2$ is an unmatched acquire and must be placed last in $\reordering$ anyway.
    Finally, if $\cfwith{e_1}{e_2}$, then the fact that $e_1 \trAfter e_2$ and that $\reordering$
    orders conflicting events the same way as $\trace$ implies that $e_1 \trAfterReorder e_2$.
\end{proof}

\subsection{Proof of \lemref{incremental-closure-comp}}

\begin{replemma}{lemma:incremental-closure-comp}
Let $e_1, e_2, e'_2 \in \events{\trace}$ be events in trace $\trace$
with $e_2 \thAfter e'_2$. 
Let  $S = \SSP{e_1, e_2}$ and let $S' = \SSP{e_1, e'_2}$.
We have the following:
\begin{enumerate*}
    \item $S \subseteq S'$.
    \item $S =$ \sspclosure{$e_1, e_2, \emptyset$}, and further this call (in \algoref{computessp}) takes $\widetilde{O}(|S|)$ time.
    \item $S' =$ \sspclosure{$e_1, e'_2, S$}, and further this call (in \algoref{computessp}) takes $\widetilde{O}(|S'| - |S|)$ time.
\end{enumerate*}
\end{replemma}

\begin{proof}[Proof Sketch]
We first show $\sspclosure(e_1, e_2, \varnothing) = S$. 
For convenience, we denote $S_i$ as $S$ after i-th iteration in \algoref{computessp}. 
Now we prove $\sspclosure(e_1, e_2, \varnothing) \subseteq S$ by induction.
\begin{itemize}
    \item Firstly, the initial set 
    $S_0 \subseteq S$ 
    \item Assuming after $i$ iterations, $S_i \subseteq S$, 
    we show $S_{i+1} \subseteq S$. 
    By definition of \sspclosure, 
    $S_{i+1} = S_i \cup \TLC{\rel}$, 
    where $\match{\trace}{\rel} \in S_i$ and 
    $\set{e_1, e_2} \cap \TLC{\rel} = \varnothing$. 
    Since $S_i \subseteq S$, 
    by definition of $\SSP{e_1, e_2}$, 
    we have $\rel \in S$ and therefore $S_{i+1} \subseteq S$
\end{itemize}
So far we proved $\sspclosure(e_1, e_2, \varnothing) \subseteq S$. 
Further we claim $\sspclosure(e_1, e_2, \varnothing)$ is optimistic lock-closed. 
Otherwise, the while loop in \algoref{computessp} will not terminate. 
This proves $S = \sspclosure(e_1, e_2, \varnothing)$.

Now we prove $S \subseteq S'$. 
We consider an arbitrary run of \algoref{computessp} on computing $S = \sspclosure(e_1, e_2, \varnothing)$, 
and construct another valid run of \algoref{computessp} on computing $S' = \sspclosure(e_1, e'_2, \varnothing)$. 
During this process, we prove after $i$ iterations, 
$S_i \subseteq S'_i$ for all $i$.
\begin{enumerate}
    \item We observe $S_0$ $\subseteq$ $S'_0$, as $\prev{\sigma}{e_2} \in \TLC{\prev{\sigma}{e_2'}}$.

    \item Assuming $S_i$ $\subseteq$ $S'_i$, we prove $S_{i+1}$ $\subseteq$ $S'_{i+1}$. 
    
    For all release event $\rel(\lk)$, if $\rel(\lk) \notin S_i$ and $\set{e_1, e_2} \cap \TLC{\rel(\lk)} = \varnothing$, we have $\set{e_1, e_2'} \cap \TLC{\rel(\lk)} = \varnothing$, because $e_2 \thAfter e_2'$. 
    Then there are two possibilities. 
    The first being that $\rel(\lk) \in S'_i$, 
    which means $S'_i$ is already a superset of $S_{i+1}$, 
    so that $S_{i+1} \subseteq S'_i \subseteq S'_{i+1}$.
    Alternatively, if  $\rel(\lk) \notin S'_i$, 
    for any update we do for $S_i$, 
    we can also do the same update for $S'_i$. 
    Therefore, after one more iteration, 
    the observation of $S_{i+1} \subseteq S'_{i+1}$ still holds.
\end{enumerate}
The observation above proves $\sspclosure(e_1, e_2, \varnothing) \subseteq \sspclosure(e_1, e'_2, \varnothing)$. 
Since $\sspclosure(e_1, e_2, \varnothing)$ 
$= S$ and 
$\sspclosure(e_1, e'_2, \varnothing) = S'$, 
we have $S \subseteq S'$.


Now we show $S$ can be computed by \sspclosure in $\Tilde{O}(|S|)$ time. 
In each iteration of \sspclosure, we need $O(\NumThreads \NumLocks)$ time to check if any updates can be done, 
and there are at most $|S|$ iterations. 
Therefore, each event is visited at most $\Tilde{O}(\NumThreads \NumLocks)$ times. 



For the third conclusion, 
we can take $S$ as a start point and call \sspclosure{$e_1, e'_2, S$} to compute $S'$, 
as $S \subseteq S'$. 
Following the proof of the second conclusion, 
this takes $\Tilde{O}(|S'| - |S|)$ time.
It remains to show \sspclosure{$e_1, e'_2, S$} returns $S'$.
We show this by induction. 
We denote $S'_i$ as $S'$ after $i$ iterations in \algoref{computessp}.
\begin{itemize}
    \item It's obvious that $S'_0 \subseteq S'$
    \item Assuming $S'_i \subseteq S'$, we show $S'_{i+1} \subseteq S'$. 
    By definition of \sspclosure, 
    $S'_{i+1} = S'_{i} \cup \TLC{\rel}$, 
    where $\match{\trace}{\rel} \in S'_i$ and 
    $\set{e_1, e'_2} \cap \TLC{\rel} = \varnothing$. 
    Since $S'_i \subseteq S'$, 
    by definition of $\SSP{e_1, e'_2}$, 
    we have $\rel \in S'$ and therefore $S'_{i+1} \subseteq S'$
\end{itemize}

This proves \sspclosure{$e_1, e'_2, S$} indeed returns $S'$. 

\end{proof}

\subsection{Proof of \thmref{osr-race-given-two-events}}
\begin{reptheorem}{theorem:osr-race-given-two-events}
Let $\trace$ be a trace and let $e_1, e_2$ be conflicting events in $\trace$.
The problem of determining if $(e_1, e_2)$ is an optimistic sync-reversal race
can be solved in time $O\big(\NumThreads(\NumThreads\NumEvents + \NumLocks)\big) = \widetilde{O}(\NumEvents)$ time.
\end{reptheorem}

\begin{proof}
    For given $e_1, e_2$, to determine if they are OSR race, 
    we firstly compute their optimistic lock closure, check for lock-feasibility and then build the abstract graph to check for cycles. 
    We have shown in 
    \secref{check-single-race} that for any given $e_1, e_2$, 
    $\SSP{e_1, e_2}$ can be computed in $O(\NumThreads^2 \NumEvents)$. 
    Lock-feasibility can be checked in $O(\NumThreads \NumLocks)$ time.

    To build the graph, we firstly add all vertices and backward edges. 
    Later, we compute earliest successors for each vertex in the graph and add forward edges correspondingly. 
   The abstract graph contains at most $2 \NumLocks$ nodes by definition. 
   Also in \secref{incremental-race}, we have shown that it takes $O(\NumLocks)$ time to add all backward edges 
   and $O(\NumThreads^2 \NumLocks)$ time to add all forward edges. 
   Checking cycles in the graph takes $O(\NumLocks + \NumLocks^2)$ time, as there are at most $O(\NumLocks)$ vertices and $O(\NumLocks^2)$ edges. 
   Therefore, building the graph and checking for cycle take $O(\NumLocks + \NumLocks + \NumThreads^2 \NumLocks + \NumLocks^2)$, 
   i.e. $O(\NumLocks (\NumThreads^2 +\NumLocks))$ in total.
   
   To do race detection on given $e_1, e_2$, 
    it takes $O(\NumEvents \NumThreads^2 + 
    \NumLocks \NumThreads + 
    \NumLocks (\NumThreads^2 +\NumLocks))$, i.e. $O(\NumThreads^2 \NumEvents + \NumLocks^2)$
\end{proof}

\subsection{Proof of \thmref{osr-race-given-event}}

\begin{reptheorem}
{theorem:osr-race-given-event}
    Let $\trace$ be an execution, $e \in \events{\trace}$
be a read or write event and let $t \in \threads{\trace}$.
The problem of checking if there is an event $e'$ 
with $\ThreadOf{e'} = t$ such that $(e, e')$ is an optimistic-sync-reversal race, can be solved
in time $O\big((\NumThreads^2 + \NumLocks) \NumLocks \NumEvents \big)$.
\end{reptheorem}

\begin{proof}
    Following \algoref{detect-races-given-one-event}, 
    the computation \sspclosure for each $(e, e')$, s.t. $\ThreadOf{e'} = t$ is equivalent to compute the \sspclosure for $e$ and the last $e'$ in thread $t$, which can be done in $O(\NumThreads^2 \NumEvents)$ time. 

    We also need to check lock-feasibility, build graph and check cycles for each $e'$ in $t$. 
    There are at most $\NumEvents$ such $e'$. 
    The total time complexity to do so is $O(\NumEvents(\NumLocks \NumThreads + \NumLocks(\NumThreads^2 + \NumLocks)))$, i.e. $O(\NumEvents \NumLocks(\NumThreads^2 + \NumLocks))$.

    In total, we need $O(\NumThreads^2 \NumEvents + \NumEvents \NumLocks(\NumThreads^2 + \NumLocks))$, 
    i.e. $O\big((\NumThreads^2 + \NumLocks) \NumLocks \NumEvents \big)$ time to check for all races between $(e, e')$, s.t. $\ThreadOf{e'} = t$
\end{proof}

\subsection{Proof of \thmref{osr-race-all}}
\begin{reptheorem}
{theorem:osr-race-all}
    Given a trace $\trace$, the problem of checking if
$\trace$ has an optimistic sync-reversal data race, can be solved in
time  $O\big(\NumThreads \NumLocks (\NumThreads^2+\NumLocks)\NumEvents^2\big) = \widetilde{O}(\NumEvents^2)$ time.
\end{reptheorem}

\begin{proof}
    Following \algoref{detect-races}, 
    we iterate over all events and for a fixed event $e$, we iterate over all threads. 
    Therefore, \algoref{detect-races-given-one-event} is called at most $O(\NumThreads \NumEvents)$ times. 
    Then the total complexity to check for races is bound by $O(\NumThreads \NumEvents \cdot (\NumThreads^2 + \NumLocks) \NumLocks \NumEvents)$, i.e. $O(\NumThreads \NumLocks (\NumThreads^2+\NumLocks)\NumEvents^2\big)$ time.
\end{proof}

\subsection{Proof of \lemref{abstract-graph-preserves-cycle}}

\begin{replemma}{lemma:abstract-graph-preserves-cycle}
Let $\trace$ be a trace and let $S \subseteq \events{\trace}$
be a $(\thAfter, \lwsingle{\trace})$-closed set.
$\optgraph_S$ has a cycle iff $\redgraph_S$ has a cycle.
\end{replemma} 

\begin{proof}[Proof Sketch]
    Firstly, we show if there is a cycle $C$ in $\redgraph_S$, then there is a cycle $C'$ in $\optgraph_S$. 
    For every edge $u \rightarrow v$ in $C$, if it is a forward edge, 
    then we replace it with the corresponding forward path from $u$ to $v$. 
    If $u \rightarrow v$ is a backward edge, then we keep as it is. 
    After this substitution, we get the replaced $C$ as a cycle $C'$ in $\optgraph_S$. 

    If there is a cycle $C'$ in $\optgraph_S$, then there is a cycle $C$ in $\redgraph_S$. 
    Considering the edges $E$ in $C'$, $E$ must contain backward edges, otherwise $C'$ cannot be a cycle. 
    Let $E_b$ be the set of backward edges in $E$ and $V_b$ be the set of nodes in $E_b$. 
    We observe that $V_b$ is a subset of the vertices in $\optgraph_S$, because the vertex set of $\optgraph_S$ is a super set over $\redgraph_S$. 
    
    Therefore, $C$ can be constructed as following. 
    First we keep all last release and open acquire event as nodes in $C'$. 
    Second we add all backward edges in $C$ to $C'$. 
    Lastly, we replace all forward paths (paths don't contain backward edges) in $C$ with a direct edge and add them into $C'$. 
    We now have successfully constructed cycle $C'$ in $\redgraph_S$. 
\end{proof}

\subsection{Proof of \thmref{n2-hardness}}
\label{appendix:n2-hardness}
\begin{reptheorem}{theorem:n2-hardness}
    Assume SETH holds.
    Given an arbitrary trace $\trace$, the problem of determining if $\trace$ has an \ssp race cannot
    be solved in time $O(\NumEvents^{2-\epsilon})$ (where  $\NumEvents = |\events{\trace}|$) for every $\epsilon>0$.
\end{reptheorem}

\input{N2Hardness}

%% file: N2Hardness.tex
\myparagraph{Orthogonal Vector Hypothesis (OV)}
The Orthogonal Vectors problem is defined as following. 
Given two sets $A, B$ each containing $\NumEvents$ $d$-dimensional 0-1 vectors, 
where $d = \widetilde{O}(log(\NumEvents))$, determine if there exists two vectors $v_1 \in A, v_2 \in B$, 
s.t. $(v_1, v_2)$ has an inner product of zero. 
OV Hypothesis is a well-known conjecture and it has been widely accepted that it's not likely to give a sub-quadratic algorithm to solve the Orthogonal Vector problem \cite{chen2019equivalence, kulkarni2021dynamic}, 
i.e. OV Hypothesis states OV Problem has a lower bound of $\widetilde{O}(\NumEvents)^2$.

We now reduce the existence problem of \ssp race to the OV problem and show that the problem of determining if there is a \ssp race in $\trace$ also has a lower bound of $\widetilde{O}(\NumEvents)^2$, unless OV Hypothesis fails.

\input{./figures/N2HardnessCons.tex}

\begin{proof}
    Given two sets $A, B$ of $\NumEvents$ $d$-dimensional 0-1 vectors, we construct a trace $\trace$ as following (shown in Figure \ref{fig:N2-Construction}). $\trace$ contains two threads $t_A, t_B$. As $A, B$ are finite sets, we enumerate elements from $A, B$ as $a_1, a_2, \dots, b_1, b_2, ...$. 
    For an arbitrary vector $v_i = (i_1, ..., i_d)$, assuming it contains $k$ non-zero bits, 
    we use a list $(j_1, ..., j_k)$ to denote the index of non-zero elements in $v_i$. 
    For example, vector $(0, 1, 0, 1)$ has non-zero bits $[2, 4]$, as its 2nd and 4th bits are 1.
    We define an event clause associated with vector $v_i$ as $C_{v_i} = \acq(\lk_{j_1}) \circ \dots \circ \acq(\lk_{j_k}) \circ  \wt(x) \circ  \rel(\lk_{j_k}) \circ  \dots \circ \rel(\lk_{j_1})$. Let $t_A = C_{a_1} \circ C_{a_2} \circ \dots \circ C_{a_\NumEvents}$ and $t_B = C_{b_1} \circ C_{b_2} \circ \dots \circ C_{b_\NumEvents}$. And we require $\forall \; e \in t_A, e' \in t_B$, $e \trAfter e'$. Then we observe a total order $\trAfter$ on $\trace$.

    Now we show there is a pair of orthogonal vectors in $A, B$, iff there is a \ssp race in $\trace$. 
    If there is a \ssp race $w_a(x), w_b(x)$ in $\trace$, they must correspond to vector $a \in A, b \in B$. 
    Since $(w_a(x), w_b(x))$ is a data race, 
    they must be from different threads and their lock set must be disjoint. 
    Therefore $\forall \; 1 \le i \le d$, either $a[i] = 0$ or $b[i] = 0$, thus $a, b$ are orthogonal. 

    If there is a pair of orthogonal vector $a, b$, then we consider their clause $C_a, C_b$. Let $w_a(x), w_b(x)$ be the two write operations in $C_a, C_b$ and now we show $w_a, w_b$ is a \ssp race. For convenience, let $S = \SSP{w_a(x), w_b(x)}$. The following observations hold.
    \begin{enumerate}
        \item $w_a(x) \in \events{t_A}$ and $w_b(x) \in \events{t_B}$, so that $a \in A$ and $b \in B$.

        \item $S = \set{e \; | \; e \thAfter \prev{\trace}{w_a(x)}} \; \cup \; \set{e \; | \; e \thAfter \prev{\trace}{w_b(x)}}$ 
        and $\forall$ lock $\lk \in \locks{S}$, there is at most one open acquire on $\lk$, because $a, b$ are orthogonal, so the clause $C_a, C_b$ don't hold the same lock. 
        This proves $S$ is potentially feasible

        \item We guarantee $S$ has no cycles, as no direct edge is from an acquire event to other events except thread order.
    \end{enumerate}

Following the definition, it's obvious to see $w_a(x)$ and $w_b(x)$ is a \ssp race, 
and thus we have proved there is a pair of orthogonal vectors in $A, B$, 
iff there is a \ssp race in $\trace$. 
If OV Hypothesis holds, then the problem of checking existence of \ssp race has a lower bound of $\widetilde{O}(\NumEvents)^2$.
 
\end{proof}

%% file: figures/N2HardnessCons.tex
\begin{figure}
    \centering
    \includegraphics[scale=0.6]{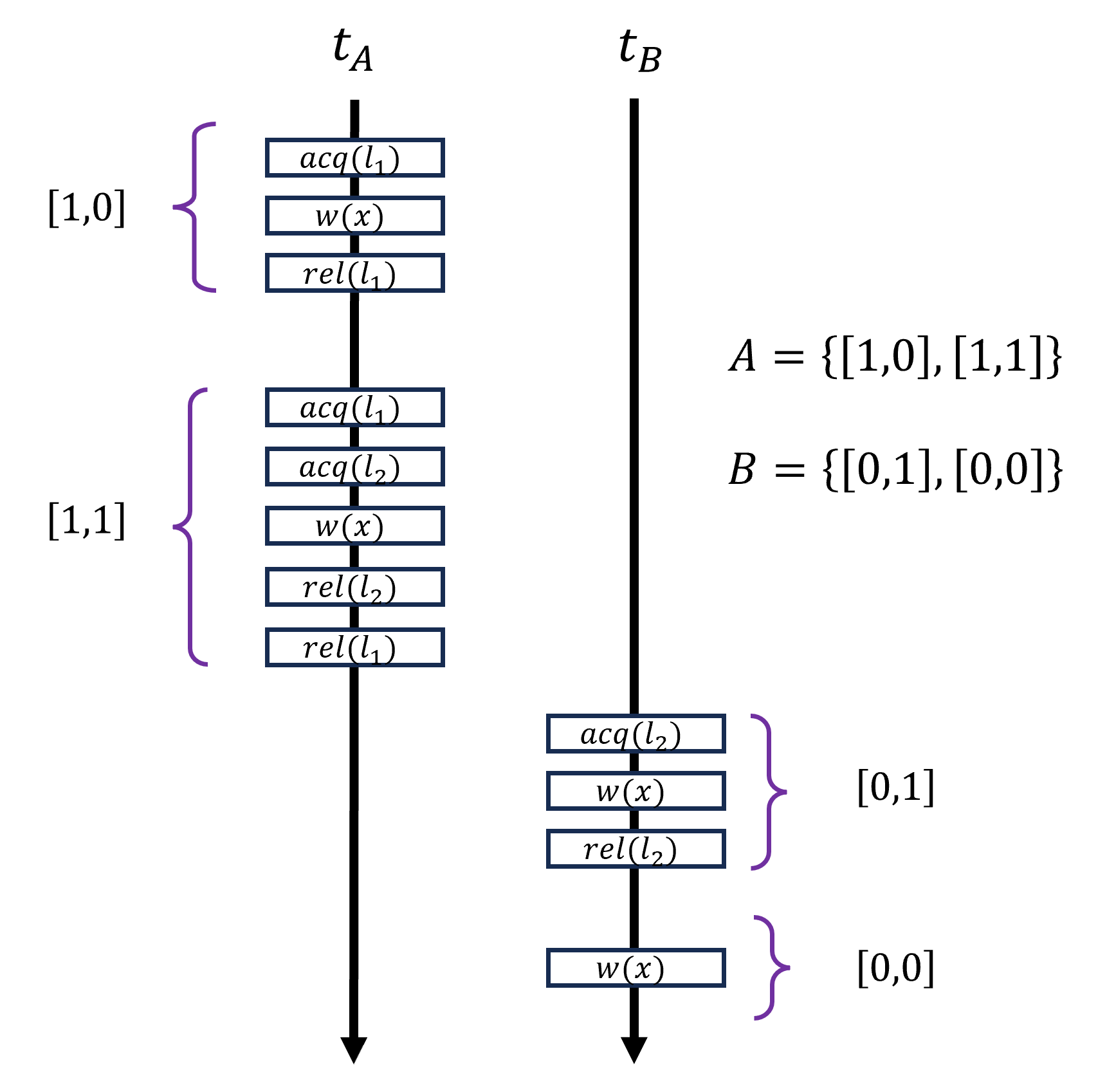}
    \caption{Given two sets $A, B$ of vectors of length 2, our construction to show quadratic hardness of \ssp race detection}
    \label{fig:N2-Construction}
\end{figure}

%% file: sub-appendix/appendix-sec5.tex
\section{Extra tables for section 5}
\applabel{sec-5}

\input{tables/SPInfoTable}

\input{tables/RaceInjectorTable}

\input{tables/ExtraNumRacesDetail}

\input{tables/traceInfoForCpp}

%% file: tables/SPInfoTable.tex
\begin{table*}

\caption{Statistics of the Java benchmarks. $\NumEvents$, $\NumThreads$, $\NumVars$, $\NumLocks$, $\NumReads$, $\NumWrites$, $\NumAcqs$ are the number of events, threads, variables, locks, read events, write events and acquire events after filtering, respectively. }
\label{tab:java-statistics}

\begin{tabular}[t]{ |c|c|c|c|c|c|c|c|  |c|c|c|c|c|c|c|c| }
 \hline
 
 \textbf{Benchmark} & $\NumEvents$ & $\NumThreads$ & $\NumVars$ & $\NumLocks$ & $\NumReads$ & $\NumWrites$ & $\NumAcqs$ & \textbf{Benchmark} & $\NumEvents$ & $\NumThreads$ & $\NumVars$ & $\NumLocks$ & $\NumReads$ & $\NumWrites$ & $\NumAcqs$ \\
 \hline
array & 11 & 3 & 2 & 1 & 1 & 4 & 2 & 
critical & 11 & 4 & 1 & 0 & 2 & 4 & 0 \\       

account & 15 & 4 & 1 & 0 & 6 & 5 & 0 &
airtickets & 18 & 5 & 1 & 0 & 9 & 5 & 0 \\

pingpong & 24 & 7 & 2 & 0 & 10 & 8 & 0 &                        
twostage & 83 & 12 & 2 & 2 & 20 & 12 & 20 \\              

wronglock & 122 & 22 & 1 & 2 & 40 & 21 & 20 &                   
bbuffer & 9 & 3 & 1 & 0 & 2 & 5 & 0 \\               

prodcons & 246 & 8 & 3 & 1 & 125 & 41 & 34 &               
clean & 867 & 8 & 2 & 2 & 286 & 96 & 239 \\

mergesort & 167 & 5 & 1 & 1 & 55 & 7 & 49 &                     
bubblesort & 1.6K & 13 & 25 & 1 & 1.1K & 263 & 119 \\        

lang & 1.8K & 7 & 100 & 0 & 1.3K & 500 & 0 &                  
readswrites & 10K & 5 & 6 & 1 & 4.2K & 2.2K & 1.7K \\   

raytracer & 526 & 3 & 3 & 0 & 514 & 9 & 0 &                      
bufwriter & 10K & 6 & 6 & 1 & 5.3K & 2.2K & 1.4K \\        

ftpserver & 17K & 11 & 135 & 143 & 7.9K & 0.8K & 4.2K &     
moldyn & 21K & 3 & 2 & 0 & 21K & 68 & 0 \\                   

linkedlist & 0.9M & 12 & 932 & 1 & 0.9M & 1.9K & 1.0K & 
derby & 75K & 4 & 190 & 133 & 19K & 12K & 22K \\

jigsaw & 3.2K & 8 & 51 & 45 & 551 & 498 & 1.1K &  
sunflow & 3.3K & 17 & 20 & 7 & 2.0K & 125 & 585 \\

cryptorsa & 1.3M & 7 & 18 & 27 & 709K & 287K & 156K &
xalan & 671K & 7 & 72 & 138 & 205K & 99K & 184K \\  

lufact & 891K & 5 & 6 & 1 & 5.3K & 2.2K & 1.4K &
batik & 131 & 7 & 5 & 0 & 115 & 10 & 0 \\

lusearch & 751K & 8 & 77 & 4 & 751K & 172 & 53 &  
tsp & 15M & 10 & 189 & 2 & 15M & 30K & 91 \\

luindex & 16K & 3 & 9 & 4 & 2.6K & 66 & 6.6K &
sor & 1.9M & 5 & 4 & 1 & 633K & 804 & 633K \\  

\hline
\end{tabular}
\end{table*}

%% file: tables/RaceInjectorTable.tex
\onecolumn
\begin{longtable}{|c|c|c|  |c|c| |c|c| |c|c| |c|c| |c|c|}

\caption{Summarized races and running time (in seconds) for RaceInjector traces}
\label{tab:race-injector}

\endfirsthead
\hline
1 & 2 & 3 & 4 & 5 & 6 & 7 & 8 & 9 & 10 & 11 & 12 & 13 \\
\hline

\textbf{Benchmark} & Trace & $\NumEvents$ & \multicolumn{2}{c||}{SHB} & \multicolumn{2}{c||}{WCP} & \multicolumn{2}{c||}{SyncP} & \multicolumn{2}{c||}{M2} & \multicolumn{2}{c|}{OSR} \\

\hline

&   &   & Races & Time & Races & Time & Races & Time & Races & Time & Races & Time \\

\hline

\endhead

\hline 

\endfoot

\endlastfoot

\hline
1 & 2 & 3 & 4 & 5 & 6 & 7 & 8 & 9 & 10 & 11 & 12 & 13 \\
\hline

\textbf{Benchmark} & Trace & $\NumEvents$ & \multicolumn{2}{c||}{SHB} & \multicolumn{2}{c||}{WCP} & \multicolumn{2}{c||}{SyncP} & \multicolumn{2}{c||}{M2} & \multicolumn{2}{c|}{OSR} \\

\hline

&   &   & Races & Time & Races & Time & Races & Time & Races & Time & Races & Time \\

\hline

\multirow{16}{*}{SHB-missed/ArrayList-27th}
 & 43 & 494 & 37 & 0.12 & 28 & 0.18 & 37 & 0.21 & 37 & 0.36 & 37 & 0.16 \\ 

 & 45 & 494 & 37 & 0.11 & 28 & 0.16 & 37 & 0.22 & 37 & 0.34 & 37 & 0.17 \\ 

 & 47 & 494 & 37 & 0.12 & 28 & 0.18 & 37 & 0.21 & 37 & 0.33 & 37 & 0.15 \\ 

 & 49 & 494 & 37 & 0.12 & 28 & 0.17 & 37 & 0.22 & 37 & 0.35 & 37 & 0.16 \\

 & 51 & 494 & 37 & 0.11 & 28 & 0.17 & 37 & 0.19 & 37 & 0.34 & 37 & 0.16 \\

 & 54 & 494 & 37 & 0.12 & 28 & 0.17 & 37 & 0.2 & 37 & 0.36 & 37 & 0.17 \\

 & 66 & 494 & 37 & 0.12 & 28 & 0.21 & 37 & 0.2 & 37 & 0.37 & 37 & 0.16 \\

 & 91 & 494 & 37 & 0.11 & 28 & 0.16 & 37 & 0.21 & 37 & 0.37 & 37 & 0.17 \\

 & 108 & 494 & 44 & 0.11 & 39 & 0.17 & 44 & 0.21 & 44 & 0.36 & 44 & 0.16 \\

 & 109 & 494 & 44 & 0.11 & 40 & 0.19 & 44 & 0.2 & 44 & 0.36 & 44 & 0.16 \\

 & 115 & 494 & 44 & 0.1 & 40 & 0.17 & 44 & 0.18 & 44 & 0.36 & 44 & 0.15 \\

 & 118 & 494 & 44 & 0.1 & 40 & 0.16 & 44 & 0.2 & 44 & 0.36 & 44 & 0.15 \\

 & 120 & 494 & 44 & 0.11 & 40 & 0.18 & 44 & 0.21 & 44 & 0.36 & 44 & 0.15 \\

 & 122 & 494 & 44 & 0.1 & 40 & 0.18 & 44 & 0.18 & 44 & 0.35 & 44 & 0.15 \\

 & 124 & 494 & 37 & 0.12 & 28 & 0.17 & 37 & 0.21 & 37 & 0.34 & 37 & 0.16 \\

 & 158 & 494 & 37 & 0.11 & 28 & 0.17 & 37 & 0.22 & 37 & 0.37 & 37 & 0.15 \\

\hline
\multirow{6}{*}{SHB-missed/Jigsaw-35th}

 & 184 & 42461 & 1129 & 0.79 & 739 & 1.09 & 1129 & 1.54 & 1129 & 50.47 & 1129 & 1.29 \\

 & 319 & 42461 & 1129 & 0.83 & 739 & 1.05 & 1129 & 1.59 & 1129 & 45.16 & 1129 & 1.21 \\

 & 414 & 42461 & 1129 & 0.87 & 739 & 1.08 & 1129 & 1.52 & 1129 & 49.41 & 1129 & 1.22 \\

 & 468 & 42461 & 1129 & 0.8 & 739 & 1.05 & 1129 & 1.53 & 1129 & 48.49 & 1129 & 1.2 \\

 & 475 & 42461 & 1129 & 0.87 & 739 & 1.07 & 1129 & 1.57 & 1129 & 48.01 & 1129 & 1.2 \\

 & 484 & 42461 & 1129 & 0.85 & 739 & 1.05 & 1129 & 1.6 & 1129 & 47.91 & 1129 & 1.23 \\

\hline
\multirow{41}{*}{SHB-missed/TreeSet-22th}

 & 97 & 635 & 42 & 0.11 & 35 & 0.17 & 42 & 0.19 & 42 & 0.3 & 42 & 0.14 \\

 & 98 & 635 & 42 & 0.12 & 34 & 0.17 & 42 & 0.19 & 42 & 0.31 & 42 & 0.14 \\

 & 99 & 635 & 42 & 0.12 & 35 & 0.16 & 42 & 0.19 & 42 & 0.33 & 42 & 0.14 \\

 & 100 & 635 & 42 & 0.13 & 34 & 0.17 & 42 & 0.21 & 42 & 0.33 & 42 & 0.14 \\

 & 101 & 635 & 42 & 0.12 & 35 & 0.17 & 42 & 0.22 & 42 & 0.32 & 42 & 0.14 \\

 & 102 & 635 & 42 & 0.12 & 34 & 0.18 & 42 & 0.2 & 42 & 0.31 & 42 & 0.14 \\

 & 105 & 635 & 42 & 0.11 & 33 & 0.17 & 42 & 0.21 & 42 & 0.31 & 42 & 0.14 \\

 & 107 & 635 & 42 & 0.12 & 33 & 0.18 & 42 & 0.2 & 42 & 0.32 & 42 & 0.15 \\

 & 109 & 635 & 42 & 0.12 & 33 & 0.18 & 42 & 0.19 & 42 & 0.33 & 42 & 0.14 \\

 & 111 & 635 & 42 & 0.12 & 34 & 0.17 & 42 & 0.19 & 42 & 0.31 & 42 & 0.14 \\

 & 113 & 635 & 42 & 0.12 & 34 & 0.18 & 42 & 0.21 & 42 & 0.31 & 42 & 0.14 \\

 & 115 & 635 & 42 & 0.12 & 34 & 0.19 & 42 & 0.2 & 42 & 0.31 & 42 & 0.19 \\

 & 117 & 635 & 42 & 0.11 & 34 & 0.18 & 42 & 0.18 & 42 & 0.37 & 42 & 0.15 \\

 & 119 & 635 & 42 & 0.12 & 34 & 0.17 & 42 & 0.21 & 42 & 0.36 & 42 & 0.14 \\

 & 120 & 635 & 42 & 0.12 & 35 & 0.17 & 42 & 0.19 & 42 & 0.31 & 42 & 0.14 \\

 & 121 & 635 & 42 & 0.11 & 34 & 0.18 & 42 & 0.21 & 42 & 0.31 & 42 & 0.14 \\

 & 122 & 635 & 42 & 0.11 & 35 & 0.16 & 42 & 0.2 & 42 & 0.32 & 42 & 0.14 \\

 & 123 & 635 & 42 & 0.11 & 35 & 0.18 & 42 & 0.2 & 42 & 0.3 & 42 & 0.15 \\

 & 126 & 635 & 42 & 0.12 & 35 & 0.17 & 42 & 0.2 & 42 & 0.31 & 42 & 0.14 \\

 & 127 & 635 & 42 & 0.11 & 34 & 0.17 & 42 & 0.18 & 42 & 0.32 & 42 & 0.14 \\

 & 128 & 635 & 42 & 0.12 & 35 & 0.18 & 42 & 0.2 & 42 & 0.32 & 42 & 0.15 \\

 & 129 & 635 & 42 & 0.13 & 34 & 0.17 & 42 & 0.2 & 42 & 0.32 & 42 & 0.14 \\

 & 130 & 635 & 42 & 0.12 & 35 & 0.18 & 42 & 0.21 & 42 & 0.3 & 42 & 0.15 \\

 & 131 & 635 & 42 & 0.12 & 34 & 0.18 & 42 & 0.19 & 42 & 0.32 & 42 & 0.14 \\

 & 132 & 635 & 42 & 0.13 & 35 & 0.17 & 42 & 0.2 & 42 & 0.31 & 42 & 0.14 \\

 & 133 & 635 & 42 & 0.13 & 35 & 0.17 & 42 & 0.2 & 42 & 0.32 & 42 & 0.14 \\

 & 134 & 635 & 42 & 0.12 & 35 & 0.18 & 42 & 0.21 & 42 & 0.31 & 42 & 0.15 \\

 & 135 & 635 & 42 & 0.11 & 35 & 0.18 & 42 & 0.19 & 42 & 0.32 & 42 & 0.15 \\

 & 136 & 635 & 42 & 0.12 & 35 & 0.19 & 42 & 0.2 & 42 & 0.33 & 42 & 0.14 \\

 & 137 & 635 & 42 & 0.11 & 35 & 0.17 & 42 & 0.19 & 42 & 0.33 & 42 & 0.14 \\

 & 138 & 635 & 42 & 0.12 & 35 & 0.17 & 42 & 0.21 & 42 & 0.31 & 42 & 0.15 \\

 & 139 & 635 & 42 & 0.12 & 35 & 0.18 & 42 & 0.17 & 42 & 0.3 & 42 & 0.15 \\

 & 140 & 635 & 42 & 0.12 & 35 & 0.19 & 42 & 0.2 & 42 & 0.32 & 42 & 0.14 \\

 & 141 & 635 & 42 & 0.13 & 35 & 0.19 & 42 & 0.19 & 42 & 0.3 & 42 & 0.14 \\

 & 142 & 635 & 42 & 0.12 & 35 & 0.18 & 42 & 0.19 & 42 & 0.32 & 42 & 0.15 \\

 & 143 & 635 & 42 & 0.12 & 35 & 0.17 & 42 & 0.19 & 42 & 0.32 & 42 & 0.15 \\

 & 144 & 635 & 42 & 0.11 & 35 & 0.16 & 42 & 0.2 & 42 & 0.31 & 42 & 0.14 \\

 & 145 & 635 & 42 & 0.12 & 35 & 0.17 & 42 & 0.21 & 42 & 0.32 & 42 & 0.15 \\

 & 149 & 635 & 42 & 0.11 & 35 & 0.17 & 42 & 0.19 & 42 & 0.32 & 42 & 0.15 \\

 & 150 & 635 & 42 & 0.12 & 34 & 0.17 & 42 & 0.22 & 42 & 0.31 & 42 & 0.14 \\

 & 151 & 635 & 42 & 0.13 & 35 & 0.2 & 42 & 0.19 & 42 & 0.32 & 42 & 0.14 \\

\hline
\multirow{21}{*}{WCP-missed/TreeSet-22th}

 & 98 & 635 & 42 & 0.12 & 34 & 0.17 & 42 & 0.19 & 42 & 0.33 & 42 & 0.16 \\

 & 100 & 635 & 42 & 0.12 & 34 & 0.17 & 42 & 0.2 & 42 & 0.31 & 42 & 0.16 \\

 & 102 & 635 & 42 & 0.12 & 34 & 0.17 & 42 & 0.2 & 42 & 0.32 & 42 & 0.15 \\

 & 109 & 635 & 42 & 0.11 & 33 & 0.18 & 42 & 0.19 & 42 & 0.33 & 42 & 0.16 \\

 & 111 & 635 & 42 & 0.12 & 34 & 0.18 & 42 & 0.2 & 42 & 0.3 & 42 & 0.16 \\

 & 113 & 635 & 42 & 0.12 & 34 & 0.17 & 42 & 0.2 & 42 & 0.32 & 42 & 0.16 \\

 & 115 & 635 & 42 & 0.12 & 34 & 0.17 & 42 & 0.2 & 42 & 0.32 & 42 & 0.16 \\

 & 117 & 635 & 42 & 0.12 & 34 & 0.17 & 42 & 0.2 & 42 & 0.34 & 42 & 0.16 \\

 & 119 & 635 & 42 & 0.12 & 34 & 0.16 & 42 & 0.19 & 42 & 0.33 & 42 & 0.16 \\

 & 121 & 635 & 42 & 0.12 & 34 & 0.17 & 42 & 0.19 & 42 & 0.32 & 42 & 0.16 \\

 & 123 & 635 & 42 & 0.11 & 35 & 0.18 & 42 & 0.2 & 42 & 0.31 & 42 & 0.16 \\

 & 127 & 635 & 42 & 0.12 & 34 & 0.18 & 42 & 0.21 & 42 & 0.32 & 42 & 0.16 \\

 & 129 & 635 & 42 & 0.11 & 34 & 0.17 & 42 & 0.2 & 42 & 0.32 & 42 & 0.15 \\

 & 131 & 635 & 42 & 0.12 & 34 & 0.18 & 42 & 0.18 & 42 & 0.34 & 42 & 0.15 \\

 & 133 & 635 & 42 & 0.12 & 35 & 0.17 & 42 & 0.2 & 42 & 0.33 & 42 & 0.16 \\

 & 135 & 635 & 42 & 0.12 & 35 & 0.18 & 42 & 0.19 & 42 & 0.31 & 42 & 0.17 \\

 & 137 & 635 & 42 & 0.12 & 35 & 0.16 & 42 & 0.2 & 42 & 0.32 & 42 & 0.16 \\

 & 139 & 635 & 42 & 0.11 & 35 & 0.19 & 42 & 0.19 & 42 & 0.3 & 42 & 0.15 \\

 & 141 & 635 & 42 & 0.11 & 35 & 0.2 & 42 & 0.21 & 42 & 0.31 & 42 & 0.15 \\

 & 143 & 635 & 42 & 0.12 & 35 & 0.19 & 42 & 0.18 & 42 & 0.31 & 42 & 0.15 \\

 & 145 & 635 & 42 & 0.12 & 35 & 0.17 & 42 & 0.19 & 42 & 0.33 & 42 & 0.16 \\

\hline
\multirow{4}{*}{SyncP-missed/ArrayList-27th}

 & 109 & 494 & 44 & 0.11 & 40 & 0.16 & 44 & 0.2 & 44 & 0.39 & 44 & 0.16 \\

 & 118 & 494 & 44 & 0.11 & 40 & 0.16 & 44 & 0.19 & 44 & 0.37 & 44 & 0.16 \\

 & 120 & 494 & 44 & 0.1 & 40 & 0.16 & 44 & 0.2 & 44 & 1.21 & 44 & 0.14 \\

 & 122 & 494 & 44 & 0.12 & 40 & 0.17 & 44 & 0.19 & 44 & 0.38 & 44 & 0.14 \\

\hline
\multirow{3}{*}{SyncP-missed/Jigsaw-35th}

 & 219 & 42461 & 1129 & 0.85 & 739 & 1.08 & 1129 & 1.47 & 1129 & 48.19 & 1129 & 1.51 \\

 & 475 & 42461 & 1129 & 0.81 & 739 & 1.07 & 1129 & 1.57 & 1129 & 48.48 & 1129 & 1.44 \\

 & 484 & 42461 & 1129 & 0.8 & 739 & 1.06 & 1129 & 1.41 & 1129 & 47.2 & 1129 & 1.43 \\

\hline
\multirow{15}{*}{SyncP-missed/TreeSet-22th}

 & 97 & 635 & 42 & 0.12 & 35 & 0.17 & 42 & 0.2 & 42 & 0.36 & 42 & 0.15 \\

 & 99 & 635 & 42 & 0.11 & 35 & 0.16 & 42 & 0.19 & 42 & 0.36 & 42 & 0.15 \\

 & 101 & 635 & 42 & 0.12 & 35 & 0.18 & 42 & 0.2 & 42 & 0.36 & 42 & 0.16 \\

 & 120 & 635 & 42 & 0.11 & 35 & 0.17 & 42 & 0.19 & 42 & 0.36 & 42 & 0.17 \\

 & 122 & 635 & 42 & 0.12 & 35 & 0.17 & 42 & 0.2 & 42 & 0.35 & 42 & 0.14 \\

 & 126 & 635 & 42 & 0.13 & 35 & 0.17 & 42 & 0.2 & 42 & 0.37 & 42 & 0.16 \\

 & 128 & 635 & 42 & 0.13 & 35 & 0.19 & 42 & 0.2 & 42 & 0.37 & 42 & 0.16 \\

 & 130 & 635 & 42 & 0.13 & 35 & 0.18 & 42 & 0.18 & 42 & 0.38 & 42 & 0.15 \\

 & 132 & 635 & 42 & 0.12 & 35 & 0.19 & 42 & 0.2 & 42 & 0.37 & 42 & 0.15 \\

 & 134 & 635 & 42 & 0.11 & 35 & 0.17 & 42 & 0.2 & 42 & 0.36 & 42 & 0.14 \\

 & 136 & 635 & 42 & 0.12 & 35 & 0.17 & 42 & 0.19 & 42 & 0.38 & 42 & 0.16 \\

 & 138 & 635 & 42 & 0.11 & 35 & 0.16 & 42 & 0.2 & 42 & 0.36 & 42 & 0.14 \\

 & 140 & 635 & 42 & 0.11 & 35 & 0.17 & 42 & 0.21 & 42 & 0.36 & 42 & 0.15 \\

 & 142 & 635 & 42 & 0.12 & 35 & 0.16 & 42 & 0.19 & 42 & 0.35 & 42 & 0.16 \\

 & 144 & 635 & 42 & 0.11 & 35 & 0.18 & 42 & 0.19 & 42 & 0.35 & 42 & 0.14 \\

\hline

\end{longtable}
\twocolumn

%% file: tables/ExtraNumRacesDetail.tex
\onecolumn
\setlength{\tabcolsep}{3pt}
\begin{longtable}{|c|c|c|  |c|c| |c|c| |c|c| |c|c| |c|c|}
\caption{Details on reported races and running time (in minute) by each algorithm on C/C++ benchmarks. Column 1-3 states the source of these benchmarks, trace name with number of threads and events number after filtering. Columns 4-13 are reported races and average running time by each algorithm.}

\label{tab:extra-results-detail} \\

\hline
1 & 2 & 3 & 4 & 5 & 6 & 7 & 8 & 9 & 10 & 11 & 12 & 13 \\

\hline

\multirow{2}{*}{\textbf{Benchmark Set}} & \multirow{2}{*}{Benchmark} & \multirow{2}{*}{$\NumEvents$} & 
\multicolumn{2}{c||}{SHB} &
\multicolumn{2}{c||}{WCP} &
\multicolumn{2}{c||}{SyncP} &
\multicolumn{2}{c||}{M2} &
\multicolumn{2}{c|}{\ssp} \\
\cline{4-13}
&   &   & Races & Time & Races & Time & Races & Time & Races & Time & Races & Time \\

\endfirsthead

\hline 

\endfoot

\hline
1 & 2 & 3 & 4 & 5 & 6 & 7 & 8 & 9 & 10 & 11 & 12 & 13 \\
\hline
\multirow{2}{*}{\textbf{Benchmark Set}} & \multirow{2}{*}{Benchmark} & \multirow{2}{*}{$\NumEvents$} & 
\multicolumn{2}{c||}{SHB} &
\multicolumn{2}{c||}{WCP} &
\multicolumn{2}{c||}{SyncP} &
\multicolumn{2}{c||}{M2} &
\multicolumn{2}{c|}{\ssp} \\
\cline{4-13}

&   &   & Races & Time & Races & Time & Races & Time & Races & Time & Races & Time \\
\hline
\endhead

\hline

\hline

\multirow{8}{*}{CoMD}
 & task-16th & 117M & 6267 & 5.5 & 5669 & 14.6 & 1 & 180.0 & 0 & 180.0 & 11757 & 6.5 \\ 

 & task-56th & 117M & 6267 & 6.1 & 5669 & 14.5 & 1 & 180.0 & 0 & 180.0 & 11757 & 5.9 \\

 & taskdeps-16th & 115M & 5627 & 2.2 & 5190 & 6.5 & 14 & 180.0 & 13 & 180.0 & 10915 & 3.4 \\

 & taskdeps-56th & 117M & 6267 & 5.0 & 5669 & 13.9 & 1 & 180.0 & 0 & 180.0 & 11757 & 5.5 \\

 & taskloop-16th & 2M & 257 & 0.1 & 168766 & 0.1 & 0 & 180.0 & 474 & 180.0 & 177566 & 0.1 \\

 & taskloop-56th & 4M & 4982 & 0.1 & 44977 & 0.2 & 0 & 180.0 & 186 & 180.0 & 194160 & 1.0 \\

 & openmp-16th & 115M & 5627 & 3.3 & 5190 & 6.8 & 14 & 180.0 & 13 & 180.0 & 10915 & 3.2 \\

 & openmp-56th & 117M & 6267 & 5.6 & 5669 & 14.3 & 1 & 180.0 & 0 & 180.0 & 11757 & 6.9 \\

\hline
\multirow{1}{*}{SimpleMOC}

 & trace-16th & 19M & 380 & 0.2 & 388 & 23.1 & 32 & 180.0 & 32 & 180.0 & 32 & 180.0 \\

\hline
\multirow{15}{*}{OMPRacer}

 & Amg2013-18th & 39M & 140541 & 0.4 & 107793 & 3.0 & 103 & 180.0 & 102 & 180.0 & 145485 & 0.7 \\

 & Amg2013-58th & 52M & 181018 & 1.9 & 133280 & 7.4 & 70 & 180.0 & 0 & 180.0 & 190994 & 4.4 \\

 & Kripke-16th & 20M & 14162 & 0.1 & 44 & 1.4 & 46 & 180.0 & 259 & 180.0 & 22481 & 0.3 \\

 & Kripke-56th & 34M & 20824 & 1.2 & 128 & 5.5 & 39 & 180.0 & 12 & 180.0 & 34,155 & 9.8 \\

 & Lulesh-16th & 10M & 51621 & 0.3 & 27472 & 0.4 & 1312 & 180.0 & 620 & 180.0 & 52940 & 0.1 \\

 & Lulesh-16th & 130M & 158081 & 3.0 & 103306 & 8.8 & 5 & 180.0 & 13 & 180.0 & 167595 & 4.5 \\

 & Lulesh-56th & 14M & 71676 & 0.3 & 40322 & 2.7 & 1 & 180.0 & 70 & 180.0 & 73432 & 0.1 \\

 & Lulesh-56th & 156M & 250954 & 7.8 & 157756 & 27.6 & 1 & 180.0 & 0 & 180.0 & 261857 & 10.7 \\

 & miniFE-18th & 44M & 148645 & 0.6 & 51478 & 2.5 & 121 & 180.0 & 77 & 180.0 & 159052 & 0.3 \\

 & miniFE-58th & 63M & 171460 & 2.7 & 74252 & 10.6 & 77 & 180.0 & 0 & 180.0 & 191862 & 4.0 \\

 & QuickSilver-56th & 1M & 20753 & 0.1 & 7288 & 0.1 & 8 & 180.0 & 610 & 180.0 & 21132 & 0.2 \\

 & XSBench-16th & 693.9K & 27 & 0.1 & 30 & 0.1 & 221 & 0.9 & 222 & 3.6 & 225 & 0.1 \\

 & XSBench-56th & 710.9K & 89 & 0.1 & 117 & 0.1 & 15 & 180.0 & 361 & 17.6 & 370 & 0.1 \\

 & RSBench-16th & 27M & 22 & 0.3 & 35 & 0.6 & 1271 & 43.7 & 199 & 180.0 & 1278 & 0.2 \\

 & RSBench-56th & 27M & 95 & 0.1 & 114 & 1.8 & 0 & 180.0 & 30 & 180.0 & 1405 & 0.3 \\

\hline
\multirow{15}{*}{DRACC-16th}

 & DRACC-009 & 70M & 16 & 0.2 & 16 & 4.1 & 31 & 180.0 & 31 & 142.1 & 32 & 3.3 \\

 & DRACC-010 & 70M & 16 & 0.4 & 16 & 4.0 & 31 & 180.0 & 31 & 144.4 & 32 & 2.7 \\

 & DRACC-011 & 0.5K & 15 & 0.1 & 15 & 0.1 & 30 & 0.1 & 30 & 0.1 & 30 & 0.1 \\

 & DRACC-012 & 103M & 527 & 1.7 & 527 & 16.0 & 542 & 180.0 & 22 & 180.0 & 543 & 13.5 \\

 & DRACC-013 & 103M & 527 & 1.7 & 527 & 22.7 & 542 & 180.0 & 22 & 180.0 & 543 & 13.7 \\

 & DRACC-014 & 0.5K & 15 & 0.1 & 15 & 0.1 & 30 & 0.1 & 30 & 0.1 & 30 & 0.1 \\

 & DRACC-015 & 70M & 16 & 0.6 & 16 & 3.7 & 31 & 180.0 & 31 & 145.3 & 32 & 3.8 \\

 & DRACC-016 & 70M & 16 & 0.7 & 16 & 4.5 & 31 & 180.0 & 31 & 144.9 & 32 & 4.2 \\

 & DRACC-017 & 0.5K & 15 & 0.1 & 15 & 0.1 & 30 & 0.1 & 30 & 0.1 & 30 & 0.1 \\

 &  & 0.5K & 15 & 0.1 & 15 & 0.1 & 30 & 0.1 & 30 & 0.1 & 30 & 0.1 \\

 & DRACC-018 & 103M & 527 & 1.5 & 527 & 19.3 & 542 & 180.0 & 21 & 180.0 & 543 & 12.4 \\

 & DRACC-019 & 103M & 527 & 1.4 & 527 & 15.8 & 542 & 180.0 & 22 & 180.0 & 543 & 12.5 \\

 & DRACC-020 & 0.5K & 15 & 0.1 & 15 & 0.1 & 30 & 0.1 & 30 & 0.1 & 30 & 0.1 \\

\hline
\multirow{21}{*}{DRB-16th}

 & DRB-062 & 70M & 31 & 0.8 & 31 & 4.4 & 46 & 4.7 & 45 & 180.0 & 46 & 1.4 \\

 & DRB-105 & 44M & 866 & 0.4 & 874 & 2.6 & 46 & 180.0 & 101 & 180.0 & 889 & 7.0 \\

 & DRB-106 & 70M & 709 & 0.7 & 732 & 4.9 & 46 & 180.0 & 46 & 180.0 & 789 & 6.8 \\

 & DRB-110 & 35M & 15 & 0.3 & 16 & 1.1 & 32 & 180.0 & 32 & 47.8 & 32 & 6.9 \\

 & DRB-122 & 0.5K & 15 & 0.1 & 15 & 0.1 & 30 & 0.1 & 30 & 0.1 & 30 & 0.1 \\

 & DRB-123 & 77M & 227 & 0.5 & 713 & 4.8 & 231 & 180.0 & 46 & 180.0 & 243 & 3.7 \\

 & DRB-144 & 70M & 16 & 0.7 & 16 & 4.1 & 30 & 180.0 & 30 & 180.0 & 30 & 180.0 \\

 & DRB-148 & 70M & 16 & 0.4 & 16 & 5.9 & 31 & 180.0 & 31 & 125.9 & 32 & 3.7 \\

 & DRB-150 & 56M & 16 & 0.4 & 16 & 2.8 & 31 & 180.0 & 31 & 104.4 & 32 & 3.3 \\

 & DRB-152 & 56M & 16 & 0.3 & 16 & 4.3 & 31 & 180.0 & 31 & 105.5 & 32 & 3.1 \\

 & DRB-154 & 0.5K & 15 & 0.1 & 15 & 0.1 & 30 & 0.1 & 30 & 0.1 & 30 & 0.1 \\

 & DRB-155 & 12M & 17 & 0.1 & 18 & 0.6 & 32 & 132.1 & 36 & 13.6 & 36 & 0.1 \\

 & DRB-176 & 47M & 1697 & 0.5 & 1899 & 2.1 & 51 & 180.0 & 77 & 180.0 & 2079 & 7.0 \\

 & DRB-176 & 272M & 1835 & 7.2 & 2005 & 20.7 & 51 & 180.0 & 0 & 180.0 & 2209 & 61.8 \\

 & DRB-176 & 782M & 2154 & 19.7 & 2385 & 61.9 & 51 & 180.0 & 0 & 180.0 & 2611 & 175.4 \\

 & DRB-177 & 45M & 1204 & 0.5 & 1238 & 2.6 & 54 & 180.0 & 78 & 180.0 & 1315 & 6.0 \\

 & DRB-177 & 191M & 1395 & 4.5 & 1424 & 18.5 & 55 & 180.0 & 0 & 180.0 & 1522 & 30.3 \\

 & DRB-177 & 106M & 982 & 1.9 & 999 & 7.4 & 53 & 180.0 & 13 & 180.0 & 1080 & 13.5 \\

 & DRB-177 & 519M & 1634 & 13.6 & 1685 & 42.0 & 54 & 180.0 & 0 & 180.0 & 1799 & 80.5 \\

 & DRB-177 & 333M & 1704 & 8.4 & 1771 & 26.4 & 54 & 180.0 & 0 & 180.0 & 1881 & 52.9 \\

 & DRB-177 & 836M & 1791 & 22.8 & 1857 & 71.7 & 53 & 180.0 & 0 & 180.0 & 1963 & 160.9 \\

\hline
\multirow{12}{*}{DRB-56th}

 & DRB-062 & 72M & 111 & 3.0 & 111 & 10.8 & 42 & 180.0 & 0 & 180.0 & 166 & 1.8 \\

 & DRB-105 & 46M & 2778 & 1.4 & 2793 & 6.0 & 42 & 180.0 & 0 & 180.0 & 2849 & 19.2 \\

 & DRB-106 & 68M & 2042 & 2.2 & 2086 & 9.4 & 42 & 180.0 & 0 & 180.0 & 2284 & 11.7 \\

 & DRB-110 & 35M & 55 & 0.5 & 56 & 5.4 & 42 & 180.0 & 15 & 180.0 & 112 & 1.3 \\

 & DRB-122 & 1.8K & 55 & 0.1 & 55 & 0.1 & 43 & 180.0 & 110 & 0.1 & 110 & 0.1 \\

 & DRB-123 & 77M & 712 & 2.5 & 228 & 10.4 & 42 & 180.0 & 0 & 180.0 & 778 & 9.5 \\

 & DRB-155 & 12M & 58 & 0.2 & 59 & 0.7 & 38 & 180.0 & 94 & 180.0 & 126 & 1.1 \\

 & DRB-176 & 49M & 5216 & 1.6 & 5989 & 7.6 & 25 & 180.0 & 0 & 180.0 & 6802 & 15.1 \\

 & DRB-176 & 348M & 6911 & 16.4 & 7985 & 70.3 & 24 & 180.0 & 0 & 180.0 & 8174 & 180.0 \\

 & DRB-176 & 900M & 7854 & 43.2 & 9110 & 169.8 & 25 & 180.0 & 0 & 180.0 & 5026 & 180.0 \\

 & DRB-177 & 43M & 3138 & 1.2 & 3115 & 5.9 & 4 & 180.0 & 0 & 180.0 & 3421 & 13.3 \\

 & DRB-177 & 326M & 5131 & 15.0 & 5197 & 57.3 & 3 & 180.0 & 0 & 180.0 & 5541 & 166.8 \\

\hline
\multirow{46}{*}{HPCBench}

 & graph500-16th & 81M & 23732 & 1.1 & 14764 & 4.5 & 40 & 180.0 & 30 & 180.0 & 113732 & 3.2 \\

 & graph500-56th & 82M & 38416 & 3.5 & 17050 & 12.0 & 38 & 180.0 & 0 & 180.0 & 119601 & 7.8 \\

 & HPCCG-16th & 55M & 9531 & 0.9 & 4480 & 3.7 & 34 & 180.0 & 60 & 180.0 & 9547 & 2.2 \\

 & HPCCG-56th & 79M & 15027 & 3.4 & 228 & 11.9 & 40 & 180.0 & 0 & 180.0 & 15083 & 3.5 \\

 & DC.S-16th & 1.0K & 78 & 0.1 & 34 & 0.1 & 97 & 0.2 & 98 & 0.1 & 99 & 0.1 \\

 & DC.S-56th & 19.8K & 570 & 0.1 & 153 & 0.1 & 60 & 180.0 & 629 & 0.2 & 629 & 0.1 \\

 & IS.W-16th & 48M & 64155 & 0.8 & 30642 & 2.3 & 31 & 180.0 & 75 & 180.0 & 64597 & 2.3 \\

 & IS.W-56th & 140M & 193236 & 9.4 & 128441 & 28.8 & 38 & 180.0 & 0 & 180.0 & 202142 & 23.7 \\

 & loopA.bad-16th & 93M & 30 & 2.0 & 30 & 7.5 & 1 & 180.0 & 28 & 180.0 & 150033 & 2.4 \\

 & loopA.bad-56th & 334M & 118 & 26.4 & 119 & 87.1 & 0 & 180.0 & 0 & 180.0 & 550125 & 21.0 \\

 & loopA.solu1-16th & 93M & 31 & 1.9 & 31 & 7.5 & 1 & 180.0 & 28 & 180.0 & 150049 & 2.0 \\

 & loopA.solu1-56th & 334M & 116 & 23.5 & 117 & 94.5 & 0 & 180.0 & 40 & 180.0 & 550181 & 23.0 \\

 & loopA.solu2-16th & 51M & 31 & 0.4 & 31 & 2.2 & 15 & 180.0 & 0 & 180.0 & 10049 & 0.6 \\

 & loopA.solu2-56th & 171M & 118 & 7.9 & 117 & 27.9 & 1 & 180.0 & 0 & 180.0 & 10180 & 10.3 \\

 & loopA.solu3-16th & 51M & 31 & 1.0 & 30 & 3.0 & 16 & 180.0 & 80 & 180.0 & 10048 & 0.6 \\

 & loopA.solu3-56th & 171M & 116 & 8.7 & 114 & 24.6 & 1 & 180.0 & 0 & 180.0 & 10177 & 8.8 \\

 & loopB.solu1-16th & 93M & 30 & 1.9 & 30 & 7.1 & 1 & 180.0 & 28 & 180.0 & 150037 & 2.1 \\

 & loopB.solu1-56th & 334M & 113 & 21.8 & 113 & 91.8 & 0 & 180.0 & 0 & 180.0 & 550127 & 21.5 \\

 & Mandelbrot-16th & 112M & 26 & 2.0 & 33 & 5.9 & 1968 & 180.0 & 13 & 180.0 & 1973 & 2.5 \\

 & Mandelbrot-56th & 114M & 87 & 4.6 & 113 & 18.0 & 2 & 180.0 & 0 & 180.0 & 2196 & 4.6 \\

 & Pi-16th & 96M & 27 & 1.4 & 35 & 5.3 & 48 & 7.5 & 28 & 180.0 & 53 & 2.5 \\

 & Pi-56th & 99M & 91 & 4.2 & 115 & 14.9 & 37 & 180.0 & 0 & 180.0 & 184 & 6.1 \\

 & QuickSort-16th & 41M & 31752 & 0.4 & 31758 & 1.5 & 2 & 180.0 & 115 & 180.0 & 91092 & 2.4 \\

 & QuickSort-56th & 41M & 31792 & 1.0 & 31798 & 6.8 & 0 & 180.0 & 2 & 180.0 & 91172 & 3.7 \\

 & fft6-16th & 0.9K & 30 & 0.1 & 31 & 0.1 & 78 & 0.2 & 81 & 0.1 & 81 & 0.1 \\

 & fft6-56th & 2.4K & 74 & 0.1 & 74 & 0.1 & 27 & 180.0 & 172 & 0.1 & 172 & 0.1 \\

 & LUReduction-16th & 45M & 89100 & 0.9 & 32 & 2.1 & 2 & 180.0 & 32 & 180.0 & 89116 & 3.5 \\

 & LUReduction-56th & 45M & 88766 & 1.5 & 112 & 7.9 & 0 & 180.0 & 0 & 180.0 & 89209 & 11.2 \\

 & MD-16th & 118M & 1499 & 2.9 & 59 & 7.9 & 1512 & 180.0 & 13 & 180.0 & 1515 & 4.5 \\

 & MD-56th & 120M & 1683 & 5.9 & 178 & 15.5 & 3 & 180.0 & 0 & 180.0 & 1747 & 8.2 \\

 & testPath-16th & 7M & 16 & 0.1 & 16 & 0.2 & 124 & 180.0 & 154 & 81.7 & 154 & 49.0 \\

 & testPath-56th & 10M & 57 & 0.2 & 57 & 0.3 & 1 & 180.0 & 44 & 180.0 & 544 & 180.0 \\

 & fft-16th & 78M & 983086 & 2.0 & 983086 & 5.7 & 98599 & 180.0 & 30 & 180.0 & 2424886 & 4.1 \\

 & fft-56th & 83M & 1030016 & 4.5 & 1030016 & 14.3 & 37 & 180.0 & 0 & 180.0 & 2565429 & 13.7 \\

 & fft-56th & 363M & 3894868 & 24.2 & 4119572 & 91.0 & 37 & 180.0 & 0 & 180.0 & 10261235 & 98.5 \\

 & qsomp1-16th & 674.6K & 16 & 0.1 & 16 & 0.1 & 282 & 180.0 & 282 & 37.7 & 282 & 0.1 \\

 & qsomp1-56th & 540.1K & 56 & 0.1 & 56 & 0.1 & 27 & 180.0 & 156 & 79.5 & 156 & 0.3 \\

 & qsomp2-16th & 870.5K & 16 & 0.1 & 16 & 0.1 & 375 & 74.1 & 375 & 41.9 & 375 & 0.2 \\

 & qsomp2-56th & 629.6K & 56 & 0.1 & 56 & 0.1 & 15 & 180.0 & 208 & 180.0 & 253 & 0.5 \\

 & qsomp3-16th & 15M & 16 & 0.1 & 17 & 0.4 & 32 & 123.4 & 32 & 15.3 & 32 & 0.3 \\

 & qsomp3-56th & 4M & 56 & 0.1 & 57 & 0.3 & 40 & 180.0 & 112 & 46.4 & 112 & 0.1 \\

 & qsomp4-16th & 19M & 17 & 0.1 & 17 & 4.3 & 37 & 160.0 & 33 & 180.0 & 37 & 0.5 \\

 & qsomp4-56th & 6M & 5269 & 0.1 & 5044 & 0.2 & 0 & 180.0 & 113 & 180.0 & 5349 & 19.7 \\

 & qsomp6-56th & 507.2K & 56 & 0.1 & 56 & 0.1 & 8 & 180.0 & 139 & 180.0 & 536 & 1.8 \\

 & qsomp7-16th & 44M & 8016 & 0.6 & 8001 & 2.7 & 6787 & 180.0 & 96 & 180.0 & 8035 & 0.9 \\

 & qsomp7-56th & 147M & 6053 & 6.4 & 6052 & 22.1 & 0 & 180.0 & 0 & 180.0 & 6119 & 20.5 \\

\hline
\multirow{7}{*}{misc}
 & biojava-4th & 0.9K & 2 & 0.1 & 3 & 0.1 & 6 & 0.1 & 6 & 0.1 & 6 & 0.1 \\

 & cassandra-132th & 28M & 5053 & 0.2 & 5026 & 180.0 & 0 & 180.0 & 0 & 180.0 & 514 & 180.0 \\

 & graphchi-20th & 206.3K & 21 & 0.1 & 21 & 0.1 & 137 & 0.1 & 138 & 0.1 & 138 & 0.1 \\

 & hsqldb-44th & 647.5K & 5 & 0.1 & 5 & 0.1 & 3 & 180.0 & 5 & 0.1 & 5 & 2.4 \\

 & tradebeans-222th & 218.9K & 170 & 0.1 & 157 & 0.1 & 0 & 180.0 & 203 & 40.9 & 205 & 0.1 \\

 & tradesoap-221th & 218.6K & 169 & 0.1 & 155 & 0.1 & 0 & 180.0 & 203 & 44.0 & 205 & 0.4 \\

 & zxing-15th & 18M & 3128 & 0.1 & 3114 & 0.9 & 333 & 180.0 & 340 & 180.0 & 3216 & 0.2 \\


\hline

\end{longtable}

\twocolumn

%% file: tables/traceInfoForCpp.tex
\onecolumn
\setlength{\tabcolsep}{3pt}
\begin{longtable}{|c|c|c|c|c|c|c|c|c|c|}
\caption{Details of C/C++ benchmarks. Columns 1-2 states the source of these benchmarks and trace name with number of threads. Columns 3-6 are number of events before filtering, number of events after filtering, number of variables, number of locks. Columns 7-10 are number of read, write, acquire and release events after filtering.}

\label{tab:cpp-traces-info} \\

\hline
1 & 2 & 3 & 4 & 5 & 6 & 7 & 8 & 9 & 10 \\

\hline

\textbf{Benchmark Set} & Benchmark & $\NumEvents'$ & $\NumEvents$ & $\NumVars$ & $\NumLocks$ & $\NumReads$ & $\NumWrites$ & $\NumAcqs$ & $\NumRels$ \\

\hline

\endfirsthead

\hline 

\endfoot

\hline
1 & 2 & 3 & 4 & 5 & 6 & 7 & 8 & 9 & 10 \\
\hline

\textbf{Benchmark Set} & Benchmark & $\NumEvents$ & $\NumEvents$ & $\NumLocks$ & $\NumVars$ & $\NumReads$ & $\NumWrites$ & $\NumAcqs$ & $\NumRels$ \\

\hline
\endhead

\multirow{8}{*}{CoMD}

 & task-16th & 174M & 117M & 11,757 & 56 & 107,864,829 & 9,345,091 & 31,555 & 31,555 \\ 

 & task-56th & 175M & 117M & 11,757 & 56 & 107,864,829 & 9,345,091 & 31,555 & 31,555 \\

 & taskdeps-16th & 174M & 115M & 10,915 & 16 & 107,576,209 & 7,749,193 & 7,065 & 7,065 \\

 & taskdeps-56th & 175M & 117M & 11,757 & 56 & 107,864,889 & 9,345,091 & 31,585 & 31,585 \\

 & taskloop-16th & 251M & 2M & 177,566 & 16 & 1,132,063 & 1,364,476 & 1,241 & 1,241 \\

 & taskloop-56th & 251M & 4M & 194,160 & 56 & 3,058,872 & 1,930,831 & 4,044 & 4,044 \\

 & openmp-16th & 174M & 115M & 10,915 & 16 & 107,576,257 & 7,749,193 & 7,089 & 7,089 \\

 & openmp-56th & 175M & 117M & 11,757 & 56 & 107,864,811 & 9,345,091 & 31,546 & 31,546 \\

\hline
\multirow{1}{*}{SimpleMOC}

 & trace-16th & 170M & 19M & 60,029 & 5,017 & 7,833,334 & 7,641,040 & 1,771,680 & 1,771,680 \\

\hline
\multirow{15}{*}{OMPRacer}

 & Amg2013-18th & 170M & 39M & 145,485 & 36 & 3,332,723 & 36,405,204 & 1,657 & 1,657 \\

 & Amg2013-58th & 190M & 52M & 190,994 & 76 & 4,306,355 & 48,531,303 & 14,039 & 14,039 \\

 & Kripke-16th & 117M & 20M & 22,482 & 17 & 12,509,801 & 8,366,557 & 9,447 & 9,447 \\

 & Kripke-56th & 119M & 34M & 34,156 & 58 & 18,994,381 & 15,904,315 & 44,773 & 44,773 \\

 & Lulesh-16th & 35M & 10M & 52,940 & 15 & 8,987,707 & 1,431,554 & 72 & 72 \\

 & Lulesh-16th & 543M & 130M & 167,595 & 16 & 123,136,205 & 7,407,451 & 2,230 & 2,230 \\

 & Lulesh-56th & 52M & 14M & 73,432 & 56 & 12,065,387 & 1,999,030 & 1,707 & 1,707 \\

 & Lulesh-56th & 569M & 156M & 261,857 & 56 & 143,236,930 & 13,541,685 & 32,046 & 32,046 \\

 & miniFE-18th & 208M & 44M & 159,052 & 36 & 6,323,648 & 37,932,412 & 1,379 & 1,379 \\

 & miniFE-58th & 207M & 63M & 191,862 & 76 & 55,826,793 & 7,658,888 & 8,823 & 8,823 \\

 & QuickSilver-56th & 133M & 1M & 21,132 & 56 & 890,633 & 650,042 & 17,387 & 17,387 \\

 & XSBench-16th & 97M & 693.9K & 225 & 15 & 691,187 & 2,467 & 114 & 114 \\

 & XSBench-56th & 97M & 710.9K & 370 & 56 & 707,006 & 2,948 & 442 & 442 \\

 & RSBench-16th & 1.2B & 27M & 1,278 & 16 & 27,005,898 & 109,606 & 123 & 123 \\

 & RSBench-56th & 1.2B & 27M & 1,405 & 56 & 27,006,574 & 121,858 & 421 & 421 \\

\hline
\multirow{13}{*}{DRACC-16th}

 & DRACC-009 & 135M & 70M & 32 & 18 & 122 & 10,000,063 & 30,000,060 & 30,000,060 \\

 & DRACC-010 & 135M & 70M & 32 & 18 & 122 & 10,000,063 & 30,000,060 & 30,000,060 \\

 & DRACC-011 & 135M & 0.5K & 30 & 15 & 204 & 60 & 102 & 102 \\

 & DRACC-012 & 105M & 103M & 543 & 18 & 632 & 102,400,575 & 600,060 & 600,060 \\

 & DRACC-013 & 105M & 103M & 543 & 18 & 632 & 102,400,575 & 600,060 & 600,060 \\

 & DRACC-014 & 105M & 0.5K & 30 & 15 & 204 & 60 & 102 & 102 \\

 & DRACC-015 & 135M & 70M & 32 & 18 & 121 & 10,000,063 & 30,000,060 & 30,000,060 \\

 & DRACC-016 & 135M & 70M & 32 & 18 & 121 & 10,000,063 & 30,000,060 & 30,000,060 \\

 & DRACC-017 & 27M & 0.5K & 30 & 15 & 204 & 60 & 102 & 102 \\

 &  & 135M & 0.5K & 30 & 15 & 204 & 60 & 102 & 102 \\

 & DRACC-018 & 105M & 103M & 543 & 18 & 632 & 102,400,575 & 600,060 & 600,060 \\

 & DRACC-019 & 105M & 103M & 543 & 18 & 632 & 102,400,575 & 600,060 & 600,060 \\

 & DRACC-020 & 105M & 0.5K & 30 & 15 & 204 & 60 & 102 & 102 \\

\hline
\multirow{33}{*}{DRB-16th}

 & DRB-062 & 184M & 70M & 46 & 15 & 36,102,121 & 33,912,061 & 60 & 60 \\

 & DRB-105 & 134M & 44M & 889 & 31 & 8,339,388 & 8,339,291 & 14,098,397 & 14,098,397 \\

 & DRB-106 & 134M & 70M & 789 & 31 & 35,341,444 & 7,144,665 & 14,098,428 & 14,098,428 \\

 & DRB-110 & 120M & 35M & 32 & 18 & 207 & 5,000,047 & 15,000,103 & 15,000,103 \\

 & DRB-122 & 112M & 0.5K & 30 & 15 & 210 & 60 & 105 & 105 \\

 & DRB-123 & 112M & 77M & 243 & 16 & 35,000,220 & 14,000,062 & 14,000,109 & 14,000,109 \\

 & DRB-144 & 140M & 70M & 32 & 17 & 121 & 10,000,063 & 30,000,060 & 30,000,060 \\

 & DRB-148 & 135M & 70M & 32 & 18 & 121 & 10,000,063 & 30,000,060 & 30,000,060 \\

 & DRB-150 & 112M & 56M & 32 & 17 & 121 & 8,000,064 & 24,000,060 & 24,000,060 \\

 & DRB-152 & 112M & 56M & 32 & 17 & 121 & 8,000,064 & 24,000,060 & 24,000,060 \\

 & DRB-154 & 112M & 0.5K & 30 & 15 & 204 & 60 & 102 & 102 \\

 & DRB-155 & 50M & 12M & 51 & 18 & 344 & 87 & 6,000,141 & 6,000,141 \\

 & DRB-176 & 90M & 47M & 2079 & 31 & 20,649,583 & 10,224,212 & 8,077,747 & 8,077,747 \\

 & DRB-176 & 341M & 272M & 2209 & 31 & 109,382,709 & 52,675,980 & 55,364,923 & 55,364,923 \\

 & DRB-176 & 1.6B & 782M & 2611 & 31 & 337,612,032 & 155,172,752 & 144,947,035 & 144,947,035 \\

 & DRB-177 & 90M & 45M & 1315 & 31 & 20,300,303 & 9,112,538 & 8,077,747 & 8,077,747 \\

 & DRB-177 & 211M & 191M & 1522 & 31 & 82,046,982 & 40,599,487 & 34,217,455 & 34,217,455 \\

 & DRB-177 & 382M & 106M & 1080 & 31 & 47,124,525 & 16,760,905 & 21,147,601 & 21,147,601 \\

 & DRB-177 & 552M & 519M & 1799 & 31 & 220,163,418 & 119,866,791 & 89,582,245 & 89,582,245 \\

 & DRB-177 & 618M & 333M & 1881 & 31 & 142,820,683 & 80,362,747 & 55,364,923 & 55,364,923 \\

 & DRB-177 & 1.6B & 836M & 1963 & 31 & 347,911,303 & 198,868,896 & 144,947,035 & 144,947,035 \\

\hline
\multirow{12}{*}{DRB-56th}

 & DRB-062 & 193M & 72M & 166 & 55 & 36,343,511 & 35,982,221 & 755 & 755 \\

 & DRB-105 & 134M & 46M & 2,849 & 111 & 9,375,994 & 9,375,741 & 14,098,575 & 14,098,575 \\

 & DRB-106 & 134M & 68M & 2,284 & 111 & 34,064,016 & 6,297,656 & 14,098,676 & 14,098,676 \\

 & DRB-110 & 120M & 35M & 112 & 58 & 765 & 5,000,167 & 15,000,382 & 15,000,382 \\

 & DRB-122 & 112M & 1.8K & 110 & 55 & 770 & 220 & 385 & 385 \\

 & DRB-123 & 112M & 77M & 778 & 56 & 35,000,781 & 14,000,276 & 14,000,389 & 14,000,389 \\

 & DRB-155 & 50M & 12M & 181 & 58 & 1296 & 307 & 6,000,538 & 6,000,538 \\

 & DRB-176 & 90M & 49M & 6802 & 111 & 22,550,856 & 10,755,451 & 8,078,107 & 8,078,107 \\

 & DRB-176 & 341M & 348M & 8946 & 111 & 160,758,955 & 76,630,057 & 55,365,283 & 55,365,283 \\

 & DRB-176 & 1.6B & 900M & 10164 & 111 & 406,628,634 & 204,003,136 & 144,947,395 & 144,947,395 \\

 & DRB-177 & 90M & 43M & 3421 & 111 & 19,743,274 & 7,490,436 & 8,078,107 & 8,078,107 \\

 & DRB-177 & 618M & 326M & 5541 & 111 & 139,688,872 & 75,806,547 & 55,365,283 & 55,365,283 \\

\hline
\multirow{46}{*}{HPCBench}

 & graph500-16th & 171M & 81M & 113,732 & 16 & 76,413,805 & 4,799,533 & 2,526 & 2,526 \\

 & graph500-56th & 172M & 82M & 119,601 & 56 & 77,444,355 & 5,086,862 & 26,472 & 26,472 \\

 & HPCCG-16th & 228M & 55M & 9,547 & 16 & 50,028,952 & 5,798,966 & 2,199 & 2,199 \\

 & HPCCG-56th & 230M & 79M & 15,083 & 56 & 72,511,642 & 6,836,516 & 17,722 & 17,722 \\

 & DC.S-16th & 12M & 1.0K & 102 & 18 & 338 & 409 & 132 & 132 \\

 & DC.S-56th & 12M & 19.8K & 633 & 57 & 13,917 & 4,731 & 524 & 524 \\

 & IS.W-16th & 153M & 48M & 64,597 & 16 & 31,324,449 & 17,038,722 & 384 & 384 \\

 & IS.W-56th & 300M & 140M & 202,142 & 56 & 119,664,913 & 20,347,145 & 4,400 & 4,400 \\

 & loopA.bad-16th & 113M & 93M & 150,033 & 16 & 78,322,852 & 15,149,966 & 825 & 825 \\

 & loopA.bad-56th & 394M & 334M & 550,125 & 56 & 279,145,300 & 55,550,150 & 9,436 & 9,436 \\

 & loopA.solu1-16th & 193M & 93M & 150,049 & 16 & 78,327,081 & 15,151,468 & 1,325 & 1,325 \\

 & loopA.solu1-56th & 674M & 334M & 550,181 & 56 & 279,184,949 & 55,555,652 & 22,485 & 22,485 \\

 & loopA.solu2-16th & 96M & 51M & 10,049 & 16 & 50,165,397 & 1,011,384 & 618 & 618 \\

 & loopA.solu2-56th & 337M & 171M & 10,180 & 56 & 170,601,797 & 1,015,606 & 11,858 & 11,858 \\

 & loopA.solu3-16th & 96M & 51M & 10,048 & 16 & 50,167,481 & 1,011,381 & 882 & 882 \\

 & loopA.solu3-56th & 337M & 171M & 10,177 & 56 & 170,604,653 & 1,015,599 & 10,840 & 10,840 \\

 & loopB.solu1-16th & 113M & 93M & 150,037 & 16 & 78,322,094 & 15,150,074 & 433 & 433 \\

 & loopB.solu1-56th & 394M & 334M & 550,127 & 56 & 279,149,802 & 55,550,254 & 11,324 & 11,324 \\

 & Mandelbrot-16th & 116M & 112M & 1,973 & 16 & 112,231,695 & 2,708 & 117 & 117 \\

 & Mandelbrot-56th & 116M & 114M & 2,196 & 56 & 114,434,245 & 3,096 & 448 & 448 \\

 & Pi-16th & 150M & 96M & 53 & 16 & 50,000,252 & 46,875,100 & 99 & 99 \\

 & Pi-56th & 150M & 99M & 184 & 56 & 50,001,048 & 49,107,502 & 420 & 420 \\

 & QuickSort-16th & 134M & 41M & 91,092 & 16 & 32,609,684 & 8,417,535 & 522 & 522 \\

 & QuickSort-56th & 134M & 41M & 91,172 & 56 & 32,612,404 & 8,417,695 & 1,882 & 1,882 \\

 & fft6-16th & 146M & 0.9K & 81 & 16 & 561 & 145 & 108 & 108 \\

 & fft6-56th & 146M & 2.4K & 172 & 55 & 1,234 & 323 & 394 & 394 \\

 & LUReduction-16th & 136M & 45M & 89,116 & 16 & 35,960,758 & 9,044,179 & 480 & 480 \\

 & LUReduction-56th & 137M & 45M & 89,209 & 56 & 36,002,402 & 9,044,365 & 15,309 & 15,309 \\

 & MD-16th & 204M & 118M & 1,515 & 16 & 113,558,997 & 5,224,843 & 522 & 522 \\

 & MD-56th & 204M & 120M & 1,747 & 56 & 115,223,309 & 5,466,221 & 3,629 & 3,629 \\

 & testPath-16th & 30M & 7M & 25,048 & 17 & 2,647,876 & 3,556,911 & 631,221 & 631,221 \\

 & testPath-56th & 37M & 10M & 69,308 & 57 & 4,219,733 & 3,671,124 & 1,303,636 & 1,303,636 \\

 & fft-16th & 496M & 78M & 2,424,886 & 17 & 53,084,901 & 25,690,379 & 159 & 159 \\

 & fft-56th & 496M & 83M & 2,565,429 & 56 & 55,988,645 & 27,188,775 & 589 & 589 \\

 & fft-56th & 2.1B & 363M & 10,261,235 & 56 & 244,470,405 & 119,014,191 & 586 & 586 \\

 & qsomp1-16th & 107M & 674.6K & 283 & 17 & 203,340 & 26,123 & 222,546 & 222,546 \\

 & qsomp1-56th & 107M & 540.1K & 157 & 57 & 143,062 & 6,387 & 195,286 & 195,286 \\

 & qsomp2-16th & 108M & 870.5K & 376 & 17 & 261,250 & 32,622 & 288,285 & 288,285 \\

 & qsomp2-56th & 107M & 629.6K & 254 & 57 & 177,428 & 14,801 & 218,671 & 218,671 \\

 & qsomp3-16th & 142M & 15M & 33 & 17 & 3,929,700 & 51 & 5,894,435 & 5,894,434 \\

 & qsomp3-56th & 115M & 4M & 113 & 57 & 1,009,842 & 171 & 1,514,348 & 1,514,347 \\

 & qsomp4-16th & 164M & 19M & 37 & 17 & 4,782,343 & 14,827 & 7,137,390 & 7,137,390 \\

 & qsomp4-56th & 114M & 6M & 5,350 & 57 & 2,707,676 & 518,055 & 1,863,535 & 1,863,535 \\

 & qsomp6-56th & 107M & 507.2K & 537 & 57 & 186,521 & 44,137 & 138,229 & 138,229 \\

 & qsomp7-16th & 89M & 44M & 8,035 & 16 & 44,130,252 & 304,813 & 123 & 123 \\

 & qsomp7-56th & 296M & 147M & 6,119 & 56 & 146,867,929 & 619,075 & 439 & 439 \\

\hline
\multirow{7}{*}{misc}

 & biojava-4th & 221M & 0.9K & 9 & 12 & 59 & 24 & 383 & 383 \\

 & cassandra-132th & 259M & 28M & 9,839 & 12,211 & 2,686,843 & 1,672,488 & 12,246,313 & 12,246,313 \\

 & graphchi-20th & 216M & 206.3K & 144 & 15 & 204,879 & 719 & 344 & 344 \\

 & hsqldb-44th & 19M & 647.5K & 318 & 51 & 260,630 & 52,096 & 167,362 & 167,362 \\

 & tradebeans-222th & 39M & 218.9K & 778 & 674 & 71,344 & 30,605 & 58,338 & 58,338 \\

 & tradesoap-221th & 39M & 218.6K & 775 & 672 & 71,264 & 30,541 & 58,266 & 58,266 \\

 & zxing-15th & 547M & 18M & 3,310 & 359 & 18,389,530 & 9,748 & 3,624 & 3,624 \\

\hline

\end{longtable}

\twocolumn